\documentclass[10pt,journal,compsoc]{IEEEtran}
\usepackage{listings}
\usepackage{gnuplot-lua-tikz}
\usepackage{subcaption}
\usepackage{algorithm}
\usepackage{algpseudocode}
\usepackage{hyperref}
\usepackage{enumitem}
\usepackage{amssymb}
\usepackage{authblk}
\usepackage{cite}

\definecolor{lbcolor}{rgb}{0.9,0.9,0.9}

\newlength{\saveparindent}
\newlength{\saveparskip}
\algblock{ParFor}{EndParFor}
\algnewcommand\algorithmicparfor{\textbf{parfor}}
\algnewcommand\algorithmicpardo{\textbf{do}}
\algnewcommand\algorithmicendparfor{\textbf{end\ parfor}}
\algrenewtext{ParFor}[1]{\algorithmicparfor\ #1\ \algorithmicpardo}
\algrenewtext{EndParFor}{\algorithmicendparfor}
\algrenewcommand\algorithmicindent{1.0em}
\let\oldReturn\Return
\renewcommand{\Return}{\State\oldReturn}

\begin{document}

\title{clusterNOR: A NUMA-Optimized Clustering Framework}
\author[1]{\rm Disa Mhembere}
\author[2]{\rm Da Zheng}
\author[3]{\rm Carey E. Priebe}
\author[4]{\rm Joshua T. Vogelstein}
\author[1]{\rm Randal Burns}
\affil[1]{Department of Computer Science, Johns Hopkins University}
\affil[2]{Amazon Inc.}
\affil[3]{Department of Applied Mathematics and Statistics, Johns Hopkins University}
\affil[4]{Institute for Computational Medicine, Department of Biomedical Engineering, Johns Hopkins University}

\maketitle

\IEEEpubid{\begin{minipage}{\textwidth}\ \\[12pt] \centering
\copyright 2021 IEEE. Personal use of this material is permitted.
    Permission from IEEE must be obtained for all other uses, in any current or
    future media, including reprinting/republishing this material for
    advertising or promotional purposes, creating new collective works,
    for resale or redistribution to servers or lists, or reuse of any
    copyrighted component of this work in other works.
\end{minipage}}

\begin{abstract}
Clustering algorithms are iterative and have complex data access patterns that
result in many small random memory accesses. Also, the performance of parallel
implementations suffer from synchronous barriers for each iteration and skewed
workloads. We rethink the parallelization of clustering for modern non-uniform
memory architectures (NUMA) to maximize independent, asynchronous computation.
We eliminate many barriers, reduce remote memory accesses, and increase cache reuse.
\textit{Clustering NUMA Optimized Routines} (\textsf{clusterNOR}) is an open-source framework
that generalizes the \textsf{knor} library for k-means clustering,
providing a uniform programming interface and expanding the scope to hierarchical
and linear algebraic algorithms. The algorithms share the Majorize-Minimization or
Minorize-Maximization (MM) pattern of computation.

We demonstrate nine modern clustering algorithms that have simple implementations
that run in-memory, with semi-external memory, or distributed.
For algorithms that rely on Euclidean distance, we develop a relaxation of
Elkan's triangle inequality algorithm that uses asymptotically less memory and halves runtime.
Our optimizations produce an order of magnitude performance improvement over
other systems, such as Spark's MLlib and Apple's Turi.

\end{abstract}


\IEEEpeerreviewmaketitle

\section{Introduction}

Clustering is a fundamental task in exploratory data analysis
and machine learning and, as datasets grow in size, scalable
and parallel algorithms for clustering emerge as a critical
capability in science and industry. Clustering is an unsupervised machine
learning task for datasets that contain no pre-existing training labels.
For example, user recommendation systems at Netflix rely on clustering
\cite{netflixprize}. Partitioning multi-billion data points
is a fundamental task for targeted advertising in organizations
such as Google \cite{cfgoogle} and Facebook
\cite{facebookanatomy}.
Clustering is widely used in the sciences.
Genetics uses clustering to infer relationships between
similar species \cite{genetic1, genetic2}.
In neuroscience, connectomics \cite{conn1, conn2, conn3}
groups anatomical regions by structural, physiological, and
functional similarity.

Although algorithms pursue diverse objectives,
most clustering algorithms follow the Majorize-Minimization or
Minorize-Maximization (MM) \cite{MM} pattern of computation.
MM algorithms optimize a surrogate function in order to minorize or majorize the
true objective function. MM algorithms have two steps that are separated by a
synchronization barrier.
In the MM pattern, the raw data are not modified. Data are processed
continuously in iterations that modify algorithmic metadata only.
Different algorithms cluster based on centroids, density, distribution,
or connectivity and generate very different clusterings.  Yet, because
they all follow the MM pattern, we implement a common framework
for parallelization, distribution and memory optimization that applies
to algorithms of all types. We optimize the core computation
kernel for MM algorithms described in Section \ref{sec:apps}. We derive this
computation kernel by generalizing the computation pattern we develop
for k-means \cite{knor} and apply this kernel to MM algorithms.

We leverage the data access and computing patterns common to MM algorithms
to overcome the fundamental performance problems faced by tool builders.
These are:
\textit{(i)} reducing the cost of the synchronization barrier between the MM steps,
\textit{(ii)} mitigating the latency of data movement through the memory hierarchy, and
\textit{(iii)} scaling to arbitrarily large datasets.
Fully asynchronous computation of both MM steps is infeasible
because each MM step updates global state.
The resulting global barriers pose a major challenge to the performance
and scalability of parallel and distributed implementations.

Popular frameworks \cite{graphlab, mahout, mllib}
emphasize scaling-out computation to the distributed
setting, neglecting to fully utilize the resources within each machine.
Data are partitioned among cluster nodes and global updates are transmitted
at the speed of the interconnect.

Our system, Clustering NUMA Optimized Routines or \textsf{clusterNOR},
maximizes resources for scale-up computation on shared-memory
multicore machines. This enables \textsf{clusterNOR} to reduce network traffic and perform
fine-grained synchronization. \textsf{clusterNOR} generalizes and expands the
core capabilities of the \textsf{knor} \cite{knor} library for k-means clustering.
\textsf{clusterNOR} supports distributed computing, but prefers to maximize
resource utilization on a single node before distribution.
Recent findings \cite{McSherry15, flashgraph, graphene},
show that the largest graph analytics tasks can be done on a small fraction of
the hardware, at less cost, and as fast on a single shared-memory node as
they can on a distributed system.
Our findings reveal that clustering has the same structure. Applications
(Section \ref{sec:apps}) are benchmarked on a
single or few machines to minimize network bottlenecks.
Single node performance (with SSDs) outperforms competitor's distributed
performance in most instances.

\subsection*{Contributions}

\textsf{clusterNOR}
improves the runtime performance of unsupervised machine learning MM algorithms.
\textsf{clusterNOR}'s major contributions are:
\begin{itemize}
    \item An extensible open-source generalized framework, C++ API,
        and R package for utilization and the development of new pre-optimized
        MM algorithms.
    \item A NUMA-aware clustering library capable of operating
    (i) in-memory, (ii) in semi-external memory, and (iii) in distributed memory.
        \textsf{clusterNOR} scales to billions of data points and improves
        performance by up to $100$x compared to popular frameworks.
    \item The minimal triangle inequality (MTI) distance computation pruning algorithm.
    	MTI relaxes Elkan's triangle inequality (TI) algorithm.
	MTI reduces the memory increment required to
        only $O(n)$ compared with TI at $O(nk)$. MTI reduces runtime by $50\%$
        or more on real world large-scale datasets despite performing $2X$
        more distance computations than TI.
\end{itemize}

\subsection*{Manuscript Organization}

Section \ref{sec:relwork} introduces frameworks tackling
clustering, and those addressing optimizing computation on NUMA architectures.
Section \ref{sec:parmm} describes computational and algorithmic advancements
\textsf{clusterNOR} introduces to perform clustering on modern multicore
NUMA machines.
Next, we review core algorithms implemented within \textsf{clusterNOR} in Section
\ref{sec:apps}. Sections \ref{sec:im-design} - \ref{sec:dist-design} focus on
design points developed to optimize in-memory, semi-external memory,
distributed clustering for both hierarchical and
non-hierarchical algorithms. In Section \ref{sec:eval}, we first conduct an
in-depth evaluation of the performance and scalability of \textsf{clusterNOR}'s
k-means in comparison to other frameworks.
Section \ref{subsec:apps-eval} evaluates the $9$ clustering algorithms implemented within
\textsf{clusterNOR}. We conclude with a discussion (Section \ref{sec:discuss})
and API description (Appendix \ref{sec:api}).

\section{Related Work} \label{sec:relwork}

Mahout \cite{mahout} is a machine learning library that combines canopy
(pre-)clustering \cite{canopyclust} alongside MM algorithms to
cluster large-scale datasets.
Mahout relies on Hadoop! an open source implementation of
MapReduce \cite{mapreduce} for parallelism and scalability.
Map/reduce allows for effortless scalability and parallelism, but offers little
flexibility in how to achieve either.
As such, Mahout is subject to load imbalance or skew in the second MM phase.
The skew occurs because data are shuffled to a smaller number of processors
than are designated for computation.

MLlib is a machine learning library for Spark \cite{spark}.
Spark imposes a functional paradigm to parallelism.
MLlib's performance is highly coupled with
Spark's ability to efficiently parallelize computation using the generic data
abstraction of the resilient distributed datasets (RDD) \cite{rdd}. The in-memory
data organization of RDDs does not currently account for NUMA architectures,
but many of the NUMA optimizations that we develop could be applied to RDDs.

Popular machine learning libraries, such as Scikit-learn \cite{sklearn},
ClusterR \cite{clusterR}, and mlpack \cite{mlpack}, support a variety of clustering
algorithms.  These frameworks perform computation on a single machine,
often serially,
without the capacity to distribute computation to the cloud or perform
computation on data larger than a machine's main-memory.
\textsf{clusterNOR} presents a lower-level API that allows users to distribute and
scale many algorithms.

Several distance computation pruning algorithms for k-means exist
\cite{hamerly2015accelerating}. Algorithms tradeoff pruning efficacy
for memory usage and runtime. Elkan developed a popular and effective pruning
algorithm, triangle inequality (TI) with bounds\cite{triineq}.
TI reduces the number of distance computations in
k-means to fewer than $\mathcal{O}(kn)$ per iteration.
The method relies on a sparse lower bound matrix of size $\mathcal{O}(kn)$.
We present the {\it minimal} triangle inequality (MTI) algorithm
that is nearly as effective as TI, but only uses $\mathcal{O}(n)$ memory,
which makes it practical for use with large-scale data.

The semi-external memory (SEM) optimizations we implement are inspired by
FlashGraph \cite{flashgraph} and implemented using the same techniques for
asynchronous I/O and overlapped computation.
FlashGraph is a SEM graph computation framework that places edge data on SSDs
and allows user-defined vertex state to be held in memory.
FlashGraph runs on top of a userspace filesystem called SAFS \cite{safs}
that merges independent I/O requests into larger transfers and
manages a \textit{page cache} to reduce I/O.
SEM computation allows \textsf{clusterNOR} to capitalize on cheap commodity SSDs
to inexpensively scale machine learning applications, often outperforming
larger, more expensive clusters of machines.

Modern multi-socket NUMA machines deliver high throughput and low latency to CPUs
local to a NUMA node while penalizing remote memory accesses to non-local nodes.
Column store databases \cite{psaroudakis2015scaling} capitalize on NUMA topologies to design
efficient main-memory allocation and scheduling policies at runtime. They
demonstrate that for column-stores with one or more tables in a multi-tenant
environment, static partitioning and task stealing is insufficient.
A scheduler that monitors resource usage and resolves imbalance \cite{psaroudakis2016adaptive}
through work stealing is needed when skew exists.
\textsf{clusterNOR} uses similar principles.

Asymsched \cite{asymsched} and AutoNUMA \cite{autonuma}
automatically reconstruct NUMA-sensitive memory policies for multithreaded
applications. AutoNUMA collocates threads that utilize a
shared region of main-memory. AutoNUMA configures several
properties that affect its efficacy. Tuning these configuration
parameters may need to be done for each individual application and performance
may even degrade if configured incorrectly. Asymsched
specifically addresses the asymmetric
bandwidth and latency properties of NUMA interconnects. Asymsched groups
threads that access the same regions of memory together then migrates memory
using custom accelerated in-kernel system calls. Kernel level instructions pose
a barrier to widespread adoption of a system. \textsf{clusterNOR} implicitly
implements high-level ideas within Asymsched and is highly adoptable
because it executes fulling in user space.

\section{Improving MM Algorithm Computation} \label{sec:parmm}

Many MM algorithms for clustering share the same computation pattern.
We isolate this pattern and optimize the core computation kernel underlying these algorithms.
To optimize the computation kernel we:
\begin{itemize}
\item Adopt a strategy to defer the synchronization barrier between the two M-steps (see Section \ref{algo})
\item Reduce the computation cost of the first M-step by algorithmically pruning computation (see Section \ref{sec:mte}).
\end{itemize}

We utilize the k-means algorithm described in Section \ref{apps:kmeans} to develop these optimizations.
The core computational kernel we optimize for k-means is then generalized for use in other MM algorithms
within \textsf{clusterNOR}.

\subsection{Notation} \label{notation}

Throughout the manuscript, we use the terms defined in Table
\ref{tbl:notation}. In a given iteration, $t$, we can cluster
any point, $\vec{v}$ into a cluster $\vec{c}^{\,t}$.
$\mathbf{d}$ denotes a generic dissimilarity metric. Additionally, we
interchangeably use $\mathbf{d}$ for Euclidean distance.

\begin{table}[!h]
\caption{Notation used throughout this manuscript}
\begin{center}
\footnotesize
\begin{tabular}{|c|c|}
\hline
    \textbf{Term} & \textbf{Definition}\\
\hline
$n$ & The number of points in the dataset\\
\hline
$P$ & The number of physical CPUs on a machine\\
\hline
$T$ & The number of threads of concurrent execution\\
\hline
$\vec{V}$ & Collection of data points with cardinality, $|\vec{V}| = n$\\
\hline
    $d$ & The dimensionality (\# features) of a data point\\
\hline
$\vec{v}$ & A $d$-dimensional vector in $\vec{V}$\\
\hline
$j$ & The number of iterations an algorithm performs\\
\hline
$t$ & The current iteration of an algorithm, $t \in \{0...j\}$\\
\hline
$\vec{c}^{\,t}$ & A $d$-dimension centroid vector at iteration $t$\\
\hline
$k$ & The number of clusters $\ni |\vec{C}^t| = k$\\
\hline
    $L$ & The number of hierarchical partitioning steps \\
\hline
    $B$ & The batch size of an iteration for an algorithm \\
\hline
    $r$ & The number of runs performed for an algorithm \\
\hline
$\mathbf{d}(\vec{v}, \vec{c}^{\,t})$ &
    A dissimilarity metric between any $\vec{v}$ and $\vec{c}^{\,t}$ \\
\hline
$\mathbf{d}$ & Euclidean distance,
    $\sqrt{\sum^n_{i=1} (\vec{v}_1 - \vec{c}^{\,t}_1)^2}$\\
\hline
$f(\vec{c}^{\,t} | t > 0)$ & Dissimilarity between, $\vec{c}^{\,t}$ \&
$\vec{c}^{\,t-1}$, i.e., $\mathbf{d}(\vec{c}^{\,t}, \vec{c}^{\,t-1})$ \\
\hline
\end{tabular}
\normalsize
\end{center}
\label{tbl:notation}
\end{table}

\subsection{Barrier Minimization}\label{algo}

\textsf{clusterNOR} minimizes synchronization barriers for algorithms in which
parts of the MM-steps can be performed simultaneously.
We generalize this pattern directly from the \textsf{knor}
library \cite{knor} in which
Lloyd's k-means algorithm \cite{lloyds} is modified, to \textit{$||$Lloyd's} by
merging the two MM-steps. \textsf{clusterNOR} employs read-only global data and
per-thread lock-free structures that are aggregated in user defined parallel
reduction procedures at the end of MM-steps. Barrier minimization
improves parallelism, at the cost of slightly higher memory consumption.
This strategy naturally leads to lock-free routines that require fewer synchronization barriers.

\subsection{Minimal Triangle Inequality (MTI) Pruning}\label{sec:mte}

We relax the constraints of Elkan's algorithm for triangle inequality pruning (TI)
\cite{triineq} and eliminate the lower
bound matrix of size $\mathcal{O}(nk)$. We tradeoff reduced pruning efficacy for
limited memory consumption and faster computation times.
Section \ref{sec:mti_vs_others} empirically demonstrates on real-world
data that MTI drastically reduces computation time in comparison to other pruning methods
while producing comparable pruning efficacy to TI.
While MTI consistently outperforms TI on large-scale real-world datasets by a factor of $2$X to
$5$X. MTI does so by minimizing time spent maintaining data structures that TI
requires for pruning distance computations.

With $\mathcal{O}(n)$ memory, MTI modifies and implements three of the
five \cite{triineq} pruning clauses performed by TI.
Let $u^t = \mathbf{d}(\vec{v}, \vec{c\_nearest}^t) + f(\vec{c\_nearest}^t)$,
be the upper bound
of the distance of a sample, $\vec{v}$, in iteration $t$
from its assigned cluster $\vec{c\_nearest}^t$.
Finally, we define $U$ to be an update function such that
$U(u^t)$ fully tightens the upper bound of $u^t$ by computing
$\mathbf{d}(\vec{v}, \vec{c\_nearest}^{t+1})$.

\begin{itemize}[label={},leftmargin=*]
\addtolength{\itemsep}{3pt}
\item\textbf{Clause 1:} if $u^t \leq \min \mathbf{d}(\vec{c\_nearest}^t,
    \vec{c}^{\,t} \, \forall \, \vec{c}^{\,t} \in \vec{C}^t)$,
		then $\vec{v}$ remains in the same cluster for the current iteration.
		For semi-external memory, this is significant because no I/O
		requests are made for data.

\item\textbf{Clause 2:} if $u^t \leq \mathbf{d}(\vec{c\_nearest}^t,
		\vec{c}^{\,t} \, \forall \, \vec{c}^{\,t} \in \vec{C}^t)$,
		then the distance computation between data point $\vec{v}$ and
		centroid $\vec{c}^{\,t}$ is pruned.

\item\textbf{Clause 3:} if $U(u^t) \leq \mathbf{d}(\vec{c\_nearest}^t,
		\vec{c}^{\,t} \, \forall \, \vec{c}^{\,t} \in \vec{C}^t)$,
		then the distance computation between data point $\vec{v}$ and
		centroid $\vec{c}^{\,t}$ is pruned.
\end{itemize}

\section{Applications (Algorithm Implementation)} \label{sec:apps}

We implement $9$ algorithms that demonstrate the utility,
extensibility and performance of \textsf{clusterNOR}. We show the
flexibility of the framework and generalized computation model that is exposed
via the C++ API (Appendix \ref{sec:api}). We include a code example in the Appendix
\ref{subsec:code} for the G-means algorithm described in
Section \ref{apps:gmeans}. Table \ref{tbl:complex} summarizes the computational and
memory complexities of all applications within the \textsf{clusterNOR} library.
Complexities are of serial implementations and do not account for
additional state or computation needed to optimize parallel performance.

\begin{table}[!htb]
    \caption{Per-iteration memory and computation complexities for algorithms.
    Brief explanations are provided where algorithms are described.}
\vspace{-15pt}
\begin{center}
\footnotesize
\begin{tabular}{|c|c|c|}
\hline
    \textbf{Algorithm} & \textbf{Memory Complexity} & \textbf{Computation
    Complexity}\\
\hline
    k-means & $\mathcal{O}(nd+kd)$ & $\mathcal{O}(knd)$ \\
\hline
    sk-means & $\mathcal{O}(nd+kd)$ & $\mathcal{O}(knd + n)$ \\
\hline
    k-means++ & $\mathcal{O}(nd+kd)$ & $\mathcal{O}(kndr)$ \\
\hline
    mbk-means & $\mathcal{O}(nd+kd)$ & $\mathcal{O}(\frac{knd}{B})$ \\
\hline
    fc-means & $\mathcal{O}(2nd + 2kd))$  & $\mathcal{O}(2nkd + nk + n + k)$ \\
\hline
    k-medoids & $\mathcal{O}(nd+kd+n)$ & $\mathcal{O}(k^3 + nk)$ \\
\hline
    H-means & $\mathcal{O}(nd + 2Ld)$   & $\mathcal{O}(2ndL)$ \\
\hline
    X-means & $\mathcal{O}(nd + 2Ld)$   & $\mathcal{O}(2ndL + kn)$ \\
\hline
    G-means & $\mathcal{O}(nd + 2Ld)$   & $\mathcal{O}(2ndL + 4kn)$ \\
\hline
\end{tabular}
\normalsize
\end{center}
\label{tbl:complex}
\end{table}

\subsection{k-means} \label{apps:kmeans}

K-means is an iterative partitioning algorithm in which data, $\vec{V}$, are assigned
to one of $k$ clusters based on the Euclidean distance, $\mathbf{d}$, from
each of the cluster means $\vec{c}^{\,t} \in \vec{C}^t$. A serial implementation
requires memory of $\mathcal{O}(nd + kd)$; $\mathcal{O}(nd)$ for the dataset
with $n$ data points of $d$ dimensions, and $\mathcal{O}(kd)$ for the $k$
centroids of $d$ dimensions. The computation complexity of k-means
both serially and parallelized within \textsf{clusterNOR} remains
$\mathcal{O}(knd)$; $\mathcal{O}(n)$ $d$-dimensional data points compute
distances to $\mathcal{O}(k)$ $d$-dimensional centroids.
The asymptotic memory consumption of k-means within \textsf{clusterNOR}
is $\mathcal{O}(nd + Tkd + n + k^2)$.
The term $T$ arises from the per-thread centroids we maintain. Likewise, the
$\mathcal{O}(n + k^2)$ terms allow us to maintain a centroid-to-centroid
distance matrix and a point-to-centroid upper bound distance vector of size
$\mathcal{O}(n)$ that we use
for computation pruning as described in Section \ref{sec:mte}.
For SEM, the computation complexity remains unchanged, but
the asymptotic memory consumption drops to $\mathcal{O}(n + Tkd + k^2)$;
dropping the $\mathcal{O}(d)$ term to disk.
K-means minimizes the following objective
function for each data point, $\vec{v}$:

\begin{equation} \label{eqn:kmeans}
min \sum_{\vec{v}\in\vec{V}}||\mathbf{d}(\vec{v}, \vec{c}^{\,t})||
\end{equation}

\subsection{Spherical k-means (sk-means)} \label{apps:skmeans}

Spherical k-means (sk-means) \cite{skmeans} projects all data points, $\vec{V}$,
to the unit sphere prior
to performing the k-means algorithm. Unlike k-means, spherical k-means uses the
cosine distance function, $\mathbf{d}_{cos} = \frac{\vec{V} \cdot \vec{C}^t}
{||\vec{V}|| ||\vec{C}^t||}$, to determine data point to
centroid proximity.

\subsection{k-means++} \label{apps:kmeanspp}

k-means++ \cite{kmeanspp} is stochastic clustering algorithm that
performs multiple runs, $r$ of the k-means algorithm then selects the best run.
The best run corresponds to the run that produces the minimum squared euclidean
distance between a centroid and constituent cluster members. The k-means++
algorithm shares memory complexity with k-means, but runs $r$ times and thus
has increased computational complexity, compared to k-means, at $\mathcal{O}(kndr)$.
k-means++ chooses each new centroid $\vec{c}^{\,t}$ from the dataset through a
weighted random selection such that:

\begin{equation} \label{eqn:kmeanspp}
    \vec{C} \leftarrow \frac{{D(\vec{v})}^2}{\sum_{\vec{v} \in \vec{V}}
    {D(\vec{v})}^2},
\end{equation}

\noindent in which $D(\vec{v})$ is the minimum distance of a datapoint to the clusters
already chosen.

\subsection{Mini-batch k-means (mbk-means)} \label{apps:mbkmeans}

Lloyd's algorithm is often referred to as batched k-means because all data points
are evaluated in every iteration. Mini-batch k-means (mbk-means) \cite{mbkmeans}
incorporates random sampling into each iteration of k-means thus reducing the
computation cost of each iteration by a factor of $B$, the batch size, to
$\mathcal{O}(\frac{nkd}{B})$ per iteration. Furthermore, a parameter
$\eta = \frac{1}{\vec{C}^t}$ is computed per centroid to determine the
learning rate and convergence. Batching does not affect the memory requirements
of k-means when run in-memory. In the SEM setting, the memory requirement is
$\mathcal{O}(\frac{knd}{B})$, a reduction by a factor of $B$. Finally, the update
function is as follows:

\begin{equation} \label{eqn:mbk}
\vec{C}^t \leftarrow (1 - \eta)C^{t-1} + \eta\vec{V}
\end{equation}

\subsection{Fuzzy C-means (fc-means)} \label{apps:fcm}

Fuzzy C-means \cite{fcm} is an iterative `soft' clustering algorithm in which
data points can belong to multiple clusters by computing a degree of association
with each centroid. A fuzziness index, $z$, is a
hyper-parameter used to control the degree of fuzziness. fc-means shares a
memory complexity with k-means. Computationally, fc-means performs
$2nkd + nk + n + k$ operations per iteration \cite{fcm}.
As such, fc-means is significantly more computationally intensive than k-means
despite retaining an identical asymptotically complexity of $\mathcal{O}(nkd)$.

Fuzzy C-means computes $J$ an association matrix representing the strength of
connectivity of a data point to a cluster. $J \in \mathbb{R}^{n \texttt{x}k}$ :
\begin{equation}\label{eqn:fcmeans}
J = \sum_{i=1}^{|N|} \sum_{k=1}^{|C|} u_{ik}^z
		||\vec{v}_i - \vec{c}_j||^2 , 1 \leq z < \inf,
\end{equation}

\noindent in which $u_{ik}$ is the degree of membership of $\vec{v}_i$ in cluster $k$.

\subsection{k-medoids} \label{apps:kmedoids}

K-medoids is a clustering algorithm that uses data point feature-vectors
as cluster representatives (medoids), instead of centroids like k-means.
In each iteration, each cluster determines whether to choose another cluster
member as the medoid. This is commonly referred to as the \textit{swap} step
with complexity $\mathcal{O}(n^2d)$.
This is followed by an MM
step to determine cluster assignment for each data point given
the updated medoids, resulting in a complexity of $\mathcal{O}(n-k)^2$.
We reduce the computation cost by
implementing a sampled variant called (CLARA)
\cite{clara} that is more practical, but still has a high asymptotic
complexity of $\mathcal{O}(k^3 + nk)$. This bound is explicitly derived
in the work of Kaufman et al. \cite{clara}.

\subsection{Hierarchical k-means (H-means)} \label{apps:hmeans}

The hierarchically divisive k-means algorithm recursively partitions the dataset
 in each iteration until a user-defined convergence criteria is achieved.
The computation complexity for $L$ hierarchical levels of partitioning is
$\mathcal{O}(2ndL)$, in which the factor $2$ is derived from
performing k-means with $k = 2$ centroids for each partition/cluster.

\subsection{X-means} \label{apps:xmeans}

X-means \cite{xmeans} is a form of divisive hierarchical clustering in which
the number of clusters is not provided a priori. Instead, X-means
determines whether or not a cluster should be split using
Bayesian Information Criterion (BIC) \cite{bic}. Computationally,
it differs from H-means (Section \ref{apps:hmeans}) by an additional
$\mathcal{O}(kn)$ term in which a decision is taken on whether or not to split
after cluster membership is accumulated.

\subsection{Gaussian Means (G-means)} \label{apps:gmeans}

G-means is a hierarchical divisive algorithm similar to X-means in its
computation complexity and in that it does not require the number of
clusters $k$ as an argument. G-means mostly varies from X-means in that it uses
the Anderson-Darling statistic \cite{adstat} as the test to decide splits.
The Anderson-Darling statistic performs roughly four times more computations
than BIC, despite having the same asymptotic complexity.

\section {In-memory design} \label{sec:im-design}

We prioritize practical performance when we implement in-memory
optimizations. We make design tradeoffs to balance
the opposing forces of minimizing memory usage and maximizing
CPU cycles spent on parallel computing.

\vspace{5pt}

\begin{figure}[t]
\centering
\includegraphics[width=\linewidth]{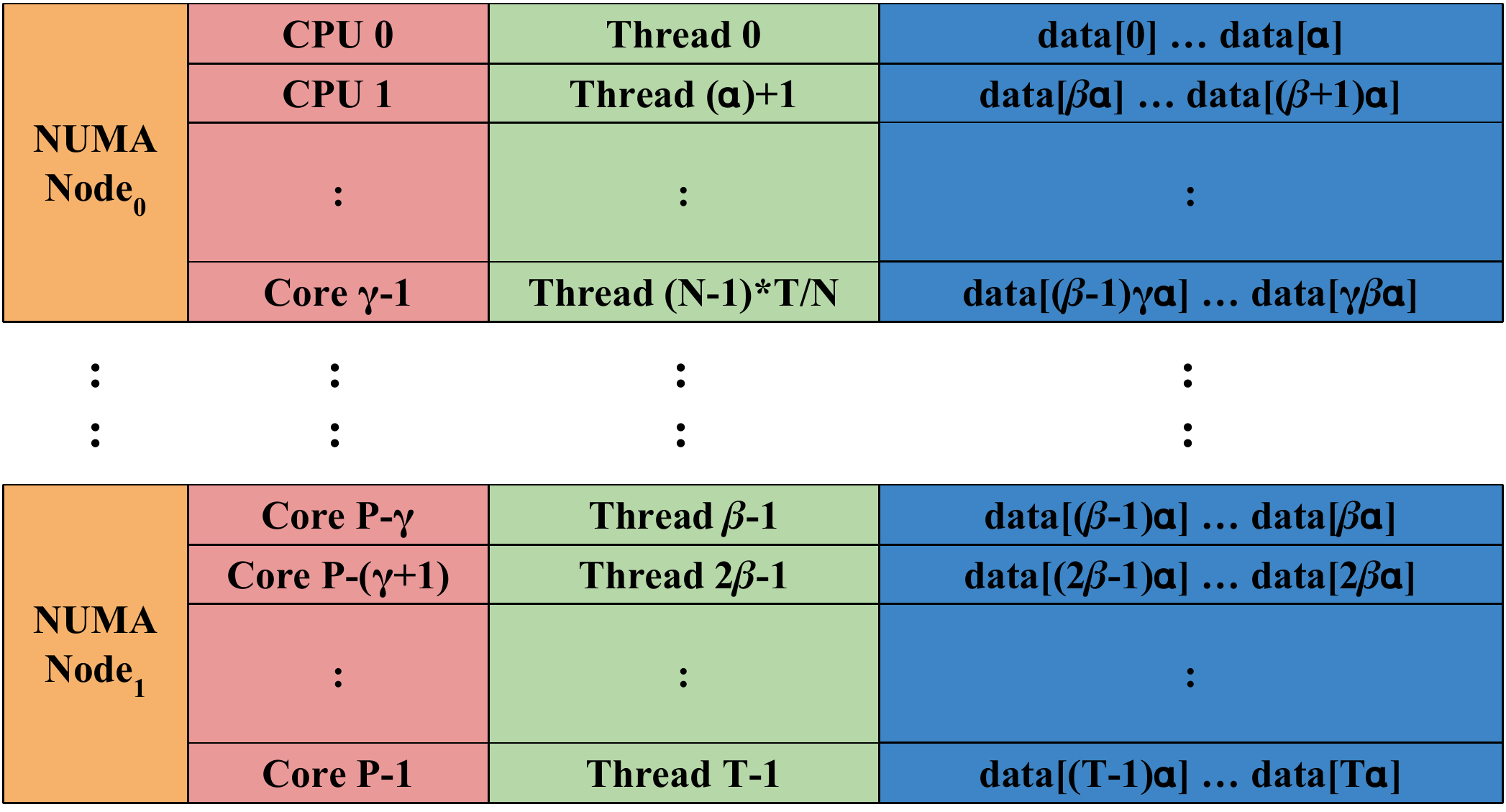}
\caption{The memory allocation and thread assignment scheme we employ.
    $\alpha = n/T$ is the amount
    of data per thread, $\beta = T/N$ is the
    number of threads per NUMA node, and $\gamma = P/N$ is the number of
    physical processors per NUMA node.
    Distributing memory across NUMA nodes maximizes memory throughput
    whereas binding threads to NUMA nodes reduces remote memory accesses.}
\label{fig:numa-mem}
\end{figure}

\textbf{Prioritize data locality for NUMA}:
To minimize remote memory accesses, we bind every thread to a single NUMA node,
equally partition the dataset across NUMA nodes, and sequentially allocate data
structures to the local NUMA node's memory. Every thread works
independently. Threads only communicate or share data to aggregate
per-thread state as required by the algorithm.
Figure \ref{fig:numa-mem} shows the data allocation and access scheme we employ.
We bind threads to NUMA nodes rather than specific CPU cores because the latter
is too restrictive to the OS scheduler. CPU thread-binding
may cause performance degradation if the number of worker threads exceeds
the number of physical cores.

\vspace{5pt}

\textbf{Customized scheduling and work stealing}:
clusterNOR
customizes scheduling for algorithm-specific
computation patterns. For example, Fuzzy C-means (Section \ref{apps:fcm})
assigns equal
work to each thread at all times, meaning it would not benefit
from dynamic scheduling and load balancing via work stealing.
As such, Fuzzy C-means invokes static scheduling. Conversely, k-means when
utilizing MTI pruning would result in skew without dynamic scheduling and
thread-level work stealing.

\begin{figure}[t]
\centering
\includegraphics[width=\linewidth]{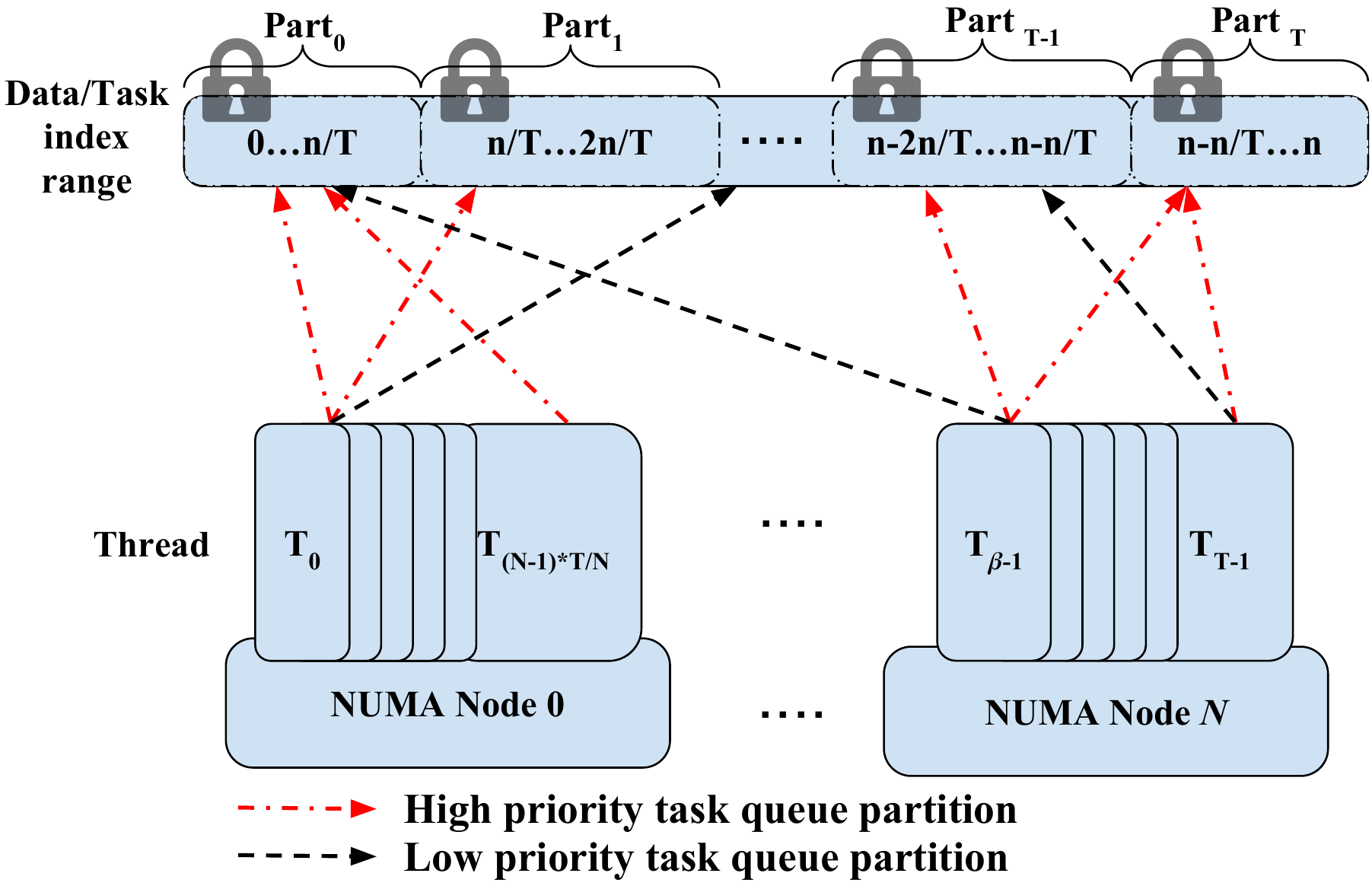}
\caption{The NUMA-aware partitioned task scheduler
 minimizes task queue lock contention and remote memory
accesses by prioritizing tasks with data in the local NUMA memory
bank.}
\label{fig:task-queue}
\vspace{-10pt}
\end{figure}

For dynamic scheduling, we develop a NUMA-aware partitioned priority task queue
(Figure \ref{fig:task-queue})
to feed worker threads, prioritizing tasks that maximize local memory access
and, consequently, limit remote memory accesses.
The task queue enables idle threads to \textit{steal} work from
threads bound to the
same NUMA node first, minimizing remote memory accesses. The queue is
partitioned into $T$ parts, each with a lock required for access.
We allow a thread to cycle through the task queue once looking for high priority
tasks before settling on another, possibly lower priority task.
This tradeoff avoids starvation and ensures threads are idle for negligible
periods of time. The result is good load balancing in addition to
optimized memory access patterns.

\vspace{5pt}
\textbf{Avoid interference and defer barriers}:
Whenever possible, per-thread data structures maintain mutable
state.
This avoids write-conflicts and elimiates locking.
Per-thread data are merged using an external-memory parallel reduction operator, much like
funnel-sort \cite{cache-obl},
when algorithms reach the end of an iteration or
the whole computation. For instance, in k-means, per-thread local centroids
contain running totals of their membership until an iteration ends when
they are finalized through a reduction.

\begin{figure}[t]
\centering
\includegraphics[width=\linewidth]{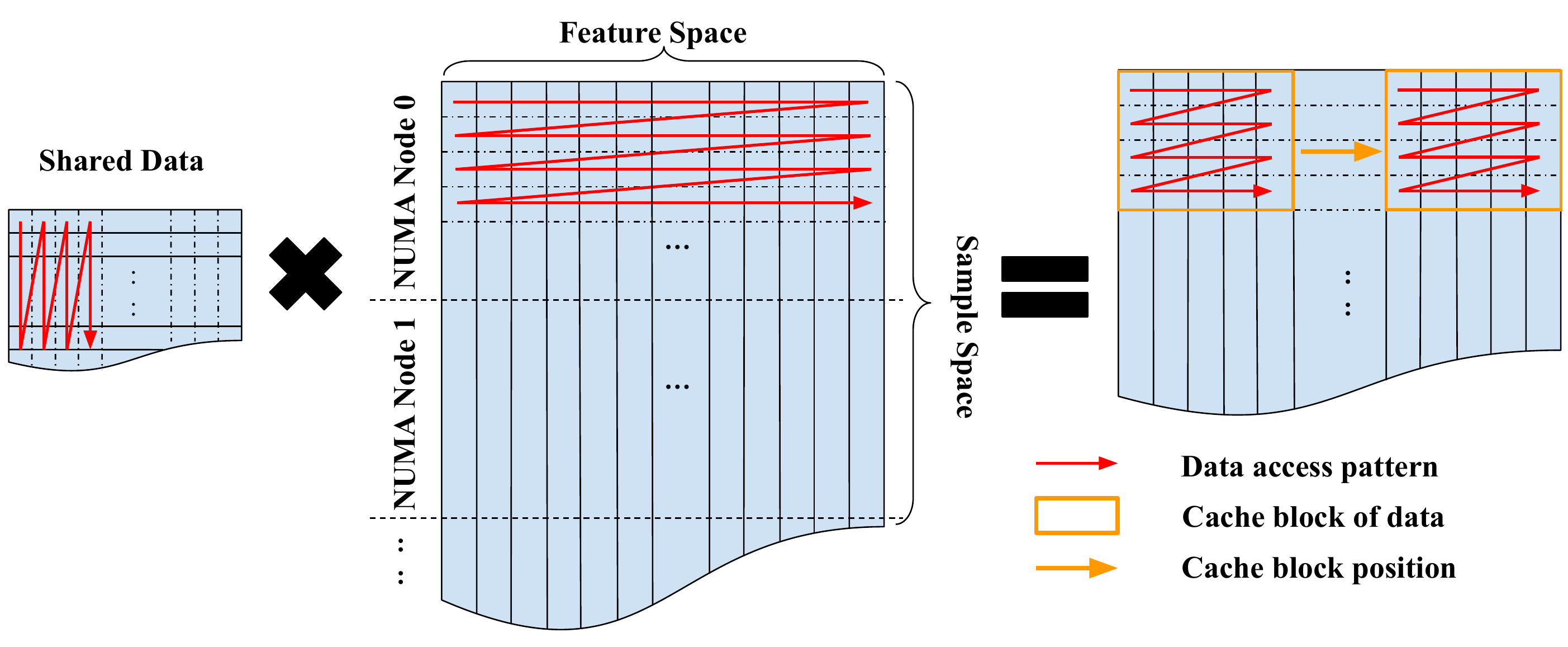}
\caption{Data access patterns support NUMA locality,
    utilize prefetched data well and
optimize cache reuse through a cache blocking scheme.}
\label{fig:numa-mult}
\vspace{-10pt}
\end{figure}

\vspace{5pt}
\textbf{Effective data layout for CPU cache exploitation and cache blocking}:
Both per-thread and
global data structures are placed in contiguously allocated chunks of memory.
Contiguous data organization and sequential access patterns
improve processor prefetching and cache line utilization. Furthermore,
we optimize access to both input and output data structures to improve performance.
In the case of a dot product operation (Figure \ref{fig:numa-mult}),
we access input data sequentially from the local NUMA memory and
write the output structure using a cache blocked scheme
for higher throughput reads and writes. The size of the block is determined based
on L1 and L2 cache specifications reported by the processor on a machine.
We utilize this optimization in Fuzzy C-means.

\subsection {Iterative Hierarchical Design} \label{sec:hclust-design}

\textsf{clusterNOR} exposes an iterative interface for recursive hierarchical algorithms.
Doing so enables \textsf{clusterNOR} to capitalize on the hardware by:

\begin{itemize}
\item eliminating stack duplication during recursive calls
\item localizing and sequentializing data access
\item improving the utility of prefetched data
\end{itemize}

\begin{figure}[!htb]
\centering
\footnotesize
\vspace{-5pt}
\begin{subfigure}{.5\textwidth}
	\includegraphics[width=\linewidth]{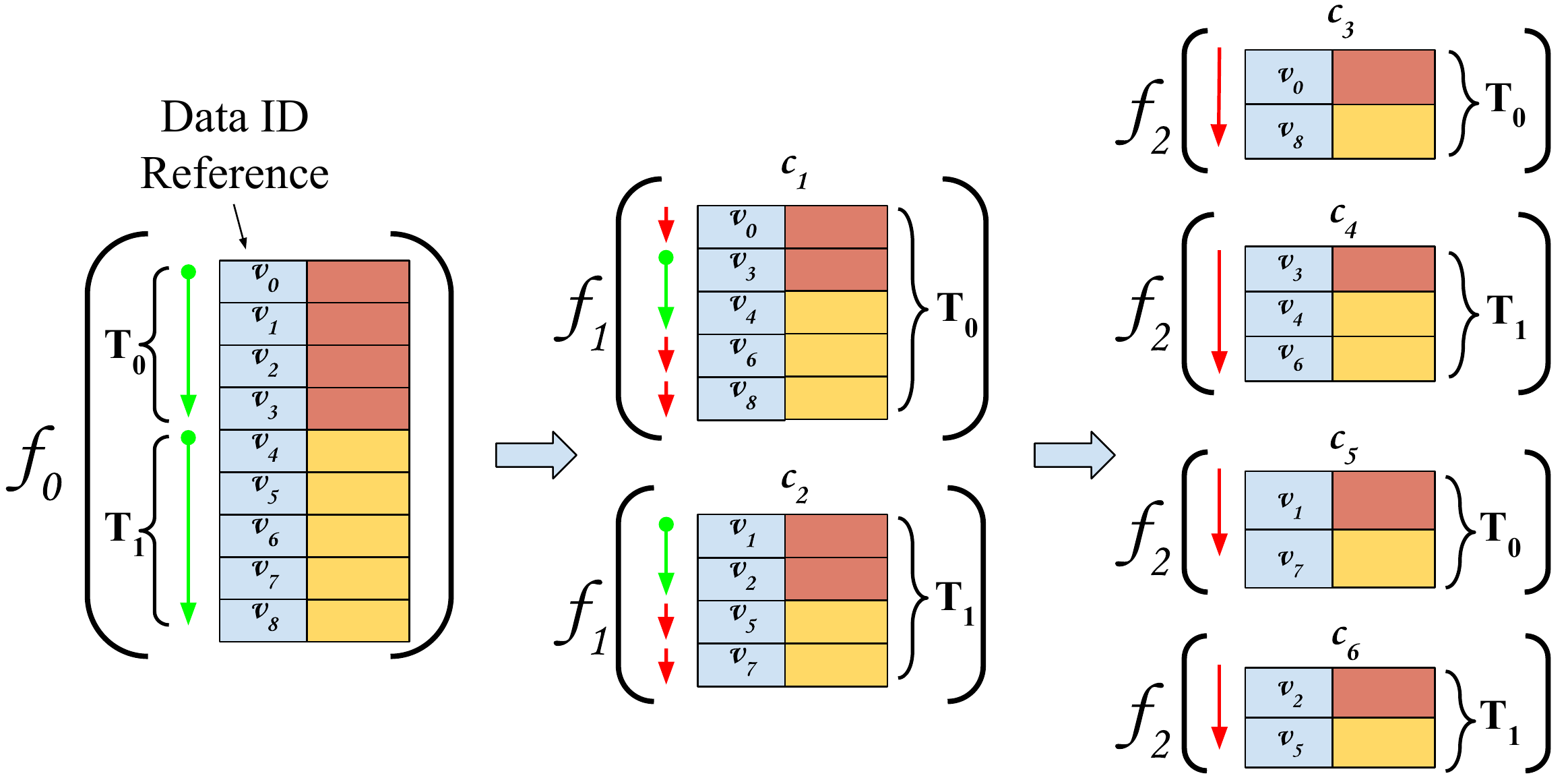}
\vspace{-10pt}
\caption{Na\"{\i}ve recursive parallel hierarchical clustering
exhibits poor data locality, and non-contiguous data access patterns.}
\label{fig:hclust-naive}
\end{subfigure}

\begin{subfigure}{.5\textwidth}
\includegraphics[width=\linewidth]{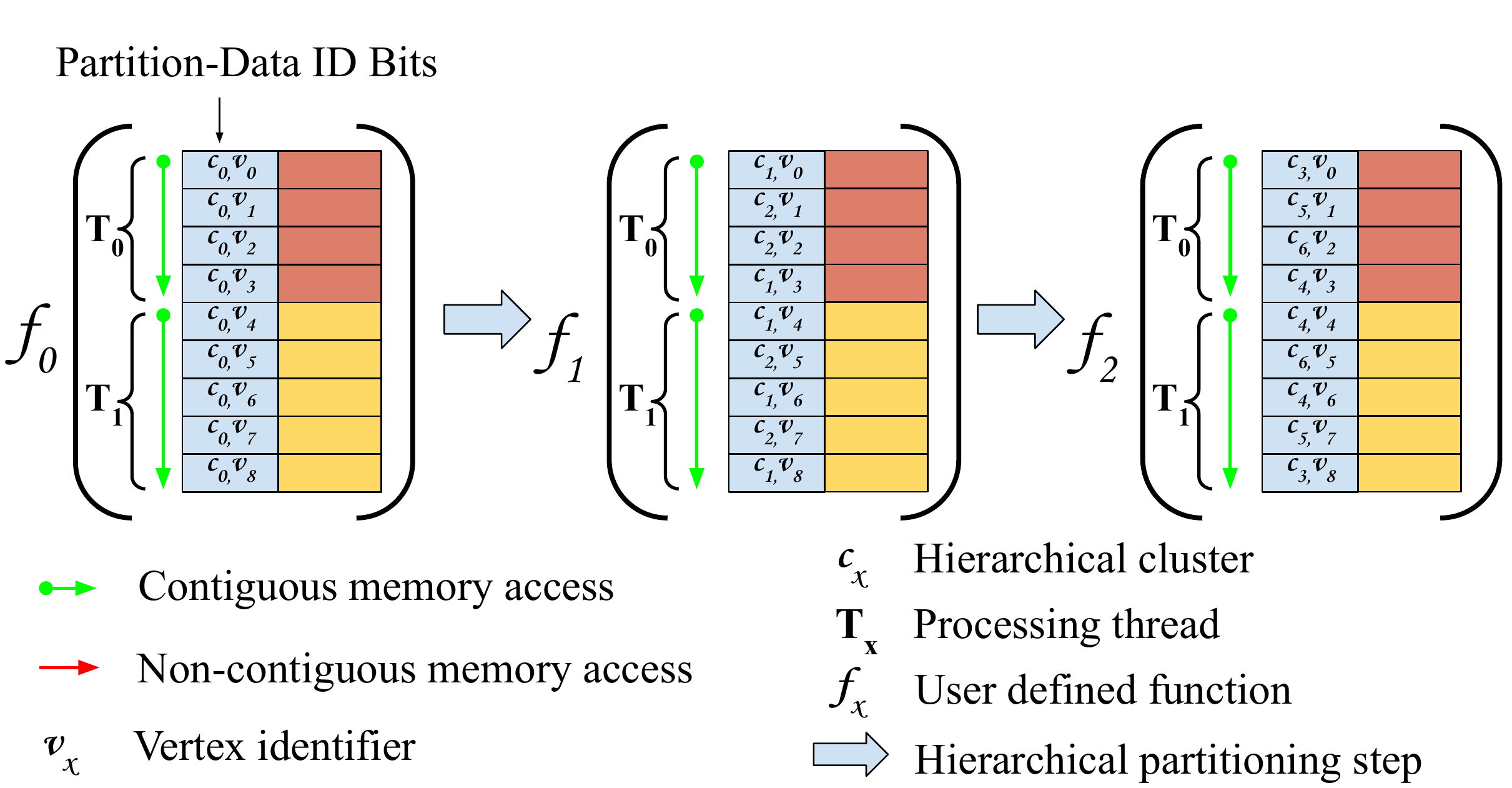}
\vspace{-10pt}
\caption{\textsf{clusterNOR} transparently provides NUMA-local,
sequential, and contiguous data access patterns.}
\label{fig:hclust-clusternor}
\end{subfigure}

\caption{A na\"{\i}ve hierarchical implementation with unfavorable
data access patterns compared to \textsf{clusterNOR}. \textsf{clusterNOR}
enforces sequential data access and maximizes cache reuse.}
\label{fig:hclust-design}
\end{figure}

Na\"{\i}ve implementations assign a thread to each cluster and shuffle data between levels of the hierarchy
 (Figure \ref{fig:hclust-naive}). This incurs a great deal of
remote memory access and non-contiguous I/O for each thread.
\textsf{clusterNOR} avoids these pitfalls by not shuffling data.
Instead, threads are assigned to contiguous regions of memory.
Figure \ref{fig:hclust-clusternor} shows the computation hierarchy in a
simple two thread computation.

\textsf{clusterNOR}'s design places computation barriers \textit{(i)} when the user
defined function, $f_x$, completes and \textit{(ii)} when clusters spawn at a
\textit{hierarchical partitioning step}.
New clusters spawn in a parallel block that only modifies membership metadata.
Computation is deferred until all clusters capable of spawning have done so.
Cluster assignment identifiers are encoded into the leading bits of the data
point identifier with a bitmap maintained to determine whether clusters have
algorithmically converged. Load balancing is performed via the
NUMA-aware scheduler (Figure \ref{fig:task-queue}).

\textsf{clusterNOR}'s design introduces a barrier at the hierarchical
partitioning step. We trade-off the introduction of this barrier for the ability to
increase memory throughput by issuing iterative contiguous tasks (blocks of
data points). A system recursively designed will incur stack creation overhead
upon spawning. Additionally, each spawned cluster will naturally operate on
increasingly smaller partitions of the data with no guarantee of contiguous
access for worker threads. Early results demonstrate at least $1.5$x
improvement in the performance of hierarchical clustering methods by
developing an iterative interface as compared to a recursive one.

\section{Semi-external Memory Design} \label{sec:sem-design}

The semi-external memory (SEM) model allows a dataset, $\vec{V} \in \mathbb{R}^{n \texttt{x} d}$
to utilize $\mathcal{O}(n)$ main-memory with the remaining $\mathcal{O}(nd)$ data in external storage devices.
External data are asynchronously streamed into main memory while out of order execution is performed.
The model reduces IO latency in comparison to a pure streaming model and reduces memory requirements
when compared to purely in-memory computation. SEM is well suited for dataset larger than main-memory
in which distributed solutions are traditionally employed.

FlashGraph \cite{flashgraph} is a SEM graph engine that supports asynchronous I/O and out of
order execution. FlashGraph targets \textbf{scale-up} computing on a multi-core NUMA machine.
We modify the FlashGraph kernel to support matrix-like computations for use within \textsf{clusterNOR}.
We find for some datasets single-node systems are faster than
distributed systems that use an order of magnitude more hardware.

We modify FlashGraph to integrate into \textsf{clusterNOR} by
altering FlashGraph's primitive data type, the \texttt{page\_vertex}.
The \texttt{page\_vertex} is interpreted as a vertex with an
index to the edge list of the \texttt{page\_vertex} on SSDs.
We define a \textit{row} of data to be equivalent to a $d$-dimension
data point, $\vec{v}_i$. Each row is composed of a unique
identifier, \textit{row-ID}, and $d$-dimension data vector, \textit{row-data}.
The row-data naturally replaces the adjacency list originally stored for FlashGraph
vertices.
We add a \texttt{page\_row} data type to FlashGraph and modify the asynchronous
I/O layer to support floating point row-data reads rather than the
numeric identifiers for graph adjacency lists.
The \texttt{page\_row} type computes its row-ID and row-data location on disk meaning
only user-defined state is stored in-memory.
The \texttt{page\_row} reduces the in-memory state necessary to use FlashGraph by $\mathcal{O}(n)$
because it does not store an index to data on SSDs unlike a
\texttt{page\_vertex}. This allows SEM applications to scale to larger
datasets than possible before on a single machine.

\subsection{I/O minimization} \label{sec:io-min-eff}

I/O bounds the performance of most well-optimized SEM applications.
Accordingly, we reduce the number of data-rows that need
to be brought into main-memory each iteration.
In the case of k-means, only Clause 1
of MTI (Section \ref{sec:mte}) facilitates the skipping of all
distance computations for a data point. Likewise for mini-batch k-means and
k-medoids that subsample the data, we perform selective IO in
each iteration. We observe the same phenomenon when data points have
converged in a cluster for H-means, G-means and X-means.
In these cases, we issue significantly fewer I/O requests but still
retrieve significantly more data than necessary from SSDs because pruning
occurs near-randomly and sampling pseudo-randomly.
Reducing the filesystem \textit{page size}, i.e. minimum read size
from SSDs alleviates this to an extent, but a small page size can lead to
a higher number of I/O requests, offsetting any gains achieved from
reduced fragmentation.
We utilize a minimum read size of $4$KB. Even with this
small value, we receive much more data from disk than we request.
To address this, we develop a lazily-updated
partitioned \textit{row cache}
that drastically reduces the amount of data brought into main-memory.

\subsubsection{Partitioned Row Cache (RC)} \label{sec:row-cache}

\begin{figure}[t]
\centering
\includegraphics[width=\linewidth]{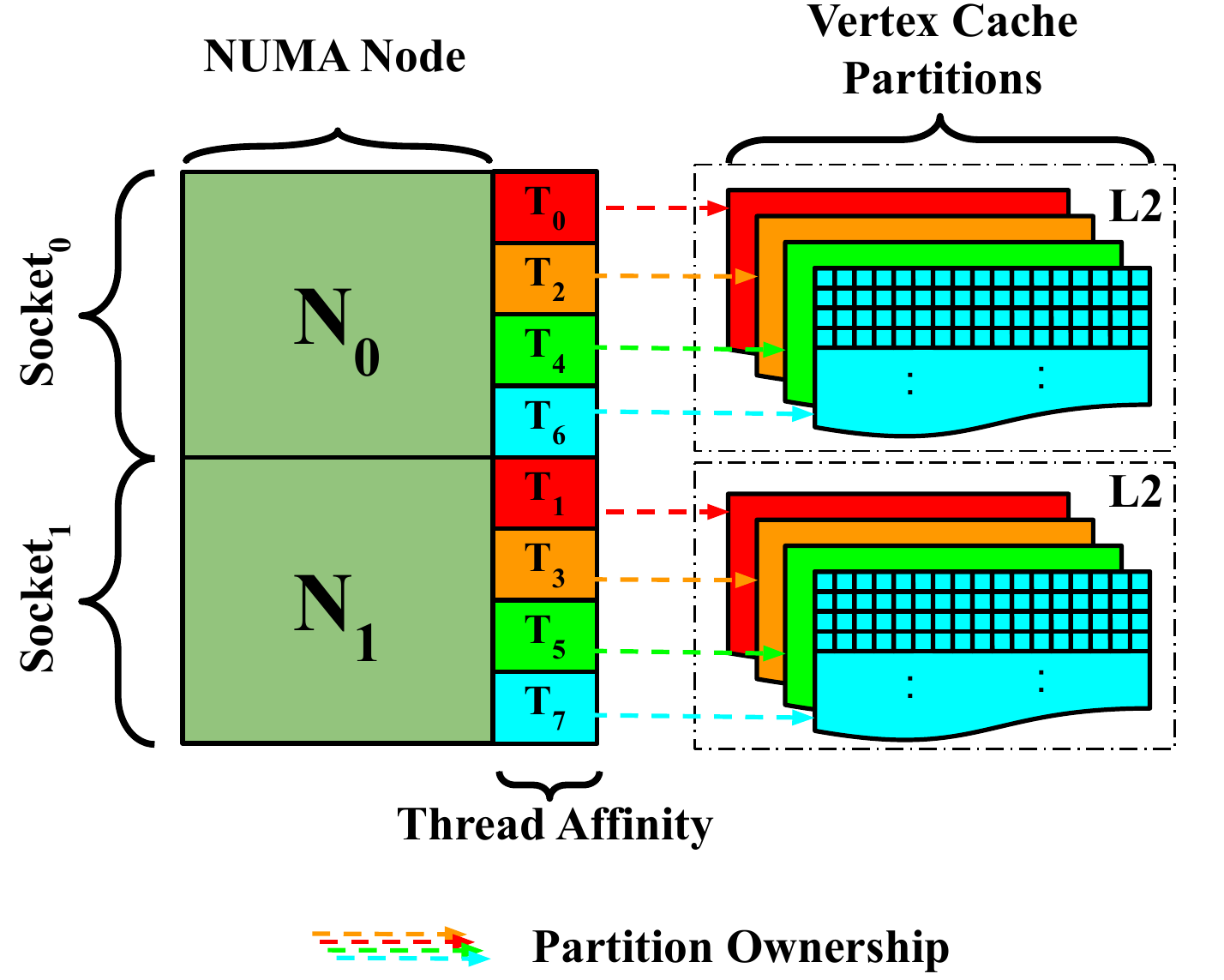}
\caption{The structure of the row cache for SEM applications in a typical
    two socket, two NUMA node machine utilizing $8$ threads on $8$ physical CPUs.
    Partitioning the row cache eliminates the need for locking during cache
    population. The aggregate size of all row cache partitions resides
    within the NUMA-node shared L2 cache.}
\label{fig:row-cache}
\vspace{-10pt}
\end{figure}

We add a layer to the memory hierarchy for SEM applications by designing a
lazily-updated row cache (Figure \ref{fig:row-cache}).
The row cache improves performance
by reducing I/O and minimizing I/O request merging and page
caching overhead in FlashGraph. A row is \textit{active} when
it performs an I/O request in the current iteration for its row-data.
The row cache places active rows to main-memory at the granularity of a row,
rather than a page, improving its effectiveness in reducing I/O compared
to a page cache.  The row cache is managed at the granularity of an entry (row)
at a time rather than a collection (page) of such entries. This is the fundamental
difference between the row cache and the page cache or similarly, buffer pools in
relation database management systems.

We partition the row cache into as many partitions as FlashGraph creates for the
underlying matrix, typically equal to the number of threads of execution.
Each partition is updated locally in a lock-free caching structure.
This vastly reduces the cache maintenance overhead, keeping the RC lightweight.
The size of the cache is user-defined, but
$1GB$ is sufficient to significantly improve the performance of
billion-point datasets.

The row cache operates in one of two modes based upon the
data access properties of the algorithm:

\textbf{Lazy update mode:} the row cache lazily updates on specified iterations
based on a user defined \textit{cache update interval} ($I_{cache}$).
The cache updates at iteration $I_{cache}$ then the update
frequency increases quadratically such that the next row cache update is
performed after $2I_{cache}$, then $4I_{cache}$ iterations and so forth.
This means that
row-data in the row cache remains static for several iterations before
the row cache is flushed then repopulated.
This tracks the row activation patterns of algorithms like
k-means, mb-kmeans, sk-means, and divisive hierarchical clustering.
In early iterations, the cache provides little benefit, because row
activations are random. As the algorithm progresses, the same data points
tend to stay active for many consecutive iterations.
As such,
much of the cache remains static for longer periods of time.

\textbf{Active update mode:} the row cache can also function as a traditional
Least Recently Used (LRU) cache. This mode simply stores the more recently
requested rows and evicts those that are less popular. This
mode has higher maintenance overhead, but is more general for cases in which data
access patterns are less predictable.

\section{Distributed Design} \label{sec:dist-design}

We scale to the distributed setting through the Message Passing Interface (MPI).
We employ modular design principles and build our distributed functionality as a
layer above our parallel in-memory framework.
Each machine maintains a decentralized \textit{driver} that
launches \textit{worker} threads that retain the NUMA performance
optimizations across its multiple processors.

We do not address load balancing between machines in the cluster.
We recognize that in some
cases it may be beneficial to dynamically dispatch tasks, but this
negatively affects the performance enhancing NUMA polices we employ. The gains in
performance of data partitioning
(Figure \ref{fig:numa-mem}) outweigh the effects of skew.
We validate this assertion empirically in
Section \ref{sec:dist-eval}.

\textsf{clusterNOR} assumes architecture homogeneity for machines identified as
cluster members via a configuration file. Data are first logically partitioned
into as many partitions as there are NUMA nodes in the cluster. Next,
partitions are randomly assigned to worker machines to reduce the potential for
data hotspots. Finally, each machine processes local data and can merge global
state based on a user defined function that operates across the cluster
virtually transparent to the programmer. The distributed design is simple and
heavily leverages the in-memory optimizations in Section \ref{sec:im-design} in
order to achieve state-of-the-art runtime performance.

\section{Experimental Evaluation} \label{sec:eval}

We begin the evaluation of \textsf{clusterNOR} by benchmarking the performance
and efficacy of our optimizations for the k-means application. k-means
is a core algorithm for the framework and a building block upon which other
applications like mini-batch k-means, H-means, X-means and
G-means are built. We refer to the \textit{k-means NUMA Optimized
Routine} as \textsf{knor}. Section \ref{subsec:apps-eval} completes our
evaluation by benchmarking all applications described in Section \ref{sec:apps}.

We evaluate \textsf{knor} optimizations and benchmark against
heavily-used scalable frameworks.
In Section \ref{sec:baseline} we evaluate the performance of the \textsf{knor}
baseline single threaded implementation and show its computation time to be
faster than competitors. We use this performance baseline to anchor the
strong speedup and scaleup results we demonstrate.

Sections \ref{sec:im-opt-eval} and \ref{sec:sem-opt-eval} evaluate the effect
of specific optimizations on our in-memory and semi-external memory tools
respectively.
Section \ref{sec:vs-others} evaluates the performance of k-means both in-memory
and in the SEM setting relative
to other popular scalable frameworks from the perspective of time
and resource
consumption. Section \ref{sec:dist-eval} specifically performs comparison
between \textsf{knor} and MLlib when run in a distributed cluster.

We evaluate \textsf{knor} optimizations on the Friendster top-8 and top-32
eigenvector datasets. The Friendster dataset represents
real-world machine learning data derived
from a graph that follows a power law distribution of edges. As such, the
resulting eigenvectors contain natural clusters with well defined centroids, which
makes MTI pruning effective, because many data points fall into
strongly rooted clusters and do not change membership. These trends hold
true for other large-scale datasets, albeit to a lesser extent on uniformly
random generated data (Section \ref{sec:vs-others}). The datasets we use for
performance and scalability evaluation are shown in Table \ref{tbl:matrices}.
A summary of \textsf{knor} routine memory bounds is shown
in Table \ref{tbl:complexity}. Table \ref{tbl:kmeans-routine-names}
defines the naming convention we assume for various k-means implementations
within figures and throughout the evaluation.

All measurements we report are an average of $10$ runs on \textbf{dedicated}
machines. Variance in timings are statistically insignificant. As such, we exclude error
bars as they do not add to the interpretation of the data. We drop all
caches between runs.

\begin{table}[!tb]
\caption{k-means routine naming convention.}
\vspace{-10pt}
\begin{center}
\footnotesize
\begin{tabular}{|c|c|}
\hline
    \textbf{Routine name} & \textbf{Definition} \\
\hline
    \textsf{knori} & \textsf{knor}, i.e., k-means, in-memory mode \\
\hline
    \textsf{knori-} & \textsf{knori}, with MTI pruning
    \textit{disabled} \\
\hline
    \textsf{knors} & \textsf{knor}, in SEM mode \\
\hline
    \textsf{knors-} & \textsf{knors}, with MTI pruning
        \textit{disabled} \\
\hline
    \textsf{knors-{}-} & \textsf{knors}, with MTI pruning
        and RC \textit{disabled} \\
\hline
    \textsf{knord} & \textsf{knor} in distributed mode in a cluster \\
\hline
    \textsf{knord-} & \textsf{knord} with MTI pruning
        \textit{disabled} \\
\hline
    MLlib-EC$2$ & MLlib's k-means, on Amazon EC$2$ \cite{aws} \\
\hline
    MPI & MPI \cite{MPI} $||$Lloyds k-means, with MTI \textit{enabled} \\
\hline
    MPI- & MPI without MTI pruning \\
\hline

\end{tabular}
\normalsize
\end{center}
\label{tbl:kmeans-routine-names}
\end{table}

\begin{table}[!htb]
\caption{Asymptotic memory complexity of \textsf{knor} routines. The
    computational complexity of all routines is identical at
    $\mathcal{O}(ndk)$.}
\vspace{-10pt}
\begin{center}
\footnotesize
\begin{tabular}{|c|c|c|}
\hline
    \textbf{Module / Routine} & \textbf{Memory complexity} \\
\hline
Na\"{\i}ve Lloyd's & $\mathcal{O} (nd + kd)$\\
\hline

\textsf{knors-}, \textsf{knors-{}-} & $\mathcal{O} (n + Tkd)$\\
\hline

\textsf{knors} & $\mathcal{O} (2n + Tkd + k^2)$\\
\hline

\textsf{knori-}, \textsf{knord-} & $\mathcal{O} (nd + Tkd) $\\
\hline

\textsf{knori}, \textsf{knord} & $\mathcal{O}(nd + Tkd + n + k^2)$\\
\hline

\end{tabular}
\normalsize
\end{center}
\label{tbl:complexity}
\end{table}

\begin{table}[!htb]
\caption{The datasets under evaluation in this study.}
\vspace{-10pt}
\begin{center}
\footnotesize
\begin{tabular}{|c|c|c|c|}
\hline
\textbf{Data Matrix} & $n$ & $d$ & \textbf{Size}\\
\hline
Friendster-8 \cite{friendster} eigenvectors & $66$M &
  $8$ & $4$GB\\
\hline
Friendster-32 \cite{friendster} eigenvectors & $66$M &
  $32$ & $16$GB\\
\hline
Rand-Multivariate (RM$_{856M}$)& $856$M & 16 & $103$GB\\
\hline
Rand-Multivariate (RM$_{1B}$)& $1.1$B & 32 & $251$GB\\
\hline
Rand-Univariate (RU$_{2B}$)& $2.1$B & 64 & $1.1$TB\\
\hline
\end{tabular}
\normalsize
\end{center}
\label{tbl:matrices}
\end{table}

For completeness we note versions of all frameworks and libraries we use for comparison
in this study; Spark v2.0.1 for MLlib, H$_2$O v3.7, Turi v2.1, R v3.3.1, MATLAB R2016b,
BLAS v3.7.0, Scikit-learn v0.18, MLpack v2.1.0.

\subsection{Single Node Evaluation Hardware}

We perform single node experiments
on a NUMA server with
four Intel Xeon E$7$-$4860$ processors clocked at $2.6$ GHz and $1$TB
of DDR$3$-$1600$ memory. Each processor has $12$ cores. The machine has three LSI
SAS $9300$-$8$e host bus adapters (HBA) connected to a SuperMicro storage
chassis, in which $24$ OCZ Intrepid $3000$ SSDs are installed. The machine
runs Linux kernel v$4.4.0$-$124$. Simultaneous multi-processing (Intel
Hyperthreading) is enabled.
Quick Path Interconnect (QPI) average idle local node latency is measured at
$115.3$ns with average remote node latency at $174$ns. For reads,
QPI average local NUMA node bandwidth is measured at $55$GB/s and average remote
node bandwidth at $11.5$GB/s.
The C++ code is compiled using mpicxx.mpich2 version $5.5.0-12$ with the
-O$3$ flag.

\subsection{Cluster Evaluation Hardware}

We perform distributed memory experiments
on Amazon EC$2$ compute optimized instances of type c$4.8$xlarge with $60$GB of
DDR$3$-$1600$ memory, running Linux kernel v$3.13.0$-$91$.
Each machine has $36$ vCPUS, corresponding to $18$-core
Intel Xeon E$5$-$2666$ v$3$ processors,
clocking $2.9$ GHz, sitting on $2$ independent sockets.
We allow no more that $18$ independent MPI processes or equivalently
$18$ Spark workers to exist on any single machine.
We constrain the cluster to a single
availability zone, subnet and placement group, maximizing
cluster-wide data locality
and minimizing network latency on the 10 Gigabit interconnect. We measure all
experiments from the point when all data is in RAM on all machines. For MLlib we
ensure that the Spark engine is configured to use the maximum available memory and
does not perform any checkpointing or I/O during computation.

\subsection{Baseline Single-Thread Performance} \label{sec:baseline}

\textsf{knori}, even with MTI pruning \textit{disabled},
performs on par with state-of-the-art implementations of Lloyd's
algorithm. This is true for implementations that utilize generalized
matrix multiplication (GEMM) techniques and vectorized operations, such as
MATLAB \cite{matlab} and BLAS \cite{BLAS}.
We find the same to be true of popular statistics
packages and frameworks such as MLpack \cite{mlpack},
Scikit-learn \cite{sklearn} and R
\cite{R} all of which use highly optimized C/C++ code,
although some use scripting language wrappers.
Table \ref{tbl:baseline} shows performance at 1 thread.
Our baseline single threaded performance tops other state-of-the-art serial routines.

\begin{table}[!htb]
\caption{Serial performance of k-means routines,
		using Lloyd's algorithm, on the Friendster-8 dataset.
		All implementations perform all distance computations.
        The \textbf{Language} column refers to the underlying language of
        implementation and not any user-facing higher level wrapper.}
\vspace{-10pt}
\begin{center}
\footnotesize
\resizebox{.475\textwidth}{!}{\begin{tabular}{|c|c|c|c|}
\hline
\textbf{Implementation} & \textbf{Type} & \textbf{Language} & \textbf{Time/iter (sec)}\\
\hline
\textbf{\textsf{knori-}} & \textbf{Iterative} & \textbf{C++} & \textbf{7.49} \\
\hline
MATLAB & GEMM & C++ & 20.68 \\
\hline
BLAS & GEMM & C++ & 20.7 \\
\hline
R & Iterative & C & 8.63 \\
\hline
Scikit-learn & Iterative & Cython & 12.84 \\
\hline
MLpack & Iterative & C++ & 13.09 \\
\hline
\end{tabular}}
\normalsize
\end{center}
\label{tbl:baseline}
\end{table}

\subsection{In-Memory Optimization} \label{sec:im-opt-eval}

\begin{figure}[!htb]
	\begin{center}
		\small
		\vspace{-10pt}
		\begin{tikzpicture}[gnuplot]
\path (0.000,0.000) rectangle (8.382,5.334);
\gpcolor{color=gp lt color border}
\gpsetlinetype{gp lt border}
\gpsetdashtype{gp dt solid}
\gpsetlinewidth{1.00}
\draw[gp path] (1.136,0.985)--(1.316,0.985);
\draw[gp path] (7.829,0.985)--(7.649,0.985);
\node[gp node right] at (0.952,0.985) {$1$};
\draw[gp path] (1.136,1.505)--(1.316,1.505);
\draw[gp path] (7.829,1.505)--(7.649,1.505);
\node[gp node right] at (0.952,1.505) {$2$};
\draw[gp path] (1.136,2.025)--(1.316,2.025);
\draw[gp path] (7.829,2.025)--(7.649,2.025);
\node[gp node right] at (0.952,2.025) {$4$};
\draw[gp path] (1.136,2.545)--(1.316,2.545);
\draw[gp path] (7.829,2.545)--(7.649,2.545);
\node[gp node right] at (0.952,2.545) {$8$};
\draw[gp path] (1.136,3.065)--(1.316,3.065);
\draw[gp path] (7.829,3.065)--(7.649,3.065);
\node[gp node right] at (0.952,3.065) {$16$};
\draw[gp path] (1.136,3.585)--(1.316,3.585);
\draw[gp path] (7.829,3.585)--(7.649,3.585);
\node[gp node right] at (0.952,3.585) {$32$};
\draw[gp path] (1.136,4.105)--(1.316,4.105);
\draw[gp path] (7.829,4.105)--(7.649,4.105);
\node[gp node right] at (0.952,4.105) {$64$};
\draw[gp path] (1.136,0.985)--(1.136,1.165);
\draw[gp path] (1.136,4.409)--(1.136,4.229);
\node[gp node center] at (1.136,0.677) {1};
\draw[gp path] (2.092,0.985)--(2.092,1.165);
\draw[gp path] (2.092,4.409)--(2.092,4.229);
\node[gp node center] at (2.092,0.677) {2};
\draw[gp path] (3.048,0.985)--(3.048,1.165);
\draw[gp path] (3.048,4.409)--(3.048,4.229);
\node[gp node center] at (3.048,0.677) {4};
\draw[gp path] (4.004,0.985)--(4.004,1.165);
\draw[gp path] (4.004,4.409)--(4.004,4.229);
\node[gp node center] at (4.004,0.677) {8};
\draw[gp path] (4.961,0.985)--(4.961,1.165);
\draw[gp path] (4.961,4.409)--(4.961,4.229);
\node[gp node center] at (4.961,0.677) {16};
\draw[gp path] (5.917,0.985)--(5.917,1.165);
\draw[gp path] (5.917,4.409)--(5.917,4.229);
\node[gp node center] at (5.917,0.677) {32};
\draw[gp path] (6.873,0.985)--(6.873,1.165);
\draw[gp path] (6.873,4.409)--(6.873,4.229);
\node[gp node center] at (6.873,0.677) {64};
\draw[gp path] (7.829,0.985)--(7.829,1.165);
\draw[gp path] (7.829,4.409)--(7.829,4.229);
\node[gp node center] at (7.829,0.677) {96};
\draw[gp path] (1.136,4.409)--(1.136,0.985)--(7.829,0.985)--(7.829,4.409)--cycle;
\node[gp node center,rotate=-270] at (0.276,2.697) {Relative Performance};
\node[gp node center] at (4.482,0.215) {No. of Threads};
\node[gp node right] at (3.198,5.000) {\textsf{knori}};
\gpcolor{rgb color={0.000,1.000,0.000}}
\gpsetlinewidth{3.00}
\draw[gp path] (3.382,5.000)--(4.298,5.000);
\draw[gp path] (1.136,0.985)--(2.092,1.505)--(3.048,1.883)--(4.004,2.363)--(4.961,2.925)%
  --(5.917,3.367)--(6.873,3.689)--(7.829,3.959);
\gpsetpointsize{4.00}
\gppoint{gp mark 1}{(1.136,0.985)}
\gppoint{gp mark 1}{(2.092,1.505)}
\gppoint{gp mark 1}{(3.048,1.883)}
\gppoint{gp mark 1}{(4.004,2.363)}
\gppoint{gp mark 1}{(4.961,2.925)}
\gppoint{gp mark 1}{(5.917,3.367)}
\gppoint{gp mark 1}{(6.873,3.689)}
\gppoint{gp mark 1}{(7.829,3.959)}
\gppoint{gp mark 1}{(3.840,5.000)}
\gpcolor{color=gp lt color border}
\node[gp node right] at (3.198,4.692) {NUMA-oblivious};
\gpcolor{rgb color={0.902,0.624,0.000}}
\draw[gp path] (3.382,4.692)--(4.298,4.692);
\draw[gp path] (1.136,0.985)--(2.092,1.024)--(3.048,1.203)--(4.004,1.417)--(4.961,1.500)%
  --(5.917,2.126)--(6.873,2.494)--(7.829,2.775);
\gppoint{gp mark 1}{(1.136,0.985)}
\gppoint{gp mark 1}{(2.092,1.024)}
\gppoint{gp mark 1}{(3.048,1.203)}
\gppoint{gp mark 1}{(4.004,1.417)}
\gppoint{gp mark 1}{(4.961,1.500)}
\gppoint{gp mark 1}{(5.917,2.126)}
\gppoint{gp mark 1}{(6.873,2.494)}
\gppoint{gp mark 1}{(7.829,2.775)}
\gppoint{gp mark 1}{(3.840,4.692)}
\gpcolor{color=gp lt color border}
\node[gp node right] at (7.058,5.000) {Linear (Ideal)};
\gpcolor{rgb color={0.000,1.000,1.000}}
\gpsetdashtype{gp dt 3}
\gpsetlinewidth{2.00}
\draw[gp path] (7.242,5.000)--(8.158,5.000);
\draw[gp path] (1.136,0.985)--(2.092,1.505)--(3.048,2.025)--(4.004,2.545)--(4.961,3.065)%
  --(5.917,3.585)--(6.873,4.105)--(7.829,4.409);
\gppoint{gp mark 2}{(1.136,0.985)}
\gppoint{gp mark 2}{(2.092,1.505)}
\gppoint{gp mark 2}{(3.048,2.025)}
\gppoint{gp mark 2}{(4.004,2.545)}
\gppoint{gp mark 2}{(4.961,3.065)}
\gppoint{gp mark 2}{(5.917,3.585)}
\gppoint{gp mark 2}{(6.873,4.105)}
\gppoint{gp mark 2}{(7.829,4.409)}
\gppoint{gp mark 2}{(7.700,5.000)}
\gpcolor{color=gp lt color border}
\gpsetdashtype{gp dt solid}
\gpsetlinewidth{1.00}
\draw[gp path] (1.136,4.409)--(1.136,0.985)--(7.829,0.985)--(7.829,4.409)--cycle;
\gpdefrectangularnode{gp plot 1}{\pgfpoint{1.136cm}{0.985cm}}{\pgfpoint{7.829cm}{4.409cm}}
\end{tikzpicture}
		\vspace{-10pt}
		\caption{NUMA-aware memory optimizations provide near-linear speedup.
		\textsf{knori} (which is NUMA-aware) vs.
			a NUMA-oblivious routine on the Friendster top-8
			eigenvector dataset, with $k=10$. The NUMA-oblivious routine is
            identical to \textsf{knori}, but we disable all NUMA optimizations.}
		\label{fig:numa-mem-eval}
	\end{center}
\end{figure}
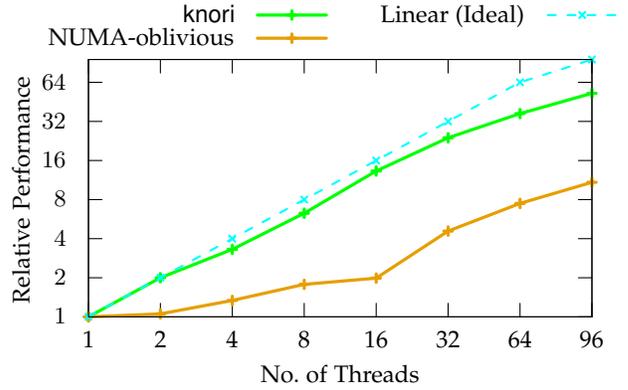

\textsf{clusterNOR} demonstrates NUMA-node thread binding, maintaining
NUMA memory locality, and NUMA-aware task scheduling
are effective strategies for improving speedup.
We achieve near-linear speedup (Figure \ref{fig:numa-mem-eval}).
Because the machine has $48$ physical cores, speedup degrades slightly between 48 and 96 threads;
additional speedup beyond 48 threads comes from simultaneous multithreading (hyperthreading).
The NUMA-aware implementation is nearly $6$x faster at $64$ threads
compared to a routine containing no NUMA optimizations, henceforth referred to as
\textit{NUMA-oblivious}. The NUMA-oblivious routine relies on the OS
to determine memory allocation, thread scheduling, and load
balancing policies.

We further show that although both the NUMA-oblivious and NUMA-aware
implementation speedup sub-linearly, the NUMA-oblivious routine has a lower linear
constant when compared with a NUMA-aware implementation (Figure \ref{fig:numa-mem-eval}).

Increased parallelism amplifies the performance degradation of
the NUMA-oblivious implementation. We identify the following as the greatest
contributors:

\begin{itemize}
\item the NUMA-oblivious allocation policies of traditional memory
allocators, such as \texttt{malloc}, place data in a contiguous
chunk within a single NUMA memory bank whenever possible. This
leads to a large number of threads performing remote memory accesses
as the number of threads increase;
\item a dynamic NUMA-oblivious task scheduler may give tasks to threads
that cause worker threads to perform many more remote memory accesses than
necessary compared to a NUMA-aware scheduler.
\end{itemize}

\begin{figure}[!tb]
	\begin{center}
		\small
		\vspace{-10pt}
		\begin{tikzpicture}[gnuplot]
\path (0.000,0.000) rectangle (8.636,3.556);
\gpcolor{color=gp lt color border}
\gpsetlinetype{gp lt border}
\gpsetlinewidth{1.00}
\draw[gp path] (1.320,0.827)--(1.500,0.827);
\draw[gp path] (8.083,0.827)--(7.903,0.827);
\node[gp node right] at (1.136,0.827) {$80$};
\draw[gp path] (1.320,1.249)--(1.500,1.249);
\draw[gp path] (8.083,1.249)--(7.903,1.249);
\node[gp node right] at (1.136,1.249) {$120$};
\draw[gp path] (1.320,1.671)--(1.500,1.671);
\draw[gp path] (8.083,1.671)--(7.903,1.671);
\node[gp node right] at (1.136,1.671) {$160$};
\draw[gp path] (1.320,2.094)--(1.500,2.094);
\draw[gp path] (8.083,2.094)--(7.903,2.094);
\node[gp node right] at (1.136,2.094) {$200$};
\draw[gp path] (1.320,2.516)--(1.500,2.516);
\draw[gp path] (8.083,2.516)--(7.903,2.516);
\node[gp node right] at (1.136,2.516) {$240$};
\draw[gp path] (2.673,0.616)--(2.673,0.796);
\draw[gp path] (2.673,2.938)--(2.673,2.758);
\node[gp node center] at (2.673,0.308) {k=10};
\draw[gp path] (4.025,0.616)--(4.025,0.796);
\draw[gp path] (4.025,2.938)--(4.025,2.758);
\node[gp node center] at (4.025,0.308) {k=20};
\draw[gp path] (5.378,0.616)--(5.378,0.796);
\draw[gp path] (5.378,2.938)--(5.378,2.758);
\node[gp node center] at (5.378,0.308) {k=50};
\draw[gp path] (6.730,0.616)--(6.730,0.796);
\draw[gp path] (6.730,2.938)--(6.730,2.758);
\node[gp node center] at (6.730,0.308) {k=100};
\draw[gp path] (1.320,2.938)--(1.320,0.616)--(8.083,0.616)--(8.083,2.938)--cycle;
\node[gp node center,rotate=-270] at (0.276,1.777) {Time/iter (msec)};
\node[gp node right] at (2.223,3.221) {\textsf{knori}};
\gpfill{rgb color={0.000,1.000,0.000},color=.!50} (2.407,3.144)--(3.323,3.144)--(3.323,3.298)--(2.407,3.298)--cycle;
\gpfill{rgb color={0.000,1.000,0.000},color=.!50} (2.402,0.616)--(2.674,0.616)--(2.674,0.907)--(2.402,0.907)--cycle;
\gpfill{rgb color={0.000,1.000,0.000},color=.!50} (3.755,0.616)--(4.026,0.616)--(4.026,0.801)--(3.755,0.801)--cycle;
\gpfill{rgb color={0.000,1.000,0.000},color=.!50} (5.107,0.616)--(5.379,0.616)--(5.379,1.431)--(5.107,1.431)--cycle;
\gpfill{rgb color={0.000,1.000,0.000},color=.!50} (6.460,0.616)--(6.731,0.616)--(6.731,1.958)--(6.460,1.958)--cycle;
\node[gp node right] at (4.611,3.221) {FIFO};
\def\gpfillpath{(4.795,3.144)--(5.711,3.144)--(5.711,3.298)--(4.795,3.298)--cycle}
\gpfill{color=gpbgfillcolor} \gpfillpath;
\gpfill{rgb color={0.000,0.000,1.000},gp pattern 1,pattern color=.} \gpfillpath;
\gpcolor{rgb color={0.000,0.000,1.000}}
\draw[gp path] (4.795,3.144)--(5.711,3.144)--(5.711,3.298)--(4.795,3.298)--cycle;
\def\gpfillpath{(2.673,0.616)--(2.944,0.616)--(2.944,0.833)--(2.673,0.833)--cycle}
\gpfill{color=gpbgfillcolor} \gpfillpath;
\gpfill{rgb color={0.000,0.000,1.000},gp pattern 1,pattern color=.} \gpfillpath;
\draw[gp path] (2.673,0.616)--(2.673,0.832)--(2.943,0.832)--(2.943,0.616)--cycle;
\def\gpfillpath{(4.025,0.616)--(4.297,0.616)--(4.297,0.964)--(4.025,0.964)--cycle}
\gpfill{color=gpbgfillcolor} \gpfillpath;
\gpfill{rgb color={0.000,0.000,1.000},gp pattern 1,pattern color=.} \gpfillpath;
\draw[gp path] (4.025,0.616)--(4.025,0.963)--(4.296,0.963)--(4.296,0.616)--cycle;
\def\gpfillpath{(5.378,0.616)--(5.649,0.616)--(5.649,1.778)--(5.378,1.778)--cycle}
\gpfill{color=gpbgfillcolor} \gpfillpath;
\gpfill{rgb color={0.000,0.000,1.000},gp pattern 1,pattern color=.} \gpfillpath;
\draw[gp path] (5.378,0.616)--(5.378,1.777)--(5.648,1.777)--(5.648,0.616)--cycle;
\def\gpfillpath{(6.730,0.616)--(7.002,0.616)--(7.002,2.314)--(6.730,2.314)--cycle}
\gpfill{color=gpbgfillcolor} \gpfillpath;
\gpfill{rgb color={0.000,0.000,1.000},gp pattern 1,pattern color=.} \gpfillpath;
\draw[gp path] (6.730,0.616)--(6.730,2.313)--(7.001,2.313)--(7.001,0.616)--cycle;
\gpcolor{color=gp lt color border}
\node[gp node right] at (6.999,3.221) {Static};
\def\gpfillpath{(7.183,3.144)--(8.099,3.144)--(8.099,3.298)--(7.183,3.298)--cycle}
\gpfill{color=gpbgfillcolor} \gpfillpath;
\gpfill{rgb color={1.000,0.000,0.000},gp pattern 2,pattern color=.} \gpfillpath;
\gpcolor{rgb color={1.000,0.000,0.000}}
\draw[gp path] (7.183,3.144)--(8.099,3.144)--(8.099,3.298)--(7.183,3.298)--cycle;
\def\gpfillpath{(2.943,0.616)--(3.215,0.616)--(3.215,0.828)--(2.943,0.828)--cycle}
\gpfill{color=gpbgfillcolor} \gpfillpath;
\gpfill{rgb color={1.000,0.000,0.000},gp pattern 2,pattern color=.} \gpfillpath;
\draw[gp path] (2.943,0.616)--(2.943,0.827)--(3.214,0.827)--(3.214,0.616)--cycle;
\def\gpfillpath{(4.296,0.616)--(4.567,0.616)--(4.567,1.006)--(4.296,1.006)--cycle}
\gpfill{color=gpbgfillcolor} \gpfillpath;
\gpfill{rgb color={1.000,0.000,0.000},gp pattern 2,pattern color=.} \gpfillpath;
\draw[gp path] (4.296,0.616)--(4.296,1.005)--(4.566,1.005)--(4.566,0.616)--cycle;
\def\gpfillpath{(5.648,0.616)--(5.920,0.616)--(5.920,2.010)--(5.648,2.010)--cycle}
\gpfill{color=gpbgfillcolor} \gpfillpath;
\gpfill{rgb color={1.000,0.000,0.000},gp pattern 2,pattern color=.} \gpfillpath;
\draw[gp path] (5.648,0.616)--(5.648,2.009)--(5.919,2.009)--(5.919,0.616)--cycle;
\def\gpfillpath{(7.001,0.616)--(7.272,0.616)--(7.272,2.796)--(7.001,2.796)--cycle}
\gpfill{color=gpbgfillcolor} \gpfillpath;
\gpfill{rgb color={1.000,0.000,0.000},gp pattern 2,pattern color=.} \gpfillpath;
\draw[gp path] (7.001,0.616)--(7.001,2.795)--(7.271,2.795)--(7.271,0.616)--cycle;
\gpcolor{color=gp lt color border}
\draw[gp path] (1.320,2.938)--(1.320,0.616)--(8.083,0.616)--(8.083,2.938)--cycle;
\gpdefrectangularnode{gp plot 1}{\pgfpoint{1.320cm}{0.616cm}}{\pgfpoint{8.083cm}{2.938cm}}
\end{tikzpicture}
		\vspace{-10pt}
		\caption{Performance of the partitioned NUMA-aware
            scheduler (\textsf{clusterNOR} default) vs. FIFO and static scheduling
            for \textsf{knori} on the Friendster-8 dataset.}
		\label{fig:sched-eval}
	\end{center}
    \vspace{-10pt}
\end{figure}
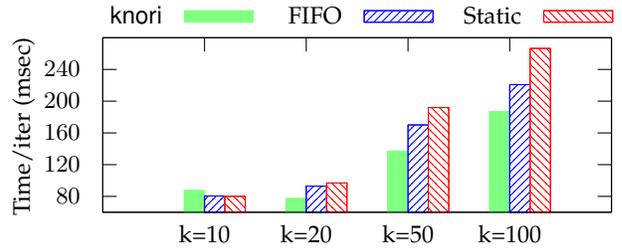

We demonstrate the effectiveness of a
NUMA-aware partitioned task scheduler for pruned computations
via \textsf{knori} (Figure \ref{fig:sched-eval}).
We define a \textit{task} as a block of
data points in contiguous memory given to a thread for computation.
We set a minimum \textit{task size}, i.e. the number of data points in the block,
to $8192$. We empirically determine that this task size is small enough
to not artificially introduce skew in billion-point datasets while
simultaneously providing enough work to amortize the cost of locking at the task
scheduler.
We compare against a static and a first in, first out (FIFO) task scheduler.
The static scheduler preassigns $n/T$ rows
to each worker thread. The FIFO scheduler first assigns
threads to tasks that are local to the thread's partition
of data, then allows threads to steal tasks from straggler
threads whose data resides on \textit{any} NUMA node.

We observe that as $k$ increases, so
does the potential for skew. When $k=10$, the NUMA-aware
scheduler performs negligibly worse than both FIFO and
static scheduling. Work stealing can result in overhead from data movement
when the application is memory-bandwidth-bound \cite{psaroudakis2016adaptive},
as we observe in this experiment. As $k$ increases, the NUMA-aware scheduler
improves performance---by more than $40\%$ when $k = 100$.
We observe similar trends in other datasets; we omit these redundant results.

\subsection{Semi-External Memory}
\label{sec:sem-opt-eval}

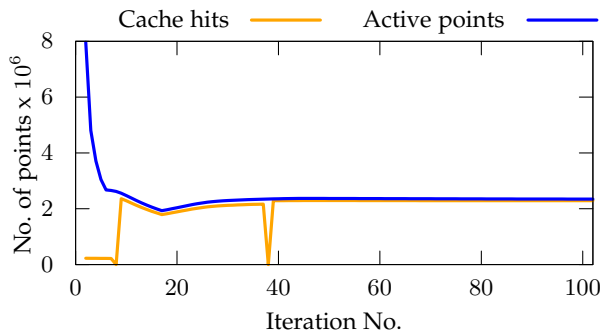
\begin{figure}[!htb]
	\begin{center}
		\small
		\vspace{-10pt}
		\begin{tikzpicture}[gnuplot]
\path (0.000,0.000) rectangle (8.382,4.572);
\gpcolor{color=gp lt color border}
\gpsetlinetype{gp lt border}
\gpsetdashtype{gp dt solid}
\gpsetlinewidth{1.00}
\draw[gp path] (0.952,0.985)--(1.132,0.985);
\draw[gp path] (7.829,0.985)--(7.649,0.985);
\node[gp node right] at (0.768,0.985) {$0$};
\draw[gp path] (0.952,1.728)--(1.132,1.728);
\draw[gp path] (7.829,1.728)--(7.649,1.728);
\node[gp node right] at (0.768,1.728) {$2$};
\draw[gp path] (0.952,2.470)--(1.132,2.470);
\draw[gp path] (7.829,2.470)--(7.649,2.470);
\node[gp node right] at (0.768,2.470) {$4$};
\draw[gp path] (0.952,3.213)--(1.132,3.213);
\draw[gp path] (7.829,3.213)--(7.649,3.213);
\node[gp node right] at (0.768,3.213) {$6$};
\draw[gp path] (0.952,3.955)--(1.132,3.955);
\draw[gp path] (7.829,3.955)--(7.649,3.955);
\node[gp node right] at (0.768,3.955) {$8$};
\draw[gp path] (0.952,0.985)--(0.952,1.165);
\draw[gp path] (0.952,3.955)--(0.952,3.775);
\node[gp node center] at (0.952,0.677) {$0$};
\draw[gp path] (2.300,0.985)--(2.300,1.165);
\draw[gp path] (2.300,3.955)--(2.300,3.775);
\node[gp node center] at (2.300,0.677) {$20$};
\draw[gp path] (3.649,0.985)--(3.649,1.165);
\draw[gp path] (3.649,3.955)--(3.649,3.775);
\node[gp node center] at (3.649,0.677) {$40$};
\draw[gp path] (4.997,0.985)--(4.997,1.165);
\draw[gp path] (4.997,3.955)--(4.997,3.775);
\node[gp node center] at (4.997,0.677) {$60$};
\draw[gp path] (6.346,0.985)--(6.346,1.165);
\draw[gp path] (6.346,3.955)--(6.346,3.775);
\node[gp node center] at (6.346,0.677) {$80$};
\draw[gp path] (7.694,0.985)--(7.694,1.165);
\draw[gp path] (7.694,3.955)--(7.694,3.775);
\node[gp node center] at (7.694,0.677) {$100$};
\draw[gp path] (0.952,3.955)--(0.952,0.985)--(7.829,0.985)--(7.829,3.955)--cycle;
\node[gp node center,rotate=-270] at (0.276,2.470) {No. of points x $10^6$};
\node[gp node center] at (4.390,0.215) {Iteration No.};
\node[gp node right] at (3.106,4.238) {Cache hits};
\gpcolor{rgb color={1.000,0.647,0.000}}
\gpsetlinewidth{3.00}
\draw[gp path] (3.290,4.238)--(4.206,4.238);
\draw[gp path] (1.087,1.069)--(1.154,1.068)--(1.222,1.068)--(1.289,1.067)--(1.357,1.067)%
  --(1.424,1.066)--(1.491,0.985)--(1.559,1.861)--(1.626,1.834)--(1.694,1.803)--(1.761,1.773)%
  --(1.828,1.745)--(1.896,1.719)--(1.963,1.694)--(2.031,1.672)--(2.098,1.650)--(2.166,1.661)%
  --(2.233,1.674)--(2.300,1.686)--(2.368,1.699)--(2.435,1.711)--(2.503,1.724)--(2.570,1.734)%
  --(2.638,1.744)--(2.705,1.753)--(2.772,1.759)--(2.840,1.765)--(2.907,1.769)--(2.975,1.773)%
  --(3.042,1.775)--(3.109,1.779)--(3.177,1.781)--(3.244,1.784)--(3.312,1.786)--(3.379,1.787)%
  --(3.447,1.789)--(3.514,0.985)--(3.581,1.835)--(3.649,1.835)--(3.716,1.835)--(3.784,1.836)%
  --(3.851,1.836)--(3.919,1.836)--(3.986,1.837)--(4.053,1.837)--(4.121,1.837)--(4.188,1.837)%
  --(4.256,1.837)--(4.323,1.837)--(4.391,1.837)--(4.458,1.837)--(4.525,1.837)--(4.593,1.837)%
  --(4.660,1.837)--(4.728,1.836)--(4.795,1.836)--(4.862,1.836)--(4.930,1.836)--(4.997,1.836)%
  --(5.065,1.836)--(5.132,1.836)--(5.200,1.836)--(5.267,1.836)--(5.334,1.836)--(5.402,1.836)%
  --(5.469,1.836)--(5.537,1.835)--(5.604,1.835)--(5.672,1.835)--(5.739,1.835)--(5.806,1.835)%
  --(5.874,1.835)--(5.941,1.835)--(6.009,1.835)--(6.076,1.835)--(6.143,1.835)--(6.211,1.835)%
  --(6.278,1.835)--(6.346,1.835)--(6.413,1.835)--(6.481,1.835)--(6.548,1.835)--(6.615,1.835)%
  --(6.683,1.835)--(6.750,1.835)--(6.818,1.835)--(6.885,1.835)--(6.953,1.835)--(7.020,1.835)%
  --(7.087,1.835)--(7.155,1.835)--(7.222,1.835)--(7.290,1.835)--(7.357,1.835)--(7.424,1.835)%
  --(7.492,1.835)--(7.559,1.835)--(7.627,1.835)--(7.694,1.835)--(7.762,1.835)--(7.829,1.835);
\gpcolor{color=gp lt color border}
\node[gp node right] at (6.782,4.238) {Active points};
\gpcolor{rgb color={0.000,0.000,1.000}}
\draw[gp path] (6.966,4.238)--(7.882,4.238);
\draw[gp path] (1.087,3.943)--(1.154,2.770)--(1.222,2.363)--(1.289,2.117)--(1.357,1.978)%
  --(1.424,1.971)--(1.491,1.958)--(1.559,1.934)--(1.626,1.903)--(1.694,1.870)--(1.761,1.837)%
  --(1.828,1.806)--(1.896,1.777)--(1.963,1.750)--(2.031,1.725)--(2.098,1.702)--(2.166,1.714)%
  --(2.233,1.728)--(2.300,1.741)--(2.368,1.755)--(2.435,1.769)--(2.503,1.783)--(2.570,1.795)%
  --(2.638,1.805)--(2.705,1.815)--(2.772,1.822)--(2.840,1.828)--(2.907,1.833)--(2.975,1.837)%
  --(3.042,1.840)--(3.109,1.843)--(3.177,1.846)--(3.244,1.849)--(3.312,1.851)--(3.379,1.853)%
  --(3.447,1.855)--(3.514,1.857)--(3.581,1.859)--(3.649,1.860)--(3.716,1.862)--(3.784,1.863)%
  --(3.851,1.864)--(3.919,1.865)--(3.986,1.865)--(4.053,1.865)--(4.121,1.865)--(4.188,1.865)%
  --(4.256,1.865)--(4.323,1.864)--(4.391,1.864)--(4.458,1.864)--(4.525,1.864)--(4.593,1.864)%
  --(4.660,1.864)--(4.728,1.863)--(4.795,1.863)--(4.862,1.863)--(4.930,1.863)--(4.997,1.863)%
  --(5.065,1.862)--(5.132,1.862)--(5.200,1.862)--(5.267,1.862)--(5.334,1.862)--(5.402,1.861)%
  --(5.469,1.861)--(5.537,1.861)--(5.604,1.861)--(5.672,1.861)--(5.739,1.860)--(5.806,1.860)%
  --(5.874,1.860)--(5.941,1.860)--(6.009,1.860)--(6.076,1.859)--(6.143,1.859)--(6.211,1.859)%
  --(6.278,1.859)--(6.346,1.859)--(6.413,1.858)--(6.481,1.858)--(6.548,1.858)--(6.615,1.858)%
  --(6.683,1.858)--(6.750,1.858)--(6.818,1.857)--(6.885,1.857)--(6.953,1.857)--(7.020,1.857)%
  --(7.087,1.857)--(7.155,1.857)--(7.222,1.857)--(7.290,1.857)--(7.357,1.857)--(7.424,1.857)%
  --(7.492,1.856)--(7.559,1.856)--(7.627,1.856)--(7.694,1.856)--(7.762,1.856)--(7.829,1.856);
\gpcolor{color=gp lt color border}
\gpsetlinewidth{1.00}
\draw[gp path] (0.952,3.955)--(0.952,0.985)--(7.829,0.985)--(7.829,3.955)--cycle;
\gpdefrectangularnode{gp plot 1}{\pgfpoint{0.952cm}{0.985cm}}{\pgfpoint{7.829cm}{3.955cm}}
\end{tikzpicture}
		\vspace{-10pt}
    \caption{Row cache hits per iteration compared
      with the maximum achievable number of hits on
        the Friendster top-32 eigenvectors dataset.}
		\label{fig:cache-hits-by-iter}
	\end{center}
\end{figure}

We evaluate \textsf{knors} optimizations, performance and scalability.
We set a small \textit{page cache} size for FlashGraph ($4$KB) to minimize the
amount of superfluous data read from disk due to fragmentation.
We set the cache refresh interval, $I_{cache}$, to $5$ for all experiments.
The choice trades-off cache freshness for reduced cache maintenance.
We demonstrate the efficacy of this strategy in Figure \ref{fig:cache-hits-by-iter}.
Finally, we disable checkpoint failure recovery during
performance evaluation for both our routines and those of our competitors.

\begin{figure}[!htb]
	\centering
	\footnotesize
	\begin{subfigure}{.5\textwidth}
		\begin{tikzpicture}[gnuplot]
\path (0.000,0.000) rectangle (8.382,4.572);
\gpcolor{color=gp lt color border}
\gpsetlinetype{gp lt border}
\gpsetdashtype{gp dt solid}
\gpsetlinewidth{1.00}
\draw[gp path] (1.688,1.227)--(1.868,1.227);
\draw[gp path] (7.829,1.227)--(7.649,1.227);
\node[gp node right] at (1.504,1.227) {0.016};
\draw[gp path] (1.688,1.711)--(1.868,1.711);
\draw[gp path] (7.829,1.711)--(7.649,1.711);
\node[gp node right] at (1.504,1.711) {0.062};
\draw[gp path] (1.688,2.195)--(1.868,2.195);
\draw[gp path] (7.829,2.195)--(7.649,2.195);
\node[gp node right] at (1.504,2.195) {0.25};
\draw[gp path] (1.688,2.679)--(1.868,2.679);
\draw[gp path] (7.829,2.679)--(7.649,2.679);
\node[gp node right] at (1.504,2.679) {1};
\draw[gp path] (1.688,3.163)--(1.868,3.163);
\draw[gp path] (7.829,3.163)--(7.649,3.163);
\node[gp node right] at (1.504,3.163) {4};
\draw[gp path] (1.688,3.647)--(1.868,3.647);
\draw[gp path] (7.829,3.647)--(7.649,3.647);
\node[gp node right] at (1.504,3.647) {16};
\draw[gp path] (1.688,0.985)--(1.688,1.165);
\draw[gp path] (1.688,3.647)--(1.688,3.467);
\node[gp node center] at (1.688,0.677) {$0$};
\draw[gp path] (2.892,0.985)--(2.892,1.165);
\draw[gp path] (2.892,3.647)--(2.892,3.467);
\node[gp node center] at (2.892,0.677) {$20$};
\draw[gp path] (4.096,0.985)--(4.096,1.165);
\draw[gp path] (4.096,3.647)--(4.096,3.467);
\node[gp node center] at (4.096,0.677) {$40$};
\draw[gp path] (5.300,0.985)--(5.300,1.165);
\draw[gp path] (5.300,3.647)--(5.300,3.467);
\node[gp node center] at (5.300,0.677) {$60$};
\draw[gp path] (6.504,0.985)--(6.504,1.165);
\draw[gp path] (6.504,3.647)--(6.504,3.467);
\node[gp node center] at (6.504,0.677) {$80$};
\draw[gp path] (7.709,0.985)--(7.709,1.165);
\draw[gp path] (7.709,3.647)--(7.709,3.467);
\node[gp node center] at (7.709,0.677) {$100$};
\draw[gp path] (1.688,3.647)--(1.688,0.985)--(7.829,0.985)--(7.829,3.647)--cycle;
\node[gp node center,rotate=-270] at (0.276,2.316) {Data (GB)};
\node[gp node center] at (4.758,0.215) {Iteration No.};
\node[gp node right] at (3.474,4.238) {No RC req};
\gpcolor{rgb color={1.000,0.000,0.000}}
\gpsetdashtype{gp dt 3}
\gpsetlinewidth{2.00}
\draw[gp path] (3.658,4.238)--(4.574,4.238);
\draw[gp path] (1.748,3.639)--(1.808,2.903)--(1.869,2.727)--(1.929,2.636)--(1.989,2.568)%
  --(2.049,2.522)--(2.109,2.519)--(2.170,2.515)--(2.230,2.506)--(2.290,2.495)--(2.350,2.482)%
  --(2.410,2.468)--(2.471,2.455)--(2.531,2.443)--(2.591,2.431)--(2.651,2.419)--(2.712,2.408)%
  --(2.772,2.414)--(2.832,2.421)--(2.892,2.427)--(2.952,2.433)--(3.013,2.439)--(3.073,2.445)%
  --(3.133,2.451)--(3.193,2.455)--(3.253,2.459)--(3.314,2.462)--(3.374,2.465)--(3.434,2.467)%
  --(3.494,2.468)--(3.554,2.470)--(3.615,2.471)--(3.675,2.472)--(3.735,2.473)--(3.795,2.474)%
  --(3.855,2.475)--(3.916,2.476)--(3.976,2.477)--(4.036,2.477)--(4.096,2.478)--(4.156,2.479)%
  --(4.217,2.479)--(4.277,2.479)--(4.337,2.480)--(4.397,2.480)--(4.457,2.480)--(4.518,2.480)%
  --(4.578,2.480)--(4.638,2.480)--(4.698,2.480)--(4.759,2.480)--(4.819,2.479)--(4.879,2.479)%
  --(4.939,2.479)--(4.999,2.479)--(5.060,2.479)--(5.120,2.479)--(5.180,2.479)--(5.240,2.479)%
  --(5.300,2.479)--(5.361,2.479)--(5.421,2.479)--(5.481,2.479)--(5.541,2.478)--(5.601,2.478)%
  --(5.662,2.478)--(5.722,2.478)--(5.782,2.478)--(5.842,2.478)--(5.902,2.478)--(5.963,2.478)%
  --(6.023,2.478)--(6.083,2.478)--(6.143,2.478)--(6.203,2.478)--(6.264,2.478)--(6.324,2.477)%
  --(6.384,2.477)--(6.444,2.477)--(6.504,2.477)--(6.565,2.477)--(6.625,2.477)--(6.685,2.477)%
  --(6.745,2.477)--(6.806,2.477)--(6.866,2.477)--(6.926,2.477)--(6.986,2.477)--(7.046,2.477)%
  --(7.107,2.477)--(7.167,2.477)--(7.227,2.476)--(7.287,2.476)--(7.347,2.476)--(7.408,2.476)%
  --(7.468,2.476)--(7.528,2.476)--(7.588,2.476)--(7.648,2.476)--(7.709,2.476)--(7.769,2.476)%
  --(7.829,2.476);
\gpcolor{color=gp lt color border}
\node[gp node right] at (3.474,3.930) {No RC read};
\gpcolor{rgb color={1.000,0.000,0.000}}
\gpsetdashtype{gp dt solid}
\gpsetlinewidth{3.00}
\draw[gp path] (3.658,3.930)--(4.574,3.930);
\draw[gp path] (1.748,3.639)--(1.808,3.540)--(1.869,3.418)--(1.929,3.331)--(1.989,3.255)%
  --(2.049,3.201)--(2.109,3.198)--(2.170,3.193)--(2.230,3.182)--(2.290,3.168)--(2.350,3.152)%
  --(2.410,3.135)--(2.471,3.119)--(2.531,3.103)--(2.591,3.087)--(2.651,3.071)--(2.712,3.056)%
  --(2.772,3.065)--(2.832,3.073)--(2.892,3.081)--(2.952,3.090)--(3.013,3.098)--(3.073,3.106)%
  --(3.133,3.112)--(3.193,3.118)--(3.253,3.123)--(3.314,3.127)--(3.374,3.130)--(3.434,3.133)%
  --(3.494,3.135)--(3.554,3.136)--(3.615,3.138)--(3.675,3.140)--(3.735,3.141)--(3.795,3.142)%
  --(3.855,3.143)--(3.916,3.144)--(3.976,3.145)--(4.036,3.146)--(4.096,3.147)--(4.156,3.148)%
  --(4.217,3.148)--(4.277,3.148)--(4.337,3.149)--(4.397,3.149)--(4.457,3.149)--(4.518,3.150)%
  --(4.578,3.149)--(4.638,3.149)--(4.698,3.149)--(4.759,3.149)--(4.819,3.149)--(4.879,3.149)%
  --(4.939,3.149)--(4.999,3.149)--(5.060,3.149)--(5.120,3.148)--(5.180,3.148)--(5.240,3.148)%
  --(5.300,3.149)--(5.361,3.148)--(5.421,3.148)--(5.481,3.148)--(5.541,3.147)--(5.601,3.148)%
  --(5.662,3.148)--(5.722,3.147)--(5.782,3.147)--(5.842,3.147)--(5.902,3.147)--(5.963,3.147)%
  --(6.023,3.147)--(6.083,3.147)--(6.143,3.147)--(6.203,3.147)--(6.264,3.146)--(6.324,3.146)%
  --(6.384,3.146)--(6.444,3.146)--(6.504,3.146)--(6.565,3.146)--(6.625,3.146)--(6.685,3.146)%
  --(6.745,3.146)--(6.806,3.146)--(6.866,3.145)--(6.926,3.146)--(6.986,3.146)--(7.046,3.145)%
  --(7.107,3.145)--(7.167,3.145)--(7.227,3.145)--(7.287,3.145)--(7.347,3.145)--(7.408,3.144)%
  --(7.468,3.145)--(7.528,3.145)--(7.588,3.145)--(7.648,3.145)--(7.709,3.145)--(7.769,3.145)%
  --(7.829,3.145);
\gpcolor{color=gp lt color border}
\node[gp node right] at (6.782,4.238) {\textsf{knors} req};
\gpcolor{rgb color={0.000,1.000,0.000}}
\gpsetdashtype{gp dt 3}
\gpsetlinewidth{2.00}
\draw[gp path] (6.966,4.238)--(7.882,4.238);
\draw[gp path] (1.748,3.638)--(1.808,2.893)--(1.869,2.710)--(1.929,2.615)--(1.989,2.541)%
  --(2.049,2.492)--(2.109,2.489)--(2.170,2.515)--(2.230,1.610)--(2.290,1.594)--(2.350,1.579)%
  --(2.410,1.564)--(2.471,1.548)--(2.531,1.533)--(2.591,1.519)--(2.651,1.504)--(2.712,1.490)%
  --(2.772,1.498)--(2.832,1.506)--(2.892,1.514)--(2.952,1.522)--(3.013,1.530)--(3.073,1.537)%
  --(3.133,1.543)--(3.193,1.549)--(3.253,1.554)--(3.314,1.558)--(3.374,1.561)--(3.434,1.563)%
  --(3.494,1.565)--(3.554,1.566)--(3.615,1.568)--(3.675,1.570)--(3.735,1.571)--(3.795,1.572)%
  --(3.855,1.573)--(3.916,1.574)--(3.976,2.477)--(4.036,1.225)--(4.096,1.241)--(4.156,1.264)%
  --(4.217,1.263)--(4.277,1.271)--(4.337,1.277)--(4.397,1.283)--(4.457,1.285)--(4.518,1.284)%
  --(4.578,1.282)--(4.638,1.277)--(4.698,1.274)--(4.759,1.273)--(4.819,1.274)--(4.879,1.271)%
  --(4.939,1.269)--(4.999,1.267)--(5.060,1.265)--(5.120,1.263)--(5.180,1.261)--(5.240,1.260)%
  --(5.300,1.258)--(5.361,1.255)--(5.421,1.254)--(5.481,1.252)--(5.541,1.249)--(5.601,1.248)%
  --(5.662,1.247)--(5.722,1.245)--(5.782,1.244)--(5.842,1.242)--(5.902,1.240)--(5.963,1.239)%
  --(6.023,1.237)--(6.083,1.235)--(6.143,1.233)--(6.203,1.231)--(6.264,1.230)--(6.324,1.227)%
  --(6.384,1.226)--(6.444,1.224)--(6.504,1.222)--(6.565,1.219)--(6.625,1.216)--(6.685,1.213)%
  --(6.745,1.211)--(6.806,1.209)--(6.866,1.207)--(6.926,1.205)--(6.986,1.203)--(7.046,1.201)%
  --(7.107,1.199)--(7.167,1.197)--(7.227,1.195)--(7.287,1.194)--(7.347,1.193)--(7.408,1.192)%
  --(7.468,1.191)--(7.528,1.190)--(7.588,1.189)--(7.648,1.187)--(7.709,1.186)--(7.769,1.185)%
  --(7.829,1.184);
\gpcolor{color=gp lt color border}
\node[gp node right] at (6.782,3.930) {\textsf{knors} read};
\gpcolor{rgb color={0.000,1.000,0.000}}
\gpsetdashtype{gp dt solid}
\gpsetlinewidth{3.00}
\draw[gp path] (6.966,3.930)--(7.882,3.930);
\draw[gp path] (1.748,3.639)--(1.808,3.536)--(1.869,3.404)--(1.929,3.308)--(1.989,3.225)%
  --(2.049,3.164)--(2.109,3.161)--(2.170,3.196)--(2.230,2.269)--(2.290,2.252)--(2.350,2.232)%
  --(2.410,2.212)--(2.471,2.191)--(2.531,2.171)--(2.591,2.150)--(2.651,2.131)--(2.712,2.112)%
  --(2.772,2.122)--(2.832,2.133)--(2.892,2.144)--(2.952,2.155)--(3.013,2.166)--(3.073,2.176)%
  --(3.133,2.184)--(3.193,2.192)--(3.253,2.198)--(3.314,2.203)--(3.374,2.207)--(3.434,2.210)%
  --(3.494,2.213)--(3.554,2.214)--(3.615,2.217)--(3.675,2.219)--(3.735,2.220)--(3.795,2.221)%
  --(3.855,2.223)--(3.916,2.224)--(3.976,3.147)--(4.036,1.807)--(4.096,1.811)--(4.156,1.817)%
  --(4.217,1.817)--(4.277,1.821)--(4.337,1.823)--(4.397,1.825)--(4.457,1.825)--(4.518,1.825)%
  --(4.578,1.824)--(4.638,1.822)--(4.698,1.821)--(4.759,1.822)--(4.819,1.821)--(4.879,1.821)%
  --(4.939,1.820)--(4.999,1.819)--(5.060,1.819)--(5.120,1.819)--(5.180,1.817)--(5.240,1.817)%
  --(5.300,1.817)--(5.361,1.816)--(5.421,1.815)--(5.481,1.814)--(5.541,1.814)--(5.601,1.813)%
  --(5.662,1.813)--(5.722,1.813)--(5.782,1.812)--(5.842,1.812)--(5.902,1.811)--(5.963,1.811)%
  --(6.023,1.810)--(6.083,1.809)--(6.143,1.809)--(6.203,1.808)--(6.264,1.808)--(6.324,1.807)%
  --(6.384,1.807)--(6.444,1.807)--(6.504,1.806)--(6.565,1.805)--(6.625,1.804)--(6.685,1.804)%
  --(6.745,1.803)--(6.806,1.803)--(6.866,1.802)--(6.926,1.802)--(6.986,1.801)--(7.046,1.801)%
  --(7.107,1.800)--(7.167,1.800)--(7.227,1.799)--(7.287,1.799)--(7.347,1.799)--(7.408,1.799)%
  --(7.468,1.799)--(7.528,1.798)--(7.588,1.798)--(7.648,1.798)--(7.709,1.798)--(7.769,1.797)%
  --(7.829,1.797);
\gpcolor{color=gp lt color border}
\gpsetlinewidth{1.00}
\draw[gp path] (1.688,3.647)--(1.688,0.985)--(7.829,0.985)--(7.829,3.647)--cycle;
\gpdefrectangularnode{gp plot 1}{\pgfpoint{1.688cm}{0.985cm}}{\pgfpoint{7.829cm}{3.647cm}}
\end{tikzpicture}
		\caption{\textsf{knors} data requested (req) vs. data read (read)
        from SSDs each iteration when the row cache (RC) is \textit{enabled}
        or \textit{disabled}. MTI pruning requests fewer data points
        from SSDs, but the file system must still read an entire block
        in which some data may not be useful. As a result, there is a
        discrepancy between the quantity of data requested and the quantity read.
        }
		\label{fig:io-by-iter}
	\end{subfigure}

\begin{subfigure}{.5\textwidth}
	\begin{tikzpicture}[gnuplot]
\path (0.000,0.000) rectangle (8.636,3.556);
\gpcolor{color=gp lt color border}
\gpsetlinetype{gp lt border}
\gpsetlinewidth{1.00}
\draw[gp path] (1.504,0.616)--(1.684,0.616);
\draw[gp path] (8.083,0.616)--(7.903,0.616);
\node[gp node right] at (1.320,0.616) {$10$};
\draw[gp path] (1.504,0.919)--(1.594,0.919);
\draw[gp path] (8.083,0.919)--(7.993,0.919);
\draw[gp path] (1.504,1.096)--(1.594,1.096);
\draw[gp path] (8.083,1.096)--(7.993,1.096);
\draw[gp path] (1.504,1.222)--(1.594,1.222);
\draw[gp path] (8.083,1.222)--(7.993,1.222);
\draw[gp path] (1.504,1.320)--(1.594,1.320);
\draw[gp path] (8.083,1.320)--(7.993,1.320);
\draw[gp path] (1.504,1.400)--(1.594,1.400);
\draw[gp path] (8.083,1.400)--(7.993,1.400);
\draw[gp path] (1.504,1.467)--(1.594,1.467);
\draw[gp path] (8.083,1.467)--(7.993,1.467);
\draw[gp path] (1.504,1.525)--(1.594,1.525);
\draw[gp path] (8.083,1.525)--(7.993,1.525);
\draw[gp path] (1.504,1.577)--(1.594,1.577);
\draw[gp path] (8.083,1.577)--(7.993,1.577);
\draw[gp path] (1.504,1.623)--(1.684,1.623);
\draw[gp path] (8.083,1.623)--(7.903,1.623);
\node[gp node right] at (1.320,1.623) {$100$};
\draw[gp path] (1.504,1.926)--(1.594,1.926);
\draw[gp path] (8.083,1.926)--(7.993,1.926);
\draw[gp path] (1.504,2.103)--(1.594,2.103);
\draw[gp path] (8.083,2.103)--(7.993,2.103);
\draw[gp path] (1.504,2.229)--(1.594,2.229);
\draw[gp path] (8.083,2.229)--(7.993,2.229);
\draw[gp path] (1.504,2.327)--(1.594,2.327);
\draw[gp path] (8.083,2.327)--(7.993,2.327);
\draw[gp path] (1.504,2.407)--(1.594,2.407);
\draw[gp path] (8.083,2.407)--(7.993,2.407);
\draw[gp path] (1.504,2.474)--(1.594,2.474);
\draw[gp path] (8.083,2.474)--(7.993,2.474);
\draw[gp path] (1.504,2.532)--(1.594,2.532);
\draw[gp path] (8.083,2.532)--(7.993,2.532);
\draw[gp path] (1.504,2.584)--(1.594,2.584);
\draw[gp path] (8.083,2.584)--(7.993,2.584);
\draw[gp path] (1.504,2.630)--(1.684,2.630);
\draw[gp path] (8.083,2.630)--(7.903,2.630);
\node[gp node right] at (1.320,2.630) {$1000$};
\draw[gp path] (3.697,0.616)--(3.697,0.796);
\draw[gp path] (3.697,2.630)--(3.697,2.450);
\node[gp node center] at (3.697,0.308) {Req I/O};
\draw[gp path] (5.890,0.616)--(5.890,0.796);
\draw[gp path] (5.890,2.630)--(5.890,2.450);
\node[gp node center] at (5.890,0.308) {Read from SSD};
\draw[gp path] (1.504,2.630)--(1.504,0.616)--(8.083,0.616)--(8.083,2.630)--cycle;
\node[gp node center,rotate=-270] at (0.276,1.623) {Log Scale Data (GB)};
\node[gp node right] at (3.509,3.221) {\textsf{knors}};
\gpfill{rgb color={0.000,1.000,0.000},color=.!50} (3.693,3.144)--(4.609,3.144)--(4.609,3.298)--(3.693,3.298)--cycle;
\gpfill{rgb color={0.000,1.000,0.000},color=.!50} (3.258,0.616)--(3.698,0.616)--(3.698,1.010)--(3.258,1.010)--cycle;
\gpfill{rgb color={0.000,1.000,0.000},color=.!50} (5.451,0.616)--(5.891,0.616)--(5.891,1.497)--(5.451,1.497)--cycle;
\node[gp node right] at (3.509,2.913) {\textsf{knors-}};
\def\gpfillpath{(3.693,2.836)--(4.609,2.836)--(4.609,2.990)--(3.693,2.990)--cycle}
\gpfill{color=gpbgfillcolor} \gpfillpath;
\gpfill{rgb color={0.000,0.000,1.000},gp pattern 1,pattern color=.} \gpfillpath;
\gpcolor{rgb color={0.000,0.000,1.000}}
\draw[gp path] (3.693,2.836)--(4.609,2.836)--(4.609,2.990)--(3.693,2.990)--cycle;
\def\gpfillpath{(3.697,0.616)--(4.137,0.616)--(4.137,1.493)--(3.697,1.493)--cycle}
\gpfill{color=gpbgfillcolor} \gpfillpath;
\gpfill{rgb color={0.000,0.000,1.000},gp pattern 1,pattern color=.} \gpfillpath;
\draw[gp path] (3.697,0.616)--(3.697,1.492)--(4.136,1.492)--(4.136,0.616)--cycle;
\def\gpfillpath{(5.890,0.616)--(6.330,0.616)--(6.330,2.243)--(5.890,2.243)--cycle}
\gpfill{color=gpbgfillcolor} \gpfillpath;
\gpfill{rgb color={0.000,0.000,1.000},gp pattern 1,pattern color=.} \gpfillpath;
\draw[gp path] (5.890,0.616)--(5.890,2.242)--(6.329,2.242)--(6.329,0.616)--cycle;
\gpcolor{color=gp lt color border}
\node[gp node right] at (6.265,3.221) {\textsf{knors-{}-}};
\def\gpfillpath{(6.449,3.144)--(7.365,3.144)--(7.365,3.298)--(6.449,3.298)--cycle}
\gpfill{color=gpbgfillcolor} \gpfillpath;
\gpfill{rgb color={1.000,0.000,0.000},gp pattern 2,pattern color=.} \gpfillpath;
\gpcolor{rgb color={1.000,0.000,0.000}}
\draw[gp path] (6.449,3.144)--(7.365,3.144)--(7.365,3.298)--(6.449,3.298)--cycle;
\def\gpfillpath{(4.136,0.616)--(4.575,0.616)--(4.575,2.419)--(4.136,2.419)--cycle}
\gpfill{color=gpbgfillcolor} \gpfillpath;
\gpfill{rgb color={1.000,0.000,0.000},gp pattern 2,pattern color=.} \gpfillpath;
\draw[gp path] (4.136,0.616)--(4.136,2.418)--(4.574,2.418)--(4.574,0.616)--cycle;
\def\gpfillpath{(6.329,0.616)--(6.768,0.616)--(6.768,2.418)--(6.329,2.418)--cycle}
\gpfill{color=gpbgfillcolor} \gpfillpath;
\gpfill{rgb color={1.000,0.000,0.000},gp pattern 2,pattern color=.} \gpfillpath;
\draw[gp path] (6.329,0.616)--(6.329,2.417)--(6.767,2.417)--(6.767,0.616)--cycle;
\gpcolor{color=gp lt color border}
\draw[gp path] (1.504,2.630)--(1.504,0.616)--(8.083,0.616)--(8.083,2.630)--cycle;
\gpdefrectangularnode{gp plot 1}{\pgfpoint{1.504cm}{0.616cm}}{\pgfpoint{8.083cm}{2.630cm}}
\end{tikzpicture}
    \caption{Total data requested (req) vs. data read from SSDs when
    \textit{(i)} both MTI and RC are \textit{disabled} (\textsf{knors}-{}-),
    \textit{(ii)} Only MTI is \textit{enabled} (\textsf{knors}-),
    \textit{(iii)} both MTI and RC are \textit{enabled} (\textsf{knors}).
		Without pruning, all data are requested and read.}
	\label{fig:tot-io}
\end{subfigure}

\caption{The effect of the row cache and MTI on I/O for the Friendster
    top-32 eigenvectors dataset. Row cache size = $512$MB, page
    cache size = $1$GB, $k=100$.}
\label{fig:io}
\end{figure}
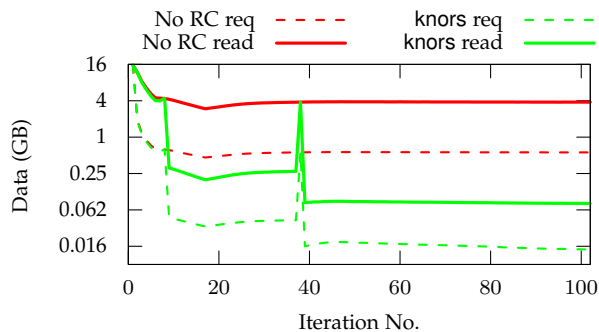
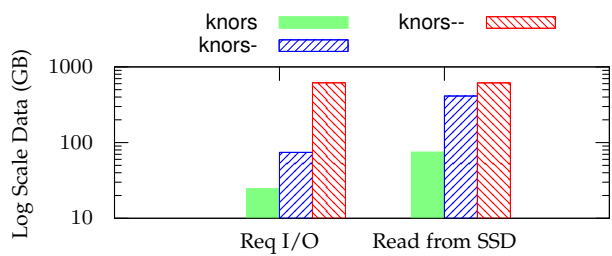

We drastically reduce the amount of data read from SSDs by utilizing
the row cache.
Figure \ref{fig:io-by-iter} shows that as the number of iterations increase,
the row cache's ability to reduce I/O and improve speed also
increases because most rows that are active are cached in memory.
Figure \ref{fig:tot-io} contrasts the total amount of data that an implementation
requests from SSDs with the amount of data SAFS actually reads and
transports into memory.

When \textsf{knors} \textit{disables} both
MTI pruning and the row cache i.e., \textsf{knors-{}-}, every
request issued for row-data
is either served by FlashGraph's page cache or
read from SSDs. When \textsf{knors} \textit{enables} MTI pruning,
but \textit{disables} the row cache, we read
an order of magnitude more data from SSDs than
when we \textit{enable} the row cache.
Figure \ref{fig:io} demonstrates that a page cache is \textbf{not}
sufficient for k-means
and that caching at the granularity of row-data is necessary to achieve
significant reductions in I/O and improvements in performance for real-world
datasets. Additionally, this observation is applicable to all computation pruning
and sub-sampling applications where selective I/O is possible.

\textsf{clusterNOR}'s lazy row update mode reduces I/O significantly for this application.
Figure \ref{fig:cache-hits-by-iter} justifies our design decision
for a lazily updated row cache. As the algorithm progresses,
we obtain nearly a $100\%$ cache hit rate, meaning that \textsf{knors} operates
at in-memory speeds for the vast majority of iterations.

\subsection{MTI Evaluation}

We begin by evaluating the pruning efficacy, runtime performance and memory
consumption of MTI when compared with other popular, effective pruning
algorithms, including TI (Section \ref{sec:mti_vs_others}). We then
show how MTI improves the performance of k-means compared to optimized
implementations without pruning in Section \ref{sec:vs-others}.

\subsubsection{MTI vs. Other State-Of-The-Art Algorithms} \label{sec:mti_vs_others}

\begin{figure}[!htb]
    \centering
    \footnotesize
    \vspace{-10pt}

    \begin{subfigure}{.5\textwidth}
        \begin{tikzpicture}[gnuplot]
\path (0.000,0.000) rectangle (8.636,3.556);
\gpcolor{color=gp lt color border}
\gpsetlinetype{gp lt border}
\draw[gp path] (1.320,0.610)--(1.500,0.610);
\draw[gp path] (8.083,0.610)--(7.903,0.610);
\node[gp node right] at (1.136,0.610) {$0$};
\draw[gp path] (1.320,1.196)--(1.500,1.196);
\draw[gp path] (8.083,1.196)--(7.903,1.196);
\node[gp node right] at (1.136,1.196) {$100$};
\draw[gp path] (1.320,1.782)--(1.500,1.782);
\draw[gp path] (8.083,1.782)--(7.903,1.782);
\node[gp node right] at (1.136,1.782) {$200$};
\draw[gp path] (1.320,2.367)--(1.500,2.367);
\draw[gp path] (8.083,2.367)--(7.903,2.367);
\node[gp node right] at (1.136,2.367) {$300$};
\draw[gp path] (1.320,2.953)--(1.500,2.953);
\draw[gp path] (8.083,2.953)--(7.903,2.953);
\node[gp node right] at (1.136,2.953) {$400$};
\draw[gp path] (3.011,0.616)--(3.011,0.796);
\draw[gp path] (3.011,3.246)--(3.011,3.066);
\node[gp node center] at (3.011,0.308) {k=8};
\draw[gp path] (4.702,0.616)--(4.702,0.796);
\draw[gp path] (4.702,3.246)--(4.702,3.066);
\node[gp node center] at (4.702,0.308) {k=16};
\draw[gp path] (6.392,0.616)--(6.392,0.796);
\draw[gp path] (6.392,3.246)--(6.392,3.066);
\node[gp node center] at (6.392,0.308) {k=32};
\draw[gp path] (1.320,3.246)--(1.320,0.616)--(8.083,0.616)--(8.083,3.246)--cycle;
\node[gp node center,rotate=-270] at (0.276,1.931) {\# Computations X $10^8$};
\def\gpfillpath{(2.419,0.616)--(2.589,0.616)--(2.589,0.668)--(2.419,0.668)--cycle}
\gpfill{color=gpbgfillcolor} \gpfillpath;
\gpfill{rgb color={0.000,0.000,0.000},gp pattern 1,pattern color=.} \gpfillpath;
\gpcolor{rgb color={0.000,0.000,0.000}}
\draw[gp path] (2.419,0.616)--(2.419,0.667)--(2.588,0.667)--(2.588,0.616)--cycle;
\def\gpfillpath{(4.110,0.616)--(4.280,0.616)--(4.280,0.994)--(4.110,0.994)--cycle}
\gpfill{color=gpbgfillcolor} \gpfillpath;
\gpfill{rgb color={0.000,0.000,0.000},gp pattern 1,pattern color=.} \gpfillpath;
\draw[gp path] (4.110,0.616)--(4.110,0.993)--(4.279,0.993)--(4.279,0.616)--cycle;
\def\gpfillpath{(5.800,0.616)--(5.971,0.616)--(5.971,1.866)--(5.800,1.866)--cycle}
\gpfill{color=gpbgfillcolor} \gpfillpath;
\gpfill{rgb color={0.000,0.000,0.000},gp pattern 1,pattern color=.} \gpfillpath;
\draw[gp path] (5.800,0.616)--(5.800,1.865)--(5.970,1.865)--(5.970,0.616)--cycle;
\def\gpfillpath{(2.588,0.616)--(2.758,0.616)--(2.758,0.629)--(2.588,0.629)--cycle}
\gpfill{color=gpbgfillcolor} \gpfillpath;
\gpfill{rgb color={0.000,0.000,1.000},gp pattern 2,pattern color=.} \gpfillpath;
\gpcolor{rgb color={0.000,0.000,1.000}}
\draw[gp path] (2.588,0.616)--(2.588,0.628)--(2.757,0.628)--(2.757,0.616)--cycle;
\def\gpfillpath{(4.279,0.616)--(4.449,0.616)--(4.449,0.683)--(4.279,0.683)--cycle}
\gpfill{color=gpbgfillcolor} \gpfillpath;
\gpfill{rgb color={0.000,0.000,1.000},gp pattern 2,pattern color=.} \gpfillpath;
\draw[gp path] (4.279,0.616)--(4.279,0.682)--(4.448,0.682)--(4.448,0.616)--cycle;
\def\gpfillpath{(5.970,0.616)--(6.140,0.616)--(6.140,0.751)--(5.970,0.751)--cycle}
\gpfill{color=gpbgfillcolor} \gpfillpath;
\gpfill{rgb color={0.000,0.000,1.000},gp pattern 2,pattern color=.} \gpfillpath;
\draw[gp path] (5.970,0.616)--(5.970,0.750)--(6.139,0.750)--(6.139,0.616)--cycle;
\def\gpfillpath{(2.757,0.616)--(2.927,0.616)--(2.927,1.226)--(2.757,1.226)--cycle}
\gpfill{color=gpbgfillcolor} \gpfillpath;
\gpfill{rgb color={1.000,0.000,0.000},gp pattern 3,pattern color=.} \gpfillpath;
\gpcolor{rgb color={1.000,0.000,0.000}}
\draw[gp path] (2.757,0.616)--(2.757,1.225)--(2.926,1.225)--(2.926,0.616)--cycle;
\def\gpfillpath{(4.448,0.616)--(4.618,0.616)--(4.618,1.841)--(4.448,1.841)--cycle}
\gpfill{color=gpbgfillcolor} \gpfillpath;
\gpfill{rgb color={1.000,0.000,0.000},gp pattern 3,pattern color=.} \gpfillpath;
\draw[gp path] (4.448,0.616)--(4.448,1.840)--(4.617,1.840)--(4.617,0.616)--cycle;
\def\gpfillpath{(6.139,0.616)--(6.309,0.616)--(6.309,3.071)--(6.139,3.071)--cycle}
\gpfill{color=gpbgfillcolor} \gpfillpath;
\gpfill{rgb color={1.000,0.000,0.000},gp pattern 3,pattern color=.} \gpfillpath;
\draw[gp path] (6.139,0.616)--(6.139,3.070)--(6.308,3.070)--(6.308,0.616)--cycle;
\def\gpfillpath{(2.926,0.616)--(3.096,0.616)--(3.096,1.226)--(2.926,1.226)--cycle}
\gpfill{color=gpbgfillcolor} \gpfillpath;
\gpfill{rgb color={1.000,0.753,0.753},gp pattern 4,pattern color=.} \gpfillpath;
\gpcolor{rgb color={1.000,0.753,0.753}}
\draw[gp path] (2.926,0.616)--(2.926,1.225)--(3.095,1.225)--(3.095,0.616)--cycle;
\def\gpfillpath{(4.617,0.616)--(4.787,0.616)--(4.787,1.841)--(4.617,1.841)--cycle}
\gpfill{color=gpbgfillcolor} \gpfillpath;
\gpfill{rgb color={1.000,0.753,0.753},gp pattern 4,pattern color=.} \gpfillpath;
\draw[gp path] (4.617,0.616)--(4.617,1.840)--(4.786,1.840)--(4.786,0.616)--cycle;
\def\gpfillpath{(6.308,0.616)--(6.478,0.616)--(6.478,3.071)--(6.308,3.071)--cycle}
\gpfill{color=gpbgfillcolor} \gpfillpath;
\gpfill{rgb color={1.000,0.753,0.753},gp pattern 4,pattern color=.} \gpfillpath;
\draw[gp path] (6.308,0.616)--(6.308,3.070)--(6.477,3.070)--(6.477,0.616)--cycle;
\def\gpfillpath{(3.095,0.616)--(3.265,0.616)--(3.265,0.704)--(3.095,0.704)--cycle}
\gpfill{color=gpbgfillcolor} \gpfillpath;
\gpfill{rgb color={0.000,1.000,1.000},gp pattern 5,pattern color=.} \gpfillpath;
\gpcolor{rgb color={0.000,1.000,1.000}}
\draw[gp path] (3.095,0.616)--(3.095,0.703)--(3.264,0.703)--(3.264,0.616)--cycle;
\def\gpfillpath{(4.786,0.616)--(4.956,0.616)--(4.956,0.777)--(4.786,0.777)--cycle}
\gpfill{color=gpbgfillcolor} \gpfillpath;
\gpfill{rgb color={0.000,1.000,1.000},gp pattern 5,pattern color=.} \gpfillpath;
\draw[gp path] (4.786,0.616)--(4.786,0.776)--(4.955,0.776)--(4.955,0.616)--cycle;
\def\gpfillpath{(6.477,0.616)--(6.647,0.616)--(6.647,0.935)--(6.477,0.935)--cycle}
\gpfill{color=gpbgfillcolor} \gpfillpath;
\gpfill{rgb color={0.000,1.000,1.000},gp pattern 5,pattern color=.} \gpfillpath;
\draw[gp path] (6.477,0.616)--(6.477,0.934)--(6.646,0.934)--(6.646,0.616)--cycle;
\def\gpfillpath{(3.264,0.616)--(3.434,0.616)--(3.434,0.705)--(3.264,0.705)--cycle}
\gpfill{color=gpbgfillcolor} \gpfillpath;
\gpfill{rgb color={0.647,0.165,0.165},gp pattern 6,pattern color=.} \gpfillpath;
\gpcolor{rgb color={0.647,0.165,0.165}}
\draw[gp path] (3.264,0.616)--(3.264,0.704)--(3.433,0.704)--(3.433,0.616)--cycle;
\def\gpfillpath{(4.955,0.616)--(5.125,0.616)--(5.125,0.774)--(4.955,0.774)--cycle}
\gpfill{color=gpbgfillcolor} \gpfillpath;
\gpfill{rgb color={0.647,0.165,0.165},gp pattern 6,pattern color=.} \gpfillpath;
\draw[gp path] (4.955,0.616)--(4.955,0.773)--(5.124,0.773)--(5.124,0.616)--cycle;
\def\gpfillpath{(6.646,0.616)--(6.816,0.616)--(6.816,0.902)--(6.646,0.902)--cycle}
\gpfill{color=gpbgfillcolor} \gpfillpath;
\gpfill{rgb color={0.647,0.165,0.165},gp pattern 6,pattern color=.} \gpfillpath;
\draw[gp path] (6.646,0.616)--(6.646,0.901)--(6.815,0.901)--(6.815,0.616)--cycle;
\def\gpfillpath{(3.433,0.616)--(3.604,0.616)--(3.604,1.085)--(3.433,1.085)--cycle}
\gpfill{color=gpbgfillcolor} \gpfillpath;
\gpfill{rgb color={0.000,0.000,1.000},gp pattern 7,pattern color=.} \gpfillpath;
\gpcolor{rgb color={0.000,0.000,1.000}}
\draw[gp path] (3.433,0.616)--(3.433,1.084)--(3.603,1.084)--(3.603,0.616)--cycle;
\def\gpfillpath{(5.124,0.616)--(5.294,0.616)--(5.294,1.845)--(5.124,1.845)--cycle}
\gpfill{color=gpbgfillcolor} \gpfillpath;
\gpfill{rgb color={0.000,0.000,1.000},gp pattern 7,pattern color=.} \gpfillpath;
\draw[gp path] (5.124,0.616)--(5.124,1.844)--(5.293,1.844)--(5.293,0.616)--cycle;
\def\gpfillpath{(6.815,0.616)--(6.985,0.616)--(6.985,3.074)--(6.815,3.074)--cycle}
\gpfill{color=gpbgfillcolor} \gpfillpath;
\gpfill{rgb color={0.000,0.000,1.000},gp pattern 7,pattern color=.} \gpfillpath;
\draw[gp path] (6.815,0.616)--(6.815,3.073)--(6.984,3.073)--(6.984,0.616)--cycle;
\gpfill{rgb color={0.000,1.000,0.000},color=.!50} (3.603,0.616)--(3.773,0.616)--(3.773,0.636)--(3.603,0.636)--cycle;
\gpcolor{rgb color={0.000,1.000,0.000}}
\draw[gp path] (3.603,0.616)--(3.603,0.635)--(3.772,0.635)--(3.772,0.616)--cycle;
\gpfill{rgb color={0.000,1.000,0.000},color=.!50} (5.293,0.616)--(5.463,0.616)--(5.463,0.727)--(5.293,0.727)--cycle;
\draw[gp path] (5.293,0.616)--(5.293,0.726)--(5.462,0.726)--(5.462,0.616)--cycle;
\gpfill{rgb color={0.000,1.000,0.000},color=.!50} (6.984,0.616)--(7.154,0.616)--(7.154,0.905)--(6.984,0.905)--cycle;
\draw[gp path] (6.984,0.616)--(6.984,0.904)--(7.153,0.904)--(7.153,0.616)--cycle;
\gpcolor{color=gp lt color border}
\draw[gp path] (1.320,3.246)--(1.320,0.616)--(8.083,0.616)--(8.083,3.246)--cycle;
\gpdefrectangularnode{gp plot 1}{\pgfpoint{1.320cm}{0.616cm}}{\pgfpoint{8.083cm}{3.246cm}}
\end{tikzpicture}
        \vspace{-10pt}
        \caption{Distance computation pruning evaluation.}
        \label{fig:prune-eval-dists}
    \end{subfigure}

    \begin{subfigure}{.5\textwidth}
        \begin{tikzpicture}[gnuplot]
\path (0.000,0.000) rectangle (8.636,3.556);
\gpcolor{color=gp lt color border}
\gpsetlinetype{gp lt border}
\gpsetlinewidth{1.00}
\draw[gp path] (1.136,0.616)--(1.316,0.616);
\draw[gp path] (8.083,0.616)--(7.903,0.616);
\node[gp node right] at (0.952,0.616) {$4$};
\draw[gp path] (1.136,1.493)--(1.316,1.493);
\draw[gp path] (8.083,1.493)--(7.903,1.493);
\node[gp node right] at (0.952,1.493) {$8$};
\draw[gp path] (1.136,2.369)--(1.316,2.369);
\draw[gp path] (8.083,2.369)--(7.903,2.369);
\node[gp node right] at (0.952,2.369) {$16$};
\draw[gp path] (1.136,3.246)--(1.316,3.246);
\draw[gp path] (8.083,3.246)--(7.903,3.246);
\node[gp node right] at (0.952,3.246) {$32$};
\draw[gp path] (2.873,0.616)--(2.873,0.796);
\draw[gp path] (2.873,3.246)--(2.873,3.066);
\node[gp node center] at (2.873,0.308) {k=8};
\draw[gp path] (4.610,0.616)--(4.610,0.796);
\draw[gp path] (4.610,3.246)--(4.610,3.066);
\node[gp node center] at (4.610,0.308) {k=16};
\draw[gp path] (6.346,0.616)--(6.346,0.796);
\draw[gp path] (6.346,3.246)--(6.346,3.066);
\node[gp node center] at (6.346,0.308) {k=32};
\draw[gp path] (1.136,3.246)--(1.136,0.616)--(8.083,0.616)--(8.083,3.246)--cycle;
\node[gp node center,rotate=-270] at (0.276,1.931) {Log Memory (GB)};
\def\gpfillpath{(2.265,0.616)--(2.440,0.616)--(2.440,0.933)--(2.265,0.933)--cycle}
\gpfill{color=gpbgfillcolor} \gpfillpath;
\gpfill{rgb color={0.000,0.000,0.000},gp pattern 1,pattern color=.} \gpfillpath;
\gpcolor{rgb color={0.000,0.000,0.000}}
\draw[gp path] (2.265,0.616)--(2.265,0.932)--(2.439,0.932)--(2.439,0.616)--cycle;
\def\gpfillpath{(4.002,0.616)--(4.176,0.616)--(4.176,0.933)--(4.002,0.933)--cycle}
\gpfill{color=gpbgfillcolor} \gpfillpath;
\gpfill{rgb color={0.000,0.000,0.000},gp pattern 1,pattern color=.} \gpfillpath;
\draw[gp path] (4.002,0.616)--(4.002,0.932)--(4.175,0.932)--(4.175,0.616)--cycle;
\def\gpfillpath{(5.738,0.616)--(5.913,0.616)--(5.913,0.963)--(5.738,0.963)--cycle}
\gpfill{color=gpbgfillcolor} \gpfillpath;
\gpfill{rgb color={0.000,0.000,0.000},gp pattern 1,pattern color=.} \gpfillpath;
\draw[gp path] (5.738,0.616)--(5.738,0.962)--(5.912,0.962)--(5.912,0.616)--cycle;
\def\gpfillpath{(2.439,0.616)--(2.613,0.616)--(2.613,1.632)--(2.439,1.632)--cycle}
\gpfill{color=gpbgfillcolor} \gpfillpath;
\gpfill{rgb color={0.000,0.000,1.000},gp pattern 2,pattern color=.} \gpfillpath;
\gpcolor{rgb color={0.000,0.000,1.000}}
\draw[gp path] (2.439,0.616)--(2.439,1.631)--(2.612,1.631)--(2.612,0.616)--cycle;
\def\gpfillpath{(4.175,0.616)--(4.350,0.616)--(4.350,2.079)--(4.175,2.079)--cycle}
\gpfill{color=gpbgfillcolor} \gpfillpath;
\gpfill{rgb color={0.000,0.000,1.000},gp pattern 2,pattern color=.} \gpfillpath;
\draw[gp path] (4.175,0.616)--(4.175,2.078)--(4.349,2.078)--(4.349,0.616)--cycle;
\def\gpfillpath{(5.912,0.616)--(6.087,0.616)--(6.087,2.678)--(5.912,2.678)--cycle}
\gpfill{color=gpbgfillcolor} \gpfillpath;
\gpfill{rgb color={0.000,0.000,1.000},gp pattern 2,pattern color=.} \gpfillpath;
\draw[gp path] (5.912,0.616)--(5.912,2.677)--(6.086,2.677)--(6.086,0.616)--cycle;
\def\gpfillpath{(2.612,0.616)--(2.787,0.616)--(2.787,0.773)--(2.612,0.773)--cycle}
\gpfill{color=gpbgfillcolor} \gpfillpath;
\gpfill{rgb color={1.000,0.000,0.000},gp pattern 3,pattern color=.} \gpfillpath;
\gpcolor{rgb color={1.000,0.000,0.000}}
\draw[gp path] (2.612,0.616)--(2.612,0.772)--(2.786,0.772)--(2.786,0.616)--cycle;
\def\gpfillpath{(4.349,0.616)--(4.524,0.616)--(4.524,0.807)--(4.349,0.807)--cycle}
\gpfill{color=gpbgfillcolor} \gpfillpath;
\gpfill{rgb color={1.000,0.000,0.000},gp pattern 3,pattern color=.} \gpfillpath;
\draw[gp path] (4.349,0.616)--(4.349,0.806)--(4.523,0.806)--(4.523,0.616)--cycle;
\def\gpfillpath{(6.086,0.616)--(6.260,0.616)--(6.260,0.839)--(6.086,0.839)--cycle}
\gpfill{color=gpbgfillcolor} \gpfillpath;
\gpfill{rgb color={1.000,0.000,0.000},gp pattern 3,pattern color=.} \gpfillpath;
\draw[gp path] (6.086,0.616)--(6.086,0.838)--(6.259,0.838)--(6.259,0.616)--cycle;
\def\gpfillpath{(2.786,0.616)--(2.961,0.616)--(2.961,0.773)--(2.786,0.773)--cycle}
\gpfill{color=gpbgfillcolor} \gpfillpath;
\gpfill{rgb color={1.000,0.753,0.753},gp pattern 4,pattern color=.} \gpfillpath;
\gpcolor{rgb color={1.000,0.753,0.753}}
\draw[gp path] (2.786,0.616)--(2.786,0.772)--(2.960,0.772)--(2.960,0.616)--cycle;
\def\gpfillpath{(4.523,0.616)--(4.697,0.616)--(4.697,0.807)--(4.523,0.807)--cycle}
\gpfill{color=gpbgfillcolor} \gpfillpath;
\gpfill{rgb color={1.000,0.753,0.753},gp pattern 4,pattern color=.} \gpfillpath;
\draw[gp path] (4.523,0.616)--(4.523,0.806)--(4.696,0.806)--(4.696,0.616)--cycle;
\def\gpfillpath{(6.259,0.616)--(6.434,0.616)--(6.434,0.839)--(6.259,0.839)--cycle}
\gpfill{color=gpbgfillcolor} \gpfillpath;
\gpfill{rgb color={1.000,0.753,0.753},gp pattern 4,pattern color=.} \gpfillpath;
\draw[gp path] (6.259,0.616)--(6.259,0.838)--(6.433,0.838)--(6.433,0.616)--cycle;
\def\gpfillpath{(2.960,0.616)--(3.134,0.616)--(3.134,0.903)--(2.960,0.903)--cycle}
\gpfill{color=gpbgfillcolor} \gpfillpath;
\gpfill{rgb color={0.000,1.000,1.000},gp pattern 5,pattern color=.} \gpfillpath;
\gpcolor{rgb color={0.000,1.000,1.000}}
\draw[gp path] (2.960,0.616)--(2.960,0.902)--(3.133,0.902)--(3.133,0.616)--cycle;
\def\gpfillpath{(4.696,0.616)--(4.871,0.616)--(4.871,0.933)--(4.696,0.933)--cycle}
\gpfill{color=gpbgfillcolor} \gpfillpath;
\gpfill{rgb color={0.000,1.000,1.000},gp pattern 5,pattern color=.} \gpfillpath;
\draw[gp path] (4.696,0.616)--(4.696,0.932)--(4.870,0.932)--(4.870,0.616)--cycle;
\def\gpfillpath{(6.433,0.616)--(6.608,0.616)--(6.608,0.963)--(6.433,0.963)--cycle}
\gpfill{color=gpbgfillcolor} \gpfillpath;
\gpfill{rgb color={0.000,1.000,1.000},gp pattern 5,pattern color=.} \gpfillpath;
\draw[gp path] (6.433,0.616)--(6.433,0.962)--(6.607,0.962)--(6.607,0.616)--cycle;
\def\gpfillpath{(3.133,0.616)--(3.308,0.616)--(3.308,0.666)--(3.133,0.666)--cycle}
\gpfill{color=gpbgfillcolor} \gpfillpath;
\gpfill{rgb color={0.647,0.165,0.165},gp pattern 6,pattern color=.} \gpfillpath;
\gpcolor{rgb color={0.647,0.165,0.165}}
\draw[gp path] (3.133,0.616)--(3.133,0.665)--(3.307,0.665)--(3.307,0.616)--cycle;
\def\gpfillpath{(4.870,0.616)--(5.045,0.616)--(5.045,0.666)--(4.870,0.666)--cycle}
\gpfill{color=gpbgfillcolor} \gpfillpath;
\gpfill{rgb color={0.647,0.165,0.165},gp pattern 6,pattern color=.} \gpfillpath;
\draw[gp path] (4.870,0.616)--(4.870,0.665)--(5.044,0.665)--(5.044,0.616)--cycle;
\def\gpfillpath{(6.607,0.616)--(6.781,0.616)--(6.781,0.704)--(6.607,0.704)--cycle}
\gpfill{color=gpbgfillcolor} \gpfillpath;
\gpfill{rgb color={0.647,0.165,0.165},gp pattern 6,pattern color=.} \gpfillpath;
\draw[gp path] (6.607,0.616)--(6.607,0.703)--(6.780,0.703)--(6.780,0.616)--cycle;
\def\gpfillpath{(3.307,0.616)--(3.482,0.616)--(3.482,1.315)--(3.307,1.315)--cycle}
\gpfill{color=gpbgfillcolor} \gpfillpath;
\gpfill{rgb color={0.000,0.000,1.000},gp pattern 7,pattern color=.} \gpfillpath;
\gpcolor{rgb color={0.000,0.000,1.000}}
\draw[gp path] (3.307,0.616)--(3.307,1.314)--(3.481,1.314)--(3.481,0.616)--cycle;
\def\gpfillpath{(5.044,0.616)--(5.218,0.616)--(5.218,1.408)--(5.044,1.408)--cycle}
\gpfill{color=gpbgfillcolor} \gpfillpath;
\gpfill{rgb color={0.000,0.000,1.000},gp pattern 7,pattern color=.} \gpfillpath;
\draw[gp path] (5.044,0.616)--(5.044,1.407)--(5.217,1.407)--(5.217,0.616)--cycle;
\def\gpfillpath{(6.780,0.616)--(6.955,0.616)--(6.955,1.552)--(6.780,1.552)--cycle}
\gpfill{color=gpbgfillcolor} \gpfillpath;
\gpfill{rgb color={0.000,0.000,1.000},gp pattern 7,pattern color=.} \gpfillpath;
\draw[gp path] (6.780,0.616)--(6.780,1.551)--(6.954,1.551)--(6.954,0.616)--cycle;
\gpfill{rgb color={0.000,1.000,0.000},color=.!50} (3.481,0.616)--(3.655,0.616)--(3.655,0.848)--(3.481,0.848)--cycle;
\gpcolor{rgb color={0.000,1.000,0.000}}
\draw[gp path] (3.481,0.616)--(3.481,0.847)--(3.654,0.847)--(3.654,0.616)--cycle;
\gpfill{rgb color={0.000,1.000,0.000},color=.!50} (5.217,0.616)--(5.392,0.616)--(5.392,0.848)--(5.217,0.848)--cycle;
\draw[gp path] (5.217,0.616)--(5.217,0.847)--(5.391,0.847)--(5.391,0.616)--cycle;
\gpfill{rgb color={0.000,1.000,0.000},color=.!50} (6.954,0.616)--(7.129,0.616)--(7.129,0.841)--(6.954,0.841)--cycle;
\draw[gp path] (6.954,0.616)--(6.954,0.840)--(7.128,0.840)--(7.128,0.616)--cycle;
\gpcolor{color=gp lt color border}
\draw[gp path] (1.136,3.246)--(1.136,0.616)--(8.083,0.616)--(8.083,3.246)--cycle;
\gpdefrectangularnode{gp plot 1}{\pgfpoint{1.136cm}{0.616cm}}{\pgfpoint{8.083cm}{3.246cm}}
\end{tikzpicture}
        \vspace{-10pt}
        \caption{Memory utilization comparison.}
        \label{fig:prune-eval-mem}
    \end{subfigure}

    \begin{subfigure}{.5\textwidth}
        \include{./charts/prune.eval.rt}
        \vspace{-10pt}
        \caption{Runtime comparison of pruning algorithms at 32 threads.}
        \label{fig:prune-eval-rt}
    \end{subfigure}

    \caption{Comparison of MTI to other pruning
         algorithms on the Friendster-8 dataset using k-means.}
    \label{fig:prune-eval-all}
\end{figure}
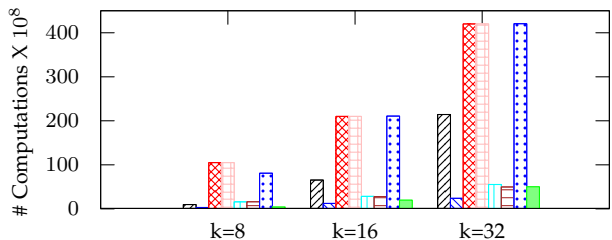
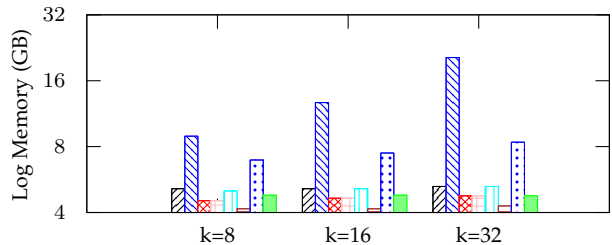
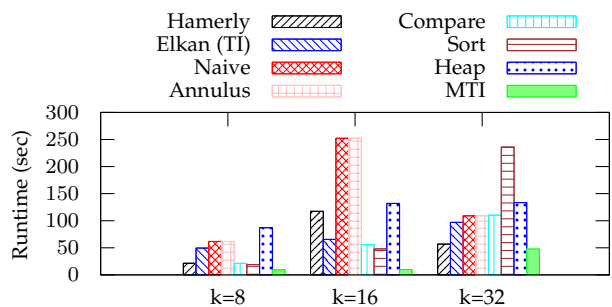

We empirically determine the runtime performance, pruning efficacy and memory
utilization of the Minimal Triangle Inequality algorithm. Figure \ref{fig:prune-eval-all} presents
findings on the real-world Friendster-8 dataset.

Figure \ref{fig:prune-eval-dists} demonstrates that MTI is
comparable to state-of-the-art for computation pruning efficacy.
MTI performs at most $2X$ more distance computations than the minimal
algorithm, most often Elkan's TI. Despite this, MTI consistently is at least $2X$
faster than competitor algorithms and uses up to $5X$ less memory
(Figure \ref{fig:prune-eval-mem}).

MTI maintains constant memory consumption with respect to the number of
clusters. TI in comparison has memory growth that is proportional to the number of
clusters, $k$. We conclude MTI is better suited to large-scale datasets with many
clusters. We recognize that Sort consistently utilizes the least amount of memory,
but this is achieved at the cost of runtime performance (Figure
\ref{fig:prune-eval-rt}), which limits scalability.

Figure \ref{fig:prune-eval-rt} demonstrates the runtime performance benefits of
MTI over competitor solutions at scale. MTI is consistently at least $2X$ faster
than other state-of-the-art algorithms. This is true despite often performing
more distance computations. MTI achieves this, by spending less runtime
maintaining data structures to reduce distance computations.
TI, for example, must spend a
large amount of time updating the $\mathcal{O}(nk)$ lower bound matrix which is
often more expensive than computations it circumvents within the k-means
algorithm.

\subsubsection{MTI Performance Characteristics}

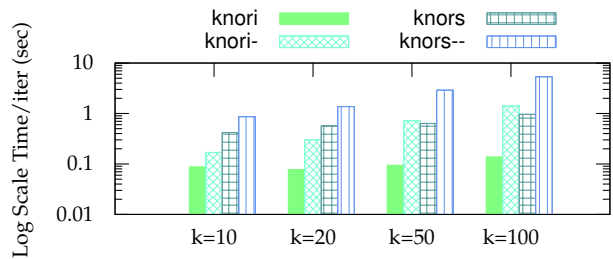
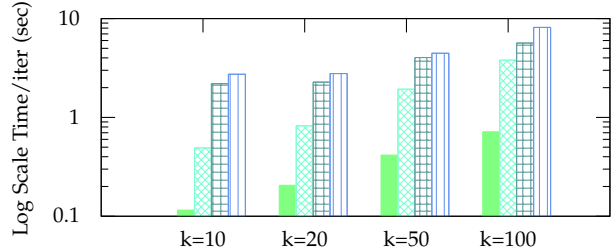
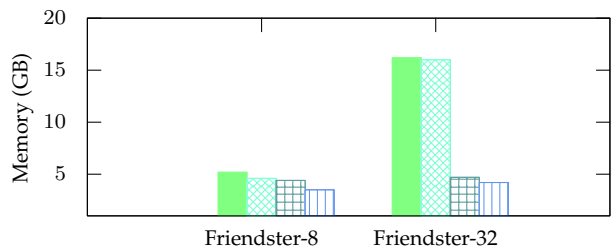
\begin{figure}[!htb]
\centering
\footnotesize
\vspace{-10pt}
\begin{subfigure}{.5\textwidth}
		\begin{tikzpicture}[gnuplot]
\path (0.000,0.000) rectangle (8.636,3.556);
\gpcolor{color=gp lt color border}
\gpsetlinetype{gp lt border}
\gpsetlinewidth{1.00}
\draw[gp path] (1.504,0.616)--(1.684,0.616);
\draw[gp path] (8.083,0.616)--(7.903,0.616);
\node[gp node right] at (1.320,0.616) {0.01};
\draw[gp path] (1.504,0.818)--(1.594,0.818);
\draw[gp path] (8.083,0.818)--(7.993,0.818);
\draw[gp path] (1.504,0.936)--(1.594,0.936);
\draw[gp path] (8.083,0.936)--(7.993,0.936);
\draw[gp path] (1.504,1.020)--(1.594,1.020);
\draw[gp path] (8.083,1.020)--(7.993,1.020);
\draw[gp path] (1.504,1.085)--(1.594,1.085);
\draw[gp path] (8.083,1.085)--(7.993,1.085);
\draw[gp path] (1.504,1.138)--(1.594,1.138);
\draw[gp path] (8.083,1.138)--(7.993,1.138);
\draw[gp path] (1.504,1.183)--(1.594,1.183);
\draw[gp path] (8.083,1.183)--(7.993,1.183);
\draw[gp path] (1.504,1.222)--(1.594,1.222);
\draw[gp path] (8.083,1.222)--(7.993,1.222);
\draw[gp path] (1.504,1.257)--(1.594,1.257);
\draw[gp path] (8.083,1.257)--(7.993,1.257);
\draw[gp path] (1.504,1.287)--(1.684,1.287);
\draw[gp path] (8.083,1.287)--(7.903,1.287);
\node[gp node right] at (1.320,1.287) {0.1};
\draw[gp path] (1.504,1.489)--(1.594,1.489);
\draw[gp path] (8.083,1.489)--(7.993,1.489);
\draw[gp path] (1.504,1.608)--(1.594,1.608);
\draw[gp path] (8.083,1.608)--(7.993,1.608);
\draw[gp path] (1.504,1.692)--(1.594,1.692);
\draw[gp path] (8.083,1.692)--(7.993,1.692);
\draw[gp path] (1.504,1.757)--(1.594,1.757);
\draw[gp path] (8.083,1.757)--(7.993,1.757);
\draw[gp path] (1.504,1.810)--(1.594,1.810);
\draw[gp path] (8.083,1.810)--(7.993,1.810);
\draw[gp path] (1.504,1.855)--(1.594,1.855);
\draw[gp path] (8.083,1.855)--(7.993,1.855);
\draw[gp path] (1.504,1.894)--(1.594,1.894);
\draw[gp path] (8.083,1.894)--(7.993,1.894);
\draw[gp path] (1.504,1.928)--(1.594,1.928);
\draw[gp path] (8.083,1.928)--(7.993,1.928);
\draw[gp path] (1.504,1.959)--(1.684,1.959);
\draw[gp path] (8.083,1.959)--(7.903,1.959);
\node[gp node right] at (1.320,1.959) {1};
\draw[gp path] (1.504,2.161)--(1.594,2.161);
\draw[gp path] (8.083,2.161)--(7.993,2.161);
\draw[gp path] (1.504,2.279)--(1.594,2.279);
\draw[gp path] (8.083,2.279)--(7.993,2.279);
\draw[gp path] (1.504,2.363)--(1.594,2.363);
\draw[gp path] (8.083,2.363)--(7.993,2.363);
\draw[gp path] (1.504,2.428)--(1.594,2.428);
\draw[gp path] (8.083,2.428)--(7.993,2.428);
\draw[gp path] (1.504,2.481)--(1.594,2.481);
\draw[gp path] (8.083,2.481)--(7.993,2.481);
\draw[gp path] (1.504,2.526)--(1.594,2.526);
\draw[gp path] (8.083,2.526)--(7.993,2.526);
\draw[gp path] (1.504,2.565)--(1.594,2.565);
\draw[gp path] (8.083,2.565)--(7.993,2.565);
\draw[gp path] (1.504,2.599)--(1.594,2.599);
\draw[gp path] (8.083,2.599)--(7.993,2.599);
\draw[gp path] (1.504,2.630)--(1.684,2.630);
\draw[gp path] (8.083,2.630)--(7.903,2.630);
\node[gp node right] at (1.320,2.630) {10};
\draw[gp path] (2.820,0.616)--(2.820,0.796);
\draw[gp path] (2.820,2.630)--(2.820,2.450);
\node[gp node center] at (2.820,0.308) {k=10};
\draw[gp path] (4.136,0.616)--(4.136,0.796);
\draw[gp path] (4.136,2.630)--(4.136,2.450);
\node[gp node center] at (4.136,0.308) {k=20};
\draw[gp path] (5.451,0.616)--(5.451,0.796);
\draw[gp path] (5.451,2.630)--(5.451,2.450);
\node[gp node center] at (5.451,0.308) {k=50};
\draw[gp path] (6.767,0.616)--(6.767,0.796);
\draw[gp path] (6.767,2.630)--(6.767,2.450);
\node[gp node center] at (6.767,0.308) {k=100};
\draw[gp path] (1.504,2.630)--(1.504,0.616)--(8.083,0.616)--(8.083,2.630)--cycle;
\node[gp node center,rotate=-270] at (0.276,1.623) {Log Scale Time/iter (sec)};
\node[gp node right] at (3.509,3.221) {\textsf{knori}};
\gpfill{rgb color={0.000,1.000,0.000},color=.!50} (3.693,3.144)--(4.609,3.144)--(4.609,3.298)--(3.693,3.298)--cycle;
\gpfill{rgb color={0.000,1.000,0.000},color=.!50} (2.491,0.616)--(2.711,0.616)--(2.711,1.249)--(2.491,1.249)--cycle;
\gpfill{rgb color={0.000,1.000,0.000},color=.!50} (3.807,0.616)--(4.027,0.616)--(4.027,1.214)--(3.807,1.214)--cycle;
\gpfill{rgb color={0.000,1.000,0.000},color=.!50} (5.122,0.616)--(5.343,0.616)--(5.343,1.271)--(5.122,1.271)--cycle;
\gpfill{rgb color={0.000,1.000,0.000},color=.!50} (6.438,0.616)--(6.659,0.616)--(6.659,1.382)--(6.438,1.382)--cycle;
\node[gp node right] at (3.509,2.913) {\textsf{knori-}};
\def\gpfillpath{(3.693,2.836)--(4.609,2.836)--(4.609,2.990)--(3.693,2.990)--cycle}
\gpfill{color=gpbgfillcolor} \gpfillpath;
\gpfill{rgb color={0.498,1.000,0.831},gp pattern 3,pattern color=.} \gpfillpath;
\gpcolor{rgb color={0.498,1.000,0.831}}
\draw[gp path] (3.693,2.836)--(4.609,2.836)--(4.609,2.990)--(3.693,2.990)--cycle;
\def\gpfillpath{(2.710,0.616)--(2.930,0.616)--(2.930,1.441)--(2.710,1.441)--cycle}
\gpfill{color=gpbgfillcolor} \gpfillpath;
\gpfill{rgb color={0.498,1.000,0.831},gp pattern 3,pattern color=.} \gpfillpath;
\draw[gp path] (2.710,0.616)--(2.710,1.440)--(2.929,1.440)--(2.929,0.616)--cycle;
\def\gpfillpath{(4.026,0.616)--(4.246,0.616)--(4.246,1.611)--(4.026,1.611)--cycle}
\gpfill{color=gpbgfillcolor} \gpfillpath;
\gpfill{rgb color={0.498,1.000,0.831},gp pattern 3,pattern color=.} \gpfillpath;
\draw[gp path] (4.026,0.616)--(4.026,1.610)--(4.245,1.610)--(4.245,0.616)--cycle;
\def\gpfillpath{(5.342,0.616)--(5.562,0.616)--(5.562,1.863)--(5.342,1.863)--cycle}
\gpfill{color=gpbgfillcolor} \gpfillpath;
\gpfill{rgb color={0.498,1.000,0.831},gp pattern 3,pattern color=.} \gpfillpath;
\draw[gp path] (5.342,0.616)--(5.342,1.862)--(5.561,1.862)--(5.561,0.616)--cycle;
\def\gpfillpath{(6.658,0.616)--(6.878,0.616)--(6.878,2.060)--(6.658,2.060)--cycle}
\gpfill{color=gpbgfillcolor} \gpfillpath;
\gpfill{rgb color={0.498,1.000,0.831},gp pattern 3,pattern color=.} \gpfillpath;
\draw[gp path] (6.658,0.616)--(6.658,2.059)--(6.877,2.059)--(6.877,0.616)--cycle;
\gpcolor{color=gp lt color border}
\node[gp node right] at (6.265,3.221) {\textsf{knors}};
\def\gpfillpath{(6.449,3.144)--(7.365,3.144)--(7.365,3.298)--(6.449,3.298)--cycle}
\gpfill{color=gpbgfillcolor} \gpfillpath;
\gpfill{rgb color={0.373,0.620,0.627},gp pattern 4,pattern color=.} \gpfillpath;
\gpcolor{rgb color={0.373,0.620,0.627}}
\draw[gp path] (6.449,3.144)--(7.365,3.144)--(7.365,3.298)--(6.449,3.298)--cycle;
\def\gpfillpath{(2.929,0.616)--(3.150,0.616)--(3.150,1.706)--(2.929,1.706)--cycle}
\gpfill{color=gpbgfillcolor} \gpfillpath;
\gpfill{rgb color={0.373,0.620,0.627},gp pattern 4,pattern color=.} \gpfillpath;
\draw[gp path] (2.929,0.616)--(2.929,1.705)--(3.149,1.705)--(3.149,0.616)--cycle;
\def\gpfillpath{(4.245,0.616)--(4.466,0.616)--(4.466,1.795)--(4.245,1.795)--cycle}
\gpfill{color=gpbgfillcolor} \gpfillpath;
\gpfill{rgb color={0.373,0.620,0.627},gp pattern 4,pattern color=.} \gpfillpath;
\draw[gp path] (4.245,0.616)--(4.245,1.794)--(4.465,1.794)--(4.465,0.616)--cycle;
\def\gpfillpath{(5.561,0.616)--(5.781,0.616)--(5.781,1.827)--(5.561,1.827)--cycle}
\gpfill{color=gpbgfillcolor} \gpfillpath;
\gpfill{rgb color={0.373,0.620,0.627},gp pattern 4,pattern color=.} \gpfillpath;
\draw[gp path] (5.561,0.616)--(5.561,1.826)--(5.780,1.826)--(5.780,0.616)--cycle;
\def\gpfillpath{(6.877,0.616)--(7.097,0.616)--(7.097,1.948)--(6.877,1.948)--cycle}
\gpfill{color=gpbgfillcolor} \gpfillpath;
\gpfill{rgb color={0.373,0.620,0.627},gp pattern 4,pattern color=.} \gpfillpath;
\draw[gp path] (6.877,0.616)--(6.877,1.947)--(7.096,1.947)--(7.096,0.616)--cycle;
\gpcolor{color=gp lt color border}
\node[gp node right] at (6.265,2.913) {\textsf{knors-{}-}};
\def\gpfillpath{(6.449,2.836)--(7.365,2.836)--(7.365,2.990)--(6.449,2.990)--cycle}
\gpfill{color=gpbgfillcolor} \gpfillpath;
\gpfill{rgb color={0.392,0.584,0.929},gp pattern 5,pattern color=.} \gpfillpath;
\gpcolor{rgb color={0.392,0.584,0.929}}
\draw[gp path] (6.449,2.836)--(7.365,2.836)--(7.365,2.990)--(6.449,2.990)--cycle;
\def\gpfillpath{(3.149,0.616)--(3.369,0.616)--(3.369,1.917)--(3.149,1.917)--cycle}
\gpfill{color=gpbgfillcolor} \gpfillpath;
\gpfill{rgb color={0.392,0.584,0.929},gp pattern 5,pattern color=.} \gpfillpath;
\draw[gp path] (3.149,0.616)--(3.149,1.916)--(3.368,1.916)--(3.368,0.616)--cycle;
\def\gpfillpath{(4.465,0.616)--(4.685,0.616)--(4.685,2.051)--(4.465,2.051)--cycle}
\gpfill{color=gpbgfillcolor} \gpfillpath;
\gpfill{rgb color={0.392,0.584,0.929},gp pattern 5,pattern color=.} \gpfillpath;
\draw[gp path] (4.465,0.616)--(4.465,2.050)--(4.684,2.050)--(4.684,0.616)--cycle;
\def\gpfillpath{(5.780,0.616)--(6.001,0.616)--(6.001,2.270)--(5.780,2.270)--cycle}
\gpfill{color=gpbgfillcolor} \gpfillpath;
\gpfill{rgb color={0.392,0.584,0.929},gp pattern 5,pattern color=.} \gpfillpath;
\draw[gp path] (5.780,0.616)--(5.780,2.269)--(6.000,2.269)--(6.000,0.616)--cycle;
\def\gpfillpath{(7.096,0.616)--(7.316,0.616)--(7.316,2.448)--(7.096,2.448)--cycle}
\gpfill{color=gpbgfillcolor} \gpfillpath;
\gpfill{rgb color={0.392,0.584,0.929},gp pattern 5,pattern color=.} \gpfillpath;
\draw[gp path] (7.096,0.616)--(7.096,2.447)--(7.315,2.447)--(7.315,0.616)--cycle;
\gpcolor{color=gp lt color border}
\draw[gp path] (1.504,2.630)--(1.504,0.616)--(8.083,0.616)--(8.083,2.630)--cycle;
\gpdefrectangularnode{gp plot 1}{\pgfpoint{1.504cm}{0.616cm}}{\pgfpoint{8.083cm}{2.630cm}}
\end{tikzpicture}
\vspace{-10pt}
\caption{Runtime performance of k-means on the Friendster-8 dataset.}
\label{fig:f8-per-iter-perf-knor}
\end{subfigure}

\begin{subfigure}{.5\textwidth}
\begin{tikzpicture}[gnuplot]
\path (0.000,0.000) rectangle (8.636,3.556);
\gpcolor{color=gp lt color border}
\gpsetlinetype{gp lt border}
\gpsetlinewidth{1.00}
\draw[gp path] (1.320,0.616)--(1.500,0.616);
\draw[gp path] (8.083,0.616)--(7.903,0.616);
\node[gp node right] at (1.136,0.616) {0.1};
\draw[gp path] (1.320,1.012)--(1.410,1.012);
\draw[gp path] (8.083,1.012)--(7.993,1.012);
\draw[gp path] (1.320,1.243)--(1.410,1.243);
\draw[gp path] (8.083,1.243)--(7.993,1.243);
\draw[gp path] (1.320,1.408)--(1.410,1.408);
\draw[gp path] (8.083,1.408)--(7.993,1.408);
\draw[gp path] (1.320,1.535)--(1.410,1.535);
\draw[gp path] (8.083,1.535)--(7.993,1.535);
\draw[gp path] (1.320,1.639)--(1.410,1.639);
\draw[gp path] (8.083,1.639)--(7.993,1.639);
\draw[gp path] (1.320,1.727)--(1.410,1.727);
\draw[gp path] (8.083,1.727)--(7.993,1.727);
\draw[gp path] (1.320,1.804)--(1.410,1.804);
\draw[gp path] (8.083,1.804)--(7.993,1.804);
\draw[gp path] (1.320,1.871)--(1.410,1.871);
\draw[gp path] (8.083,1.871)--(7.993,1.871);
\draw[gp path] (1.320,1.931)--(1.500,1.931);
\draw[gp path] (8.083,1.931)--(7.903,1.931);
\node[gp node right] at (1.136,1.931) {1};
\draw[gp path] (1.320,2.327)--(1.410,2.327);
\draw[gp path] (8.083,2.327)--(7.993,2.327);
\draw[gp path] (1.320,2.558)--(1.410,2.558);
\draw[gp path] (8.083,2.558)--(7.993,2.558);
\draw[gp path] (1.320,2.723)--(1.410,2.723);
\draw[gp path] (8.083,2.723)--(7.993,2.723);
\draw[gp path] (1.320,2.850)--(1.410,2.850);
\draw[gp path] (8.083,2.850)--(7.993,2.850);
\draw[gp path] (1.320,2.954)--(1.410,2.954);
\draw[gp path] (8.083,2.954)--(7.993,2.954);
\draw[gp path] (1.320,3.042)--(1.410,3.042);
\draw[gp path] (8.083,3.042)--(7.993,3.042);
\draw[gp path] (1.320,3.119)--(1.410,3.119);
\draw[gp path] (8.083,3.119)--(7.993,3.119);
\draw[gp path] (1.320,3.186)--(1.410,3.186);
\draw[gp path] (8.083,3.186)--(7.993,3.186);
\draw[gp path] (1.320,3.246)--(1.500,3.246);
\draw[gp path] (8.083,3.246)--(7.903,3.246);
\node[gp node right] at (1.136,3.246) {10};
\draw[gp path] (2.673,0.616)--(2.673,0.796);
\draw[gp path] (2.673,3.246)--(2.673,3.066);
\node[gp node center] at (2.673,0.308) {k=10};
\draw[gp path] (4.025,0.616)--(4.025,0.796);
\draw[gp path] (4.025,3.246)--(4.025,3.066);
\node[gp node center] at (4.025,0.308) {k=20};
\draw[gp path] (5.378,0.616)--(5.378,0.796);
\draw[gp path] (5.378,3.246)--(5.378,3.066);
\node[gp node center] at (5.378,0.308) {k=50};
\draw[gp path] (6.730,0.616)--(6.730,0.796);
\draw[gp path] (6.730,3.246)--(6.730,3.066);
\node[gp node center] at (6.730,0.308) {k=100};
\draw[gp path] (1.320,3.246)--(1.320,0.616)--(8.083,0.616)--(8.083,3.246)--cycle;
\node[gp node center,rotate=-270] at (0.276,1.931) {Log Scale Time/iter (sec)};
\gpfill{rgb color={0.000,1.000,0.000},color=.!50} (2.334,0.616)--(2.561,0.616)--(2.561,0.698)--(2.334,0.698)--cycle;
\gpfill{rgb color={0.000,1.000,0.000},color=.!50} (3.687,0.616)--(3.913,0.616)--(3.913,1.028)--(3.687,1.028)--cycle;
\gpfill{rgb color={0.000,1.000,0.000},color=.!50} (5.040,0.616)--(5.266,0.616)--(5.266,1.430)--(5.040,1.430)--cycle;
\gpfill{rgb color={0.000,1.000,0.000},color=.!50} (6.392,0.616)--(6.619,0.616)--(6.619,1.740)--(6.392,1.740)--cycle;
\def\gpfillpath{(2.560,0.616)--(2.786,0.616)--(2.786,1.525)--(2.560,1.525)--cycle}
\gpfill{color=gpbgfillcolor} \gpfillpath;
\gpfill{rgb color={0.498,1.000,0.831},gp pattern 3,pattern color=.} \gpfillpath;
\gpcolor{rgb color={0.498,1.000,0.831}}
\draw[gp path] (2.560,0.616)--(2.560,1.524)--(2.785,1.524)--(2.785,0.616)--cycle;
\def\gpfillpath{(3.912,0.616)--(4.139,0.616)--(4.139,1.820)--(3.912,1.820)--cycle}
\gpfill{color=gpbgfillcolor} \gpfillpath;
\gpfill{rgb color={0.498,1.000,0.831},gp pattern 3,pattern color=.} \gpfillpath;
\draw[gp path] (3.912,0.616)--(3.912,1.819)--(4.138,1.819)--(4.138,0.616)--cycle;
\def\gpfillpath{(5.265,0.616)--(5.492,0.616)--(5.492,2.307)--(5.265,2.307)--cycle}
\gpfill{color=gpbgfillcolor} \gpfillpath;
\gpfill{rgb color={0.498,1.000,0.831},gp pattern 3,pattern color=.} \gpfillpath;
\draw[gp path] (5.265,0.616)--(5.265,2.306)--(5.491,2.306)--(5.491,0.616)--cycle;
\def\gpfillpath{(6.618,0.616)--(6.844,0.616)--(6.844,2.694)--(6.618,2.694)--cycle}
\gpfill{color=gpbgfillcolor} \gpfillpath;
\gpfill{rgb color={0.498,1.000,0.831},gp pattern 3,pattern color=.} \gpfillpath;
\draw[gp path] (6.618,0.616)--(6.618,2.693)--(6.843,2.693)--(6.843,0.616)--cycle;
\def\gpfillpath{(2.785,0.616)--(3.012,0.616)--(3.012,2.380)--(2.785,2.380)--cycle}
\gpfill{color=gpbgfillcolor} \gpfillpath;
\gpfill{rgb color={0.373,0.620,0.627},gp pattern 4,pattern color=.} \gpfillpath;
\gpcolor{rgb color={0.373,0.620,0.627}}
\draw[gp path] (2.785,0.616)--(2.785,2.379)--(3.011,2.379)--(3.011,0.616)--cycle;
\def\gpfillpath{(4.138,0.616)--(4.364,0.616)--(4.364,2.401)--(4.138,2.401)--cycle}
\gpfill{color=gpbgfillcolor} \gpfillpath;
\gpfill{rgb color={0.373,0.620,0.627},gp pattern 4,pattern color=.} \gpfillpath;
\draw[gp path] (4.138,0.616)--(4.138,2.400)--(4.363,2.400)--(4.363,0.616)--cycle;
\def\gpfillpath{(5.491,0.616)--(5.717,0.616)--(5.717,2.728)--(5.491,2.728)--cycle}
\gpfill{color=gpbgfillcolor} \gpfillpath;
\gpfill{rgb color={0.373,0.620,0.627},gp pattern 4,pattern color=.} \gpfillpath;
\draw[gp path] (5.491,0.616)--(5.491,2.727)--(5.716,2.727)--(5.716,0.616)--cycle;
\def\gpfillpath{(6.843,0.616)--(7.070,0.616)--(7.070,2.921)--(6.843,2.921)--cycle}
\gpfill{color=gpbgfillcolor} \gpfillpath;
\gpfill{rgb color={0.373,0.620,0.627},gp pattern 4,pattern color=.} \gpfillpath;
\draw[gp path] (6.843,0.616)--(6.843,2.920)--(7.069,2.920)--(7.069,0.616)--cycle;
\def\gpfillpath{(3.011,0.616)--(3.237,0.616)--(3.237,2.506)--(3.011,2.506)--cycle}
\gpfill{color=gpbgfillcolor} \gpfillpath;
\gpfill{rgb color={0.392,0.584,0.929},gp pattern 5,pattern color=.} \gpfillpath;
\gpcolor{rgb color={0.392,0.584,0.929}}
\draw[gp path] (3.011,0.616)--(3.011,2.505)--(3.236,2.505)--(3.236,0.616)--cycle;
\def\gpfillpath{(4.363,0.616)--(4.590,0.616)--(4.590,2.515)--(4.363,2.515)--cycle}
\gpfill{color=gpbgfillcolor} \gpfillpath;
\gpfill{rgb color={0.392,0.584,0.929},gp pattern 5,pattern color=.} \gpfillpath;
\draw[gp path] (4.363,0.616)--(4.363,2.514)--(4.589,2.514)--(4.589,0.616)--cycle;
\def\gpfillpath{(5.716,0.616)--(5.942,0.616)--(5.942,2.785)--(5.716,2.785)--cycle}
\gpfill{color=gpbgfillcolor} \gpfillpath;
\gpfill{rgb color={0.392,0.584,0.929},gp pattern 5,pattern color=.} \gpfillpath;
\draw[gp path] (5.716,0.616)--(5.716,2.784)--(5.941,2.784)--(5.941,0.616)--cycle;
\def\gpfillpath{(7.069,0.616)--(7.295,0.616)--(7.295,3.129)--(7.069,3.129)--cycle}
\gpfill{color=gpbgfillcolor} \gpfillpath;
\gpfill{rgb color={0.392,0.584,0.929},gp pattern 5,pattern color=.} \gpfillpath;
\draw[gp path] (7.069,0.616)--(7.069,3.128)--(7.294,3.128)--(7.294,0.616)--cycle;
\gpcolor{color=gp lt color border}
\draw[gp path] (1.320,3.246)--(1.320,0.616)--(8.083,0.616)--(8.083,3.246)--cycle;
\gpdefrectangularnode{gp plot 1}{\pgfpoint{1.320cm}{0.616cm}}{\pgfpoint{8.083cm}{3.246cm}}
\end{tikzpicture}
\vspace{-10pt}
\caption{Runtime performance of k-means on the Friendster-32 dataset.}
\label{fig:f32-per-iter-perf-knor}
\end{subfigure}

\begin{subfigure}{.5\textwidth}
\begin{tikzpicture}[gnuplot]
\path (0.000,0.000) rectangle (8.636,3.556);
\gpcolor{color=gp lt color border}
\gpsetlinetype{gp lt border}
\gpsetlinewidth{1.00}
\draw[gp path] (1.136,1.170)--(1.316,1.170);
\draw[gp path] (8.083,1.170)--(7.903,1.170);
\node[gp node right] at (0.952,1.170) {$5$};
\draw[gp path] (1.136,1.862)--(1.316,1.862);
\draw[gp path] (8.083,1.862)--(7.903,1.862);
\node[gp node right] at (0.952,1.862) {$10$};
\draw[gp path] (1.136,2.554)--(1.316,2.554);
\draw[gp path] (8.083,2.554)--(7.903,2.554);
\node[gp node right] at (0.952,2.554) {$15$};
\draw[gp path] (1.136,3.246)--(1.316,3.246);
\draw[gp path] (8.083,3.246)--(7.903,3.246);
\node[gp node right] at (0.952,3.246) {$20$};
\draw[gp path] (3.452,0.616)--(3.452,0.796);
\draw[gp path] (3.452,3.246)--(3.452,3.066);
\node[gp node center] at (3.452,0.308) {Friendster-8};
\draw[gp path] (5.767,0.616)--(5.767,0.796);
\draw[gp path] (5.767,3.246)--(5.767,3.066);
\node[gp node center] at (5.767,0.308) {Friendster-32};
\draw[gp path] (1.136,3.246)--(1.136,0.616)--(8.083,0.616)--(8.083,3.246)--cycle;
\node[gp node center,rotate=-270] at (0.276,1.931) {Memory (GB)};
\gpfill{rgb color={0.000,1.000,0.000},color=.!50} (2.873,0.616)--(3.260,0.616)--(3.260,1.198)--(2.873,1.198)--cycle;
\gpfill{rgb color={0.000,1.000,0.000},color=.!50} (5.188,0.616)--(5.575,0.616)--(5.575,2.721)--(5.188,2.721)--cycle;
\def\gpfillpath{(3.259,0.616)--(3.646,0.616)--(3.646,1.115)--(3.259,1.115)--cycle}
\gpfill{color=gpbgfillcolor} \gpfillpath;
\gpfill{rgb color={0.498,1.000,0.831},gp pattern 3,pattern color=.} \gpfillpath;
\gpcolor{rgb color={0.498,1.000,0.831}}
\draw[gp path] (3.259,0.616)--(3.259,1.114)--(3.645,1.114)--(3.645,0.616)--cycle;
\def\gpfillpath{(5.574,0.616)--(5.961,0.616)--(5.961,2.693)--(5.574,2.693)--cycle}
\gpfill{color=gpbgfillcolor} \gpfillpath;
\gpfill{rgb color={0.498,1.000,0.831},gp pattern 3,pattern color=.} \gpfillpath;
\draw[gp path] (5.574,0.616)--(5.574,2.692)--(5.960,2.692)--(5.960,0.616)--cycle;
\def\gpfillpath{(3.645,0.616)--(4.032,0.616)--(4.032,1.088)--(3.645,1.088)--cycle}
\gpfill{color=gpbgfillcolor} \gpfillpath;
\gpfill{rgb color={0.373,0.620,0.627},gp pattern 4,pattern color=.} \gpfillpath;
\gpcolor{rgb color={0.373,0.620,0.627}}
\draw[gp path] (3.645,0.616)--(3.645,1.087)--(4.031,1.087)--(4.031,0.616)--cycle;
\def\gpfillpath{(5.960,0.616)--(6.347,0.616)--(6.347,1.129)--(5.960,1.129)--cycle}
\gpfill{color=gpbgfillcolor} \gpfillpath;
\gpfill{rgb color={0.373,0.620,0.627},gp pattern 4,pattern color=.} \gpfillpath;
\draw[gp path] (5.960,0.616)--(5.960,1.128)--(6.346,1.128)--(6.346,0.616)--cycle;
\def\gpfillpath{(4.031,0.616)--(4.418,0.616)--(4.418,0.963)--(4.031,0.963)--cycle}
\gpfill{color=gpbgfillcolor} \gpfillpath;
\gpfill{rgb color={0.392,0.584,0.929},gp pattern 5,pattern color=.} \gpfillpath;
\gpcolor{rgb color={0.392,0.584,0.929}}
\draw[gp path] (4.031,0.616)--(4.031,0.962)--(4.417,0.962)--(4.417,0.616)--cycle;
\def\gpfillpath{(6.346,0.616)--(6.733,0.616)--(6.733,1.060)--(6.346,1.060)--cycle}
\gpfill{color=gpbgfillcolor} \gpfillpath;
\gpfill{rgb color={0.392,0.584,0.929},gp pattern 5,pattern color=.} \gpfillpath;
\draw[gp path] (6.346,0.616)--(6.346,1.059)--(6.732,1.059)--(6.732,0.616)--cycle;
\gpcolor{color=gp lt color border}
\draw[gp path] (1.136,3.246)--(1.136,0.616)--(8.083,0.616)--(8.083,3.246)--cycle;
\gpdefrectangularnode{gp plot 1}{\pgfpoint{1.136cm}{0.616cm}}{\pgfpoint{8.083cm}{3.246cm}}
\end{tikzpicture}
\vspace{-10pt}
\caption{Memory comparison of fully optimized \textsf{knor} routines
(\textsf{knori}, \textsf{knors}) compared to more vanilla \textsf{knor}
routines (\textsf{knori-}, \textsf{knors-{}-}).}
\label{fig:mem-knor}
\end{subfigure}

\caption{Performance and memory usage comparison of \textsf{knor} modules
		on matrices from the Friendster graph top-8 and top-32 eigenvectors.}
\label{fig:knor-comp}
\vspace{-10pt}
\end{figure}

Figures \ref{fig:f8-per-iter-perf-knor} and \ref{fig:f32-per-iter-perf-knor}
highlight the performance improvement of
\textsf{knor} modules with MTI \textit{enabled} over MTI \textit{disabled}
counterparts. We show that MTI provides a few factors of improvement in
time when enabled.
Figure \ref{fig:mem-knor} highlights that MTI increases the memory load by
negligible amounts compared to non-pruning modules.
We conclude that MTI (unlike TI) is a viable optimization for large-scale datasets.

\subsection{{\large \textsf{knor}} vs. Other Frameworks} \label{sec:vs-others}

We evaluate the performance of \textsf{knor} in comparison with other
frameworks on the datasets in Table \ref{tbl:matrices}. We demonstrate the
following:

\begin{itemize}
    \item \textsf{knori} achieves greater than an order of magnitude runtime
        improvement over competitor frameworks.
    \item \textsf{knors} in many settings outperforms competitor frameworks by
        several factors.
    \item \textsf{knord} achieves up to an order of magnitude better runtime
        than competitor solutions.
\end{itemize}

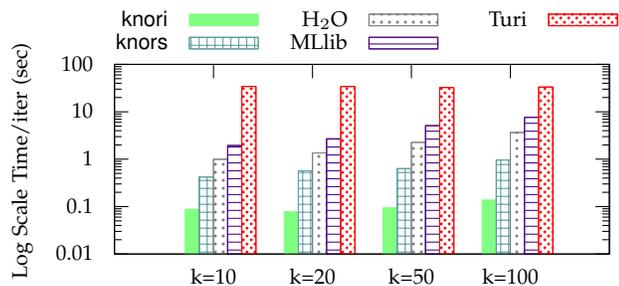
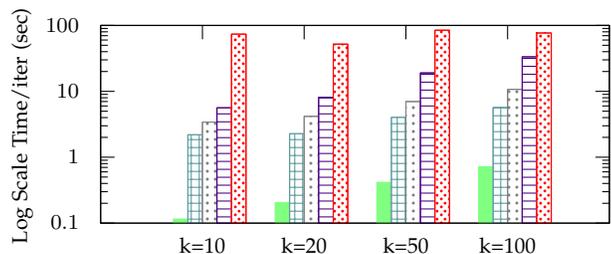
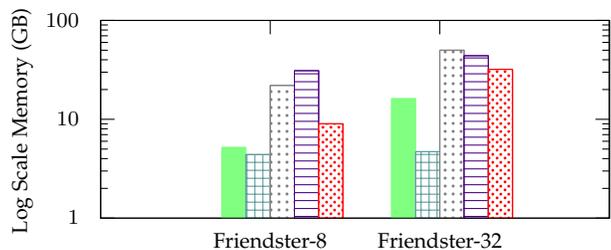
\begin{figure}[!htb]
\centering
\footnotesize
\begin{subfigure}{.5\textwidth}
		\include{./charts/perf.iter.friendster8}
\caption{Runtime performance of k-means on the Friendster-8 dataset.}
\label{fig:f8-per-iter-perf}
\end{subfigure}

\begin{subfigure}{.5\textwidth}
\begin{tikzpicture}[gnuplot]
\path (0.000,0.000) rectangle (8.636,3.556);
\gpcolor{color=gp lt color border}
\gpsetlinetype{gp lt border}
\gpsetlinewidth{1.00}
\draw[gp path] (1.320,0.616)--(1.500,0.616);
\draw[gp path] (8.083,0.616)--(7.903,0.616);
\node[gp node right] at (1.136,0.616) {$0.1$};
\draw[gp path] (1.320,0.880)--(1.410,0.880);
\draw[gp path] (8.083,0.880)--(7.993,0.880);
\draw[gp path] (1.320,1.034)--(1.410,1.034);
\draw[gp path] (8.083,1.034)--(7.993,1.034);
\draw[gp path] (1.320,1.144)--(1.410,1.144);
\draw[gp path] (8.083,1.144)--(7.993,1.144);
\draw[gp path] (1.320,1.229)--(1.410,1.229);
\draw[gp path] (8.083,1.229)--(7.993,1.229);
\draw[gp path] (1.320,1.298)--(1.410,1.298);
\draw[gp path] (8.083,1.298)--(7.993,1.298);
\draw[gp path] (1.320,1.357)--(1.410,1.357);
\draw[gp path] (8.083,1.357)--(7.993,1.357);
\draw[gp path] (1.320,1.408)--(1.410,1.408);
\draw[gp path] (8.083,1.408)--(7.993,1.408);
\draw[gp path] (1.320,1.453)--(1.410,1.453);
\draw[gp path] (8.083,1.453)--(7.993,1.453);
\draw[gp path] (1.320,1.493)--(1.500,1.493);
\draw[gp path] (8.083,1.493)--(7.903,1.493);
\node[gp node right] at (1.136,1.493) {$1$};
\draw[gp path] (1.320,1.757)--(1.410,1.757);
\draw[gp path] (8.083,1.757)--(7.993,1.757);
\draw[gp path] (1.320,1.911)--(1.410,1.911);
\draw[gp path] (8.083,1.911)--(7.993,1.911);
\draw[gp path] (1.320,2.020)--(1.410,2.020);
\draw[gp path] (8.083,2.020)--(7.993,2.020);
\draw[gp path] (1.320,2.105)--(1.410,2.105);
\draw[gp path] (8.083,2.105)--(7.993,2.105);
\draw[gp path] (1.320,2.175)--(1.410,2.175);
\draw[gp path] (8.083,2.175)--(7.993,2.175);
\draw[gp path] (1.320,2.234)--(1.410,2.234);
\draw[gp path] (8.083,2.234)--(7.993,2.234);
\draw[gp path] (1.320,2.284)--(1.410,2.284);
\draw[gp path] (8.083,2.284)--(7.993,2.284);
\draw[gp path] (1.320,2.329)--(1.410,2.329);
\draw[gp path] (8.083,2.329)--(7.993,2.329);
\draw[gp path] (1.320,2.369)--(1.500,2.369);
\draw[gp path] (8.083,2.369)--(7.903,2.369);
\node[gp node right] at (1.136,2.369) {$10$};
\draw[gp path] (1.320,2.633)--(1.410,2.633);
\draw[gp path] (8.083,2.633)--(7.993,2.633);
\draw[gp path] (1.320,2.788)--(1.410,2.788);
\draw[gp path] (8.083,2.788)--(7.993,2.788);
\draw[gp path] (1.320,2.897)--(1.410,2.897);
\draw[gp path] (8.083,2.897)--(7.993,2.897);
\draw[gp path] (1.320,2.982)--(1.410,2.982);
\draw[gp path] (8.083,2.982)--(7.993,2.982);
\draw[gp path] (1.320,3.052)--(1.410,3.052);
\draw[gp path] (8.083,3.052)--(7.993,3.052);
\draw[gp path] (1.320,3.110)--(1.410,3.110);
\draw[gp path] (8.083,3.110)--(7.993,3.110);
\draw[gp path] (1.320,3.161)--(1.410,3.161);
\draw[gp path] (8.083,3.161)--(7.993,3.161);
\draw[gp path] (1.320,3.206)--(1.410,3.206);
\draw[gp path] (8.083,3.206)--(7.993,3.206);
\draw[gp path] (1.320,3.246)--(1.500,3.246);
\draw[gp path] (8.083,3.246)--(7.903,3.246);
\node[gp node right] at (1.136,3.246) {$100$};
\draw[gp path] (2.673,0.616)--(2.673,0.796);
\draw[gp path] (2.673,3.246)--(2.673,3.066);
\node[gp node center] at (2.673,0.308) {k=10};
\draw[gp path] (4.025,0.616)--(4.025,0.796);
\draw[gp path] (4.025,3.246)--(4.025,3.066);
\node[gp node center] at (4.025,0.308) {k=20};
\draw[gp path] (5.378,0.616)--(5.378,0.796);
\draw[gp path] (5.378,3.246)--(5.378,3.066);
\node[gp node center] at (5.378,0.308) {k=50};
\draw[gp path] (6.730,0.616)--(6.730,0.796);
\draw[gp path] (6.730,3.246)--(6.730,3.066);
\node[gp node center] at (6.730,0.308) {k=100};
\draw[gp path] (1.320,3.246)--(1.320,0.616)--(8.083,0.616)--(8.083,3.246)--cycle;
\node[gp node center,rotate=-270] at (0.276,1.931) {Log Scale Time/iter (sec)};
\gpfill{rgb color={0.000,1.000,0.000},color=.!50} (2.286,0.616)--(2.480,0.616)--(2.480,0.671)--(2.286,0.671)--cycle;
\gpfill{rgb color={0.000,1.000,0.000},color=.!50} (3.639,0.616)--(3.833,0.616)--(3.833,0.891)--(3.639,0.891)--cycle;
\gpfill{rgb color={0.000,1.000,0.000},color=.!50} (4.991,0.616)--(5.186,0.616)--(5.186,1.159)--(4.991,1.159)--cycle;
\gpfill{rgb color={0.000,1.000,0.000},color=.!50} (6.344,0.616)--(6.538,0.616)--(6.538,1.366)--(6.344,1.366)--cycle;
\def\gpfillpath{(2.479,0.616)--(2.674,0.616)--(2.674,1.792)--(2.479,1.792)--cycle}
\gpfill{color=gpbgfillcolor} \gpfillpath;
\gpfill{rgb color={0.373,0.620,0.627},gp pattern 4,pattern color=.} \gpfillpath;
\gpcolor{rgb color={0.373,0.620,0.627}}
\draw[gp path] (2.479,0.616)--(2.479,1.791)--(2.673,1.791)--(2.673,0.616)--cycle;
\def\gpfillpath{(3.832,0.616)--(4.026,0.616)--(4.026,1.806)--(3.832,1.806)--cycle}
\gpfill{color=gpbgfillcolor} \gpfillpath;
\gpfill{rgb color={0.373,0.620,0.627},gp pattern 4,pattern color=.} \gpfillpath;
\draw[gp path] (3.832,0.616)--(3.832,1.805)--(4.025,1.805)--(4.025,0.616)--cycle;
\def\gpfillpath{(5.185,0.616)--(5.379,0.616)--(5.379,2.024)--(5.185,2.024)--cycle}
\gpfill{color=gpbgfillcolor} \gpfillpath;
\gpfill{rgb color={0.373,0.620,0.627},gp pattern 4,pattern color=.} \gpfillpath;
\draw[gp path] (5.185,0.616)--(5.185,2.023)--(5.378,2.023)--(5.378,0.616)--cycle;
\def\gpfillpath{(6.537,0.616)--(6.731,0.616)--(6.731,2.153)--(6.537,2.153)--cycle}
\gpfill{color=gpbgfillcolor} \gpfillpath;
\gpfill{rgb color={0.373,0.620,0.627},gp pattern 4,pattern color=.} \gpfillpath;
\draw[gp path] (6.537,0.616)--(6.537,2.152)--(6.730,2.152)--(6.730,0.616)--cycle;
\def\gpfillpath{(2.673,0.616)--(2.867,0.616)--(2.867,1.961)--(2.673,1.961)--cycle}
\gpfill{color=gpbgfillcolor} \gpfillpath;
\gpfill{rgb color={0.498,0.498,0.498},gp pattern 7,pattern color=.} \gpfillpath;
\gpcolor{rgb color={0.498,0.498,0.498}}
\draw[gp path] (2.673,0.616)--(2.673,1.960)--(2.866,1.960)--(2.866,0.616)--cycle;
\def\gpfillpath{(4.025,0.616)--(4.219,0.616)--(4.219,2.037)--(4.025,2.037)--cycle}
\gpfill{color=gpbgfillcolor} \gpfillpath;
\gpfill{rgb color={0.498,0.498,0.498},gp pattern 7,pattern color=.} \gpfillpath;
\draw[gp path] (4.025,0.616)--(4.025,2.036)--(4.218,2.036)--(4.218,0.616)--cycle;
\def\gpfillpath{(5.378,0.616)--(5.572,0.616)--(5.572,2.234)--(5.378,2.234)--cycle}
\gpfill{color=gpbgfillcolor} \gpfillpath;
\gpfill{rgb color={0.498,0.498,0.498},gp pattern 7,pattern color=.} \gpfillpath;
\draw[gp path] (5.378,0.616)--(5.378,2.233)--(5.571,2.233)--(5.571,0.616)--cycle;
\def\gpfillpath{(6.730,0.616)--(6.925,0.616)--(6.925,2.396)--(6.730,2.396)--cycle}
\gpfill{color=gpbgfillcolor} \gpfillpath;
\gpfill{rgb color={0.498,0.498,0.498},gp pattern 7,pattern color=.} \gpfillpath;
\draw[gp path] (6.730,0.616)--(6.730,2.395)--(6.924,2.395)--(6.924,0.616)--cycle;
\def\gpfillpath{(2.866,0.616)--(3.060,0.616)--(3.060,2.151)--(2.866,2.151)--cycle}
\gpfill{color=gpbgfillcolor} \gpfillpath;
\gpfill{rgb color={0.294,0.000,0.510},gp pattern 6,pattern color=.} \gpfillpath;
\gpcolor{rgb color={0.294,0.000,0.510}}
\draw[gp path] (2.866,0.616)--(2.866,2.150)--(3.059,2.150)--(3.059,0.616)--cycle;
\def\gpfillpath{(4.218,0.616)--(4.413,0.616)--(4.413,2.290)--(4.218,2.290)--cycle}
\gpfill{color=gpbgfillcolor} \gpfillpath;
\gpfill{rgb color={0.294,0.000,0.510},gp pattern 6,pattern color=.} \gpfillpath;
\draw[gp path] (4.218,0.616)--(4.218,2.289)--(4.412,2.289)--(4.412,0.616)--cycle;
\def\gpfillpath{(5.571,0.616)--(5.765,0.616)--(5.765,2.615)--(5.571,2.615)--cycle}
\gpfill{color=gpbgfillcolor} \gpfillpath;
\gpfill{rgb color={0.294,0.000,0.510},gp pattern 6,pattern color=.} \gpfillpath;
\draw[gp path] (5.571,0.616)--(5.571,2.614)--(5.764,2.614)--(5.764,0.616)--cycle;
\def\gpfillpath{(6.924,0.616)--(7.118,0.616)--(7.118,2.829)--(6.924,2.829)--cycle}
\gpfill{color=gpbgfillcolor} \gpfillpath;
\gpfill{rgb color={0.294,0.000,0.510},gp pattern 6,pattern color=.} \gpfillpath;
\draw[gp path] (6.924,0.616)--(6.924,2.828)--(7.117,2.828)--(7.117,0.616)--cycle;
\def\gpfillpath{(3.059,0.616)--(3.253,0.616)--(3.253,3.131)--(3.059,3.131)--cycle}
\gpfill{color=gpbgfillcolor} \gpfillpath;
\gpfill{rgb color={1.000,0.000,0.000},gp pattern 8,pattern color=.} \gpfillpath;
\gpcolor{rgb color={1.000,0.000,0.000}}
\draw[gp path] (3.059,0.616)--(3.059,3.130)--(3.252,3.130)--(3.252,0.616)--cycle;
\def\gpfillpath{(4.412,0.616)--(4.606,0.616)--(4.606,2.998)--(4.412,2.998)--cycle}
\gpfill{color=gpbgfillcolor} \gpfillpath;
\gpfill{rgb color={1.000,0.000,0.000},gp pattern 8,pattern color=.} \gpfillpath;
\draw[gp path] (4.412,0.616)--(4.412,2.997)--(4.605,2.997)--(4.605,0.616)--cycle;
\def\gpfillpath{(5.764,0.616)--(5.958,0.616)--(5.958,3.183)--(5.764,3.183)--cycle}
\gpfill{color=gpbgfillcolor} \gpfillpath;
\gpfill{rgb color={1.000,0.000,0.000},gp pattern 8,pattern color=.} \gpfillpath;
\draw[gp path] (5.764,0.616)--(5.764,3.182)--(5.957,3.182)--(5.957,0.616)--cycle;
\def\gpfillpath{(7.117,0.616)--(7.311,0.616)--(7.311,3.146)--(7.117,3.146)--cycle}
\gpfill{color=gpbgfillcolor} \gpfillpath;
\gpfill{rgb color={1.000,0.000,0.000},gp pattern 8,pattern color=.} \gpfillpath;
\draw[gp path] (7.117,0.616)--(7.117,3.145)--(7.310,3.145)--(7.310,0.616)--cycle;
\gpcolor{color=gp lt color border}
\draw[gp path] (1.320,3.246)--(1.320,0.616)--(8.083,0.616)--(8.083,3.246)--cycle;
\gpdefrectangularnode{gp plot 1}{\pgfpoint{1.320cm}{0.616cm}}{\pgfpoint{8.083cm}{3.246cm}}
\end{tikzpicture}
\caption{Runtime performance of k-means on the Friendster-32 dataset.}
\label{fig:f32-per-iter-perf}
\end{subfigure}

\begin{subfigure}{.5\textwidth}
\begin{tikzpicture}[gnuplot]
\path (0.000,0.000) rectangle (8.636,3.556);
\gpcolor{color=gp lt color border}
\gpsetlinetype{gp lt border}
\gpsetlinewidth{1.00}
\draw[gp path] (1.320,0.616)--(1.500,0.616);
\draw[gp path] (8.083,0.616)--(7.903,0.616);
\node[gp node right] at (1.136,0.616) {$1$};
\draw[gp path] (1.320,1.012)--(1.410,1.012);
\draw[gp path] (8.083,1.012)--(7.993,1.012);
\draw[gp path] (1.320,1.243)--(1.410,1.243);
\draw[gp path] (8.083,1.243)--(7.993,1.243);
\draw[gp path] (1.320,1.408)--(1.410,1.408);
\draw[gp path] (8.083,1.408)--(7.993,1.408);
\draw[gp path] (1.320,1.535)--(1.410,1.535);
\draw[gp path] (8.083,1.535)--(7.993,1.535);
\draw[gp path] (1.320,1.639)--(1.410,1.639);
\draw[gp path] (8.083,1.639)--(7.993,1.639);
\draw[gp path] (1.320,1.727)--(1.410,1.727);
\draw[gp path] (8.083,1.727)--(7.993,1.727);
\draw[gp path] (1.320,1.804)--(1.410,1.804);
\draw[gp path] (8.083,1.804)--(7.993,1.804);
\draw[gp path] (1.320,1.871)--(1.410,1.871);
\draw[gp path] (8.083,1.871)--(7.993,1.871);
\draw[gp path] (1.320,1.931)--(1.500,1.931);
\draw[gp path] (8.083,1.931)--(7.903,1.931);
\node[gp node right] at (1.136,1.931) {$10$};
\draw[gp path] (1.320,2.327)--(1.410,2.327);
\draw[gp path] (8.083,2.327)--(7.993,2.327);
\draw[gp path] (1.320,2.558)--(1.410,2.558);
\draw[gp path] (8.083,2.558)--(7.993,2.558);
\draw[gp path] (1.320,2.723)--(1.410,2.723);
\draw[gp path] (8.083,2.723)--(7.993,2.723);
\draw[gp path] (1.320,2.850)--(1.410,2.850);
\draw[gp path] (8.083,2.850)--(7.993,2.850);
\draw[gp path] (1.320,2.954)--(1.410,2.954);
\draw[gp path] (8.083,2.954)--(7.993,2.954);
\draw[gp path] (1.320,3.042)--(1.410,3.042);
\draw[gp path] (8.083,3.042)--(7.993,3.042);
\draw[gp path] (1.320,3.119)--(1.410,3.119);
\draw[gp path] (8.083,3.119)--(7.993,3.119);
\draw[gp path] (1.320,3.186)--(1.410,3.186);
\draw[gp path] (8.083,3.186)--(7.993,3.186);
\draw[gp path] (1.320,3.246)--(1.500,3.246);
\draw[gp path] (8.083,3.246)--(7.903,3.246);
\node[gp node right] at (1.136,3.246) {$100$};
\draw[gp path] (3.574,0.616)--(3.574,0.796);
\draw[gp path] (3.574,3.246)--(3.574,3.066);
\node[gp node center] at (3.574,0.308) {Friendster-8};
\draw[gp path] (5.829,0.616)--(5.829,0.796);
\draw[gp path] (5.829,3.246)--(5.829,3.066);
\node[gp node center] at (5.829,0.308) {Friendster-32};
\draw[gp path] (1.320,3.246)--(1.320,0.616)--(8.083,0.616)--(8.083,3.246)--cycle;
\node[gp node center,rotate=-270] at (0.276,1.931) {Log Scale Memory (GB)};
\gpfill{rgb color={0.000,1.000,0.000},color=.!50} (2.930,0.616)--(3.253,0.616)--(3.253,1.559)--(2.930,1.559)--cycle;
\gpfill{rgb color={0.000,1.000,0.000},color=.!50} (5.185,0.616)--(5.508,0.616)--(5.508,2.208)--(5.185,2.208)--cycle;
\def\gpfillpath{(3.252,0.616)--(3.575,0.616)--(3.575,1.463)--(3.252,1.463)--cycle}
\gpfill{color=gpbgfillcolor} \gpfillpath;
\gpfill{rgb color={0.373,0.620,0.627},gp pattern 4,pattern color=.} \gpfillpath;
\gpcolor{rgb color={0.373,0.620,0.627}}
\draw[gp path] (3.252,0.616)--(3.252,1.462)--(3.574,1.462)--(3.574,0.616)--cycle;
\def\gpfillpath{(5.507,0.616)--(5.830,0.616)--(5.830,1.501)--(5.507,1.501)--cycle}
\gpfill{color=gpbgfillcolor} \gpfillpath;
\gpfill{rgb color={0.373,0.620,0.627},gp pattern 4,pattern color=.} \gpfillpath;
\draw[gp path] (5.507,0.616)--(5.507,1.500)--(5.829,1.500)--(5.829,0.616)--cycle;
\def\gpfillpath{(3.574,0.616)--(3.897,0.616)--(3.897,2.382)--(3.574,2.382)--cycle}
\gpfill{color=gpbgfillcolor} \gpfillpath;
\gpfill{rgb color={0.498,0.498,0.498},gp pattern 7,pattern color=.} \gpfillpath;
\gpcolor{rgb color={0.498,0.498,0.498}}
\draw[gp path] (3.574,0.616)--(3.574,2.381)--(3.896,2.381)--(3.896,0.616)--cycle;
\def\gpfillpath{(5.829,0.616)--(6.152,0.616)--(6.152,2.851)--(5.829,2.851)--cycle}
\gpfill{color=gpbgfillcolor} \gpfillpath;
\gpfill{rgb color={0.498,0.498,0.498},gp pattern 7,pattern color=.} \gpfillpath;
\draw[gp path] (5.829,0.616)--(5.829,2.850)--(6.151,2.850)--(6.151,0.616)--cycle;
\def\gpfillpath{(3.896,0.616)--(4.219,0.616)--(4.219,2.578)--(3.896,2.578)--cycle}
\gpfill{color=gpbgfillcolor} \gpfillpath;
\gpfill{rgb color={0.294,0.000,0.510},gp pattern 6,pattern color=.} \gpfillpath;
\gpcolor{rgb color={0.294,0.000,0.510}}
\draw[gp path] (3.896,0.616)--(3.896,2.577)--(4.218,2.577)--(4.218,0.616)--cycle;
\def\gpfillpath{(6.151,0.616)--(6.474,0.616)--(6.474,2.778)--(6.151,2.778)--cycle}
\gpfill{color=gpbgfillcolor} \gpfillpath;
\gpfill{rgb color={0.294,0.000,0.510},gp pattern 6,pattern color=.} \gpfillpath;
\draw[gp path] (6.151,0.616)--(6.151,2.777)--(6.473,2.777)--(6.473,0.616)--cycle;
\def\gpfillpath{(4.218,0.616)--(4.541,0.616)--(4.541,1.872)--(4.218,1.872)--cycle}
\gpfill{color=gpbgfillcolor} \gpfillpath;
\gpfill{rgb color={1.000,0.000,0.000},gp pattern 8,pattern color=.} \gpfillpath;
\gpcolor{rgb color={1.000,0.000,0.000}}
\draw[gp path] (4.218,0.616)--(4.218,1.871)--(4.540,1.871)--(4.540,0.616)--cycle;
\def\gpfillpath{(6.473,0.616)--(6.796,0.616)--(6.796,2.596)--(6.473,2.596)--cycle}
\gpfill{color=gpbgfillcolor} \gpfillpath;
\gpfill{rgb color={1.000,0.000,0.000},gp pattern 8,pattern color=.} \gpfillpath;
\draw[gp path] (6.473,0.616)--(6.473,2.595)--(6.795,2.595)--(6.795,0.616)--cycle;
\gpcolor{color=gp lt color border}
\draw[gp path] (1.320,3.246)--(1.320,0.616)--(8.083,0.616)--(8.083,3.246)--cycle;
\gpdefrectangularnode{gp plot 1}{\pgfpoint{1.320cm}{0.616cm}}{\pgfpoint{8.083cm}{3.246cm}}
\end{tikzpicture}
\caption{Peak memory consumption on the Friendster datasets,
    with $k=10$. Row cache size = $512$MB, page cache size = $1$GB.
}
\label{fig:mem}
\end{subfigure}

\caption{\textsf{knor} routines outperform competitor solutions in
runtime performance and memory consumption.}
\label{fig:per-iter-perf}
\end{figure}

Both our in-memory and semi-external memory modules incur
little memory overhead. Figure \ref{fig:mem} shows memory consumption.
We note that MLlib requires the placement of temporary
Spark block manager files. Because the block manager cannot be disabled,
we provide an in-memory RAM-disk so as to not influence MLlib's performance
negatively. We configure MLlib, H$_2$O and Turi to
use the minimum amount of memory necessary to achieve their highest
performance. We acknowledge that a reduction in memory for these frameworks
is possible,
but would degrade runtime and lead to unfair comparisons.

We demonstrate that \textsf{knori}
is no less than an order of magnitude faster than
competitors (Figure \ref{fig:per-iter-perf}).
\textsf{knori} is often hundreds of
times faster than Turi.

\textsf{knors} is consistently
twice as fast as competitor in-memory frameworks.
We further demonstrate performance improvements over competitor
frameworks on algorithmically identical implementations by \textit{disabling} MTI.
\textsf{knori-} is nearly $10$x
faster than competitor solutions, whereas \textsf{knors-}
is comparable and often faster than competitor in-memory solutions.
We have shown performance gains over other frameworks when MTI is \textit{disabled} is due
to $||$Lloyd's parallelization scheme and NUMA optimizations.
Lastly, Figure \ref{fig:knor-comp} demonstrates a
consistent $30\%$ improvement in \textsf{knors} when we utilize the row cache. This is
evidence that the design of our lazily updated row cache provides a performance boost.

Comparing \textsf{knori-} and \textsf{knors-{}-} to MLlib, H$_2$O and Turi
(Figures \ref{fig:knor-comp} and \ref{fig:per-iter-perf}) reveals \textsf{knor}
to be several times faster and to use significantly less memory.
This is relevant because \textsf{knori-} and \textsf{knors-{}-} are
algorithmically identical to k-means within MLlib, Turi and H$_2$O.

\subsection{Single-node Scalability}

To demonstrate scalability, we compare the performance of k-means on
synthetic datasets drawn from
random distributions that contain hundreds of
millions to billions of data points. Uniformly random data are typically
the worst case scenario for the convergence of k-means, because many
data points tend to be near several centroids.

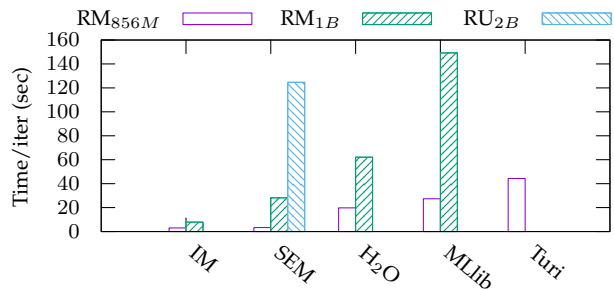
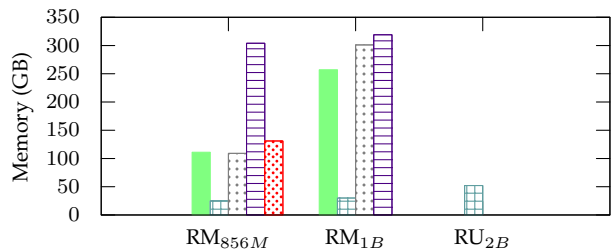
\begin{figure}[!htb]
\centering
\footnotesize
\vspace{-5pt}
\begin{subfigure}{.5\textwidth}
		\begin{tikzpicture}[gnuplot]
\path (0.000,0.000) rectangle (8.636,4.064);
\gpcolor{color=gp lt color border}
\gpsetlinetype{gp lt border}
\gpsetlinewidth{1.00}
\draw[gp path] (1.320,0.899)--(1.500,0.899);
\draw[gp path] (8.083,0.899)--(7.903,0.899);
\node[gp node right] at (1.136,0.899) {$0$};
\draw[gp path] (1.320,1.218)--(1.500,1.218);
\draw[gp path] (8.083,1.218)--(7.903,1.218);
\node[gp node right] at (1.136,1.218) {$20$};
\draw[gp path] (1.320,1.536)--(1.500,1.536);
\draw[gp path] (8.083,1.536)--(7.903,1.536);
\node[gp node right] at (1.136,1.536) {$40$};
\draw[gp path] (1.320,1.855)--(1.500,1.855);
\draw[gp path] (8.083,1.855)--(7.903,1.855);
\node[gp node right] at (1.136,1.855) {$60$};
\draw[gp path] (1.320,2.173)--(1.500,2.173);
\draw[gp path] (8.083,2.173)--(7.903,2.173);
\node[gp node right] at (1.136,2.173) {$80$};
\draw[gp path] (1.320,2.492)--(1.500,2.492);
\draw[gp path] (8.083,2.492)--(7.903,2.492);
\node[gp node right] at (1.136,2.492) {$100$};
\draw[gp path] (1.320,2.810)--(1.500,2.810);
\draw[gp path] (8.083,2.810)--(7.903,2.810);
\node[gp node right] at (1.136,2.810) {$120$};
\draw[gp path] (1.320,3.129)--(1.500,3.129);
\draw[gp path] (8.083,3.129)--(7.903,3.129);
\node[gp node right] at (1.136,3.129) {$140$};
\draw[gp path] (1.320,3.447)--(1.500,3.447);
\draw[gp path] (8.083,3.447)--(7.903,3.447);
\node[gp node right] at (1.136,3.447) {$160$};
\draw[gp path] (2.447,0.899)--(2.447,1.079);
\draw[gp path] (2.447,3.447)--(2.447,3.267);
\node[gp node left,rotate=-40] at (2.447,0.715) {IM};
\draw[gp path] (3.574,0.899)--(3.574,1.079);
\draw[gp path] (3.574,3.447)--(3.574,3.267);
\node[gp node left,rotate=-40] at (3.574,0.715) {SEM};
\draw[gp path] (4.702,0.899)--(4.702,1.079);
\draw[gp path] (4.702,3.447)--(4.702,3.267);
\node[gp node left,rotate=-40] at (4.702,0.715) {H$_2$O};
\draw[gp path] (5.829,0.899)--(5.829,1.079);
\draw[gp path] (5.829,3.447)--(5.829,3.267);
\node[gp node left,rotate=-40] at (5.829,0.715) {MLlib};
\draw[gp path] (6.956,0.899)--(6.956,1.079);
\draw[gp path] (6.956,3.447)--(6.956,3.267);
\node[gp node left,rotate=-40] at (6.956,0.715) {Turi};
\draw[gp path] (1.320,3.447)--(1.320,0.899)--(8.083,0.899)--(8.083,3.447)--cycle;
\node[gp node center,rotate=-270] at (0.276,2.173) {Time/iter (sec)};
\node[gp node right] at (2.223,3.730) {RM$_{856M}$};
\def\gpfillpath{(2.407,3.653)--(3.323,3.653)--(3.323,3.807)--(2.407,3.807)--cycle}
\gpfill{color=gpbgfillcolor} \gpfillpath;
\gpfill{rgb color={0.580,0.000,0.827},gp pattern 0,pattern color=.} \gpfillpath;
\gpcolor{rgb color={0.580,0.000,0.827}}
\draw[gp path] (2.407,3.653)--(3.323,3.653)--(3.323,3.807)--(2.407,3.807)--cycle;
\def\gpfillpath{(2.222,0.899)--(2.448,0.899)--(2.448,0.948)--(2.222,0.948)--cycle}
\gpfill{color=gpbgfillcolor} \gpfillpath;
\gpfill{rgb color={0.580,0.000,0.827},gp pattern 0,pattern color=.} \gpfillpath;
\draw[gp path] (2.222,0.899)--(2.222,0.947)--(2.447,0.947)--(2.447,0.899)--cycle;
\def\gpfillpath{(3.349,0.899)--(3.575,0.899)--(3.575,0.953)--(3.349,0.953)--cycle}
\gpfill{color=gpbgfillcolor} \gpfillpath;
\gpfill{rgb color={0.580,0.000,0.827},gp pattern 0,pattern color=.} \gpfillpath;
\draw[gp path] (3.349,0.899)--(3.349,0.952)--(3.574,0.952)--(3.574,0.899)--cycle;
\def\gpfillpath{(4.476,0.899)--(4.703,0.899)--(4.703,1.215)--(4.476,1.215)--cycle}
\gpfill{color=gpbgfillcolor} \gpfillpath;
\gpfill{rgb color={0.580,0.000,0.827},gp pattern 0,pattern color=.} \gpfillpath;
\draw[gp path] (4.476,0.899)--(4.476,1.214)--(4.702,1.214)--(4.702,0.899)--cycle;
\def\gpfillpath{(5.603,0.899)--(5.830,0.899)--(5.830,1.336)--(5.603,1.336)--cycle}
\gpfill{color=gpbgfillcolor} \gpfillpath;
\gpfill{rgb color={0.580,0.000,0.827},gp pattern 0,pattern color=.} \gpfillpath;
\draw[gp path] (5.603,0.899)--(5.603,1.335)--(5.829,1.335)--(5.829,0.899)--cycle;
\def\gpfillpath{(6.730,0.899)--(6.957,0.899)--(6.957,1.606)--(6.730,1.606)--cycle}
\gpfill{color=gpbgfillcolor} \gpfillpath;
\gpfill{rgb color={0.580,0.000,0.827},gp pattern 0,pattern color=.} \gpfillpath;
\draw[gp path] (6.730,0.899)--(6.730,1.605)--(6.956,1.605)--(6.956,0.899)--cycle;
\gpcolor{color=gp lt color border}
\node[gp node right] at (4.611,3.730) {RM$_{1B}$};
\def\gpfillpath{(4.795,3.653)--(5.711,3.653)--(5.711,3.807)--(4.795,3.807)--cycle}
\gpfill{color=gpbgfillcolor} \gpfillpath;
\gpfill{rgb color={0.000,0.620,0.451},gp pattern 1,pattern color=.} \gpfillpath;
\gpcolor{rgb color={0.000,0.620,0.451}}
\draw[gp path] (4.795,3.653)--(5.711,3.653)--(5.711,3.807)--(4.795,3.807)--cycle;
\def\gpfillpath{(2.447,0.899)--(2.674,0.899)--(2.674,1.025)--(2.447,1.025)--cycle}
\gpfill{color=gpbgfillcolor} \gpfillpath;
\gpfill{rgb color={0.000,0.620,0.451},gp pattern 1,pattern color=.} \gpfillpath;
\draw[gp path] (2.447,0.899)--(2.447,1.024)--(2.673,1.024)--(2.673,0.899)--cycle;
\def\gpfillpath{(3.574,0.899)--(3.801,0.899)--(3.801,1.348)--(3.574,1.348)--cycle}
\gpfill{color=gpbgfillcolor} \gpfillpath;
\gpfill{rgb color={0.000,0.620,0.451},gp pattern 1,pattern color=.} \gpfillpath;
\draw[gp path] (3.574,0.899)--(3.574,1.347)--(3.800,1.347)--(3.800,0.899)--cycle;
\def\gpfillpath{(4.702,0.899)--(4.928,0.899)--(4.928,1.889)--(4.702,1.889)--cycle}
\gpfill{color=gpbgfillcolor} \gpfillpath;
\gpfill{rgb color={0.000,0.620,0.451},gp pattern 1,pattern color=.} \gpfillpath;
\draw[gp path] (4.702,0.899)--(4.702,1.888)--(4.927,1.888)--(4.927,0.899)--cycle;
\def\gpfillpath{(5.829,0.899)--(6.055,0.899)--(6.055,3.276)--(5.829,3.276)--cycle}
\gpfill{color=gpbgfillcolor} \gpfillpath;
\gpfill{rgb color={0.000,0.620,0.451},gp pattern 1,pattern color=.} \gpfillpath;
\draw[gp path] (5.829,0.899)--(5.829,3.275)--(6.054,3.275)--(6.054,0.899)--cycle;
\gpcolor{color=gp lt color border}
\node[gp node right] at (6.999,3.730) {RU$_{2B}$};
\def\gpfillpath{(7.183,3.653)--(8.099,3.653)--(8.099,3.807)--(7.183,3.807)--cycle}
\gpfill{color=gpbgfillcolor} \gpfillpath;
\gpfill{rgb color={0.337,0.706,0.914},gp pattern 2,pattern color=.} \gpfillpath;
\gpcolor{rgb color={0.337,0.706,0.914}}
\draw[gp path] (7.183,3.653)--(8.099,3.653)--(8.099,3.807)--(7.183,3.807)--cycle;
\def\gpfillpath{(3.800,0.899)--(4.026,0.899)--(4.026,2.885)--(3.800,2.885)--cycle}
\gpfill{color=gpbgfillcolor} \gpfillpath;
\gpfill{rgb color={0.337,0.706,0.914},gp pattern 2,pattern color=.} \gpfillpath;
\draw[gp path] (3.800,0.899)--(3.800,2.884)--(4.025,2.884)--(4.025,0.899)--cycle;
\gpcolor{color=gp lt color border}
\draw[gp path] (1.320,3.447)--(1.320,0.899)--(8.083,0.899)--(8.083,3.447)--cycle;
\gpdefrectangularnode{gp plot 1}{\pgfpoint{1.320cm}{0.899cm}}{\pgfpoint{8.083cm}{3.447cm}}
\end{tikzpicture}
\vspace{-10pt}
\caption{Per iteration runtime of each routine.}
\label{fig:perf-rmvm}
\end{subfigure}

\begin{subfigure}{.5\textwidth}
\begin{tikzpicture}[gnuplot]
\path (0.000,0.000) rectangle (8.636,3.556);
\gpcolor{color=gp lt color border}
\gpsetlinetype{gp lt border}
\gpsetlinewidth{1.00}
\draw[gp path] (1.320,0.616)--(1.500,0.616);
\draw[gp path] (8.083,0.616)--(7.903,0.616);
\node[gp node right] at (1.136,0.616) {$0$};
\draw[gp path] (1.320,0.992)--(1.500,0.992);
\draw[gp path] (8.083,0.992)--(7.903,0.992);
\node[gp node right] at (1.136,0.992) {$50$};
\draw[gp path] (1.320,1.367)--(1.500,1.367);
\draw[gp path] (8.083,1.367)--(7.903,1.367);
\node[gp node right] at (1.136,1.367) {$100$};
\draw[gp path] (1.320,1.743)--(1.500,1.743);
\draw[gp path] (8.083,1.743)--(7.903,1.743);
\node[gp node right] at (1.136,1.743) {$150$};
\draw[gp path] (1.320,2.119)--(1.500,2.119);
\draw[gp path] (8.083,2.119)--(7.903,2.119);
\node[gp node right] at (1.136,2.119) {$200$};
\draw[gp path] (1.320,2.495)--(1.500,2.495);
\draw[gp path] (8.083,2.495)--(7.903,2.495);
\node[gp node right] at (1.136,2.495) {$250$};
\draw[gp path] (1.320,2.870)--(1.500,2.870);
\draw[gp path] (8.083,2.870)--(7.903,2.870);
\node[gp node right] at (1.136,2.870) {$300$};
\draw[gp path] (1.320,3.246)--(1.500,3.246);
\draw[gp path] (8.083,3.246)--(7.903,3.246);
\node[gp node right] at (1.136,3.246) {$350$};
\draw[gp path] (3.011,0.616)--(3.011,0.796);
\draw[gp path] (3.011,3.246)--(3.011,3.066);
\node[gp node center] at (3.011,0.308) {RM$_{856M}$};
\draw[gp path] (4.702,0.616)--(4.702,0.796);
\draw[gp path] (4.702,3.246)--(4.702,3.066);
\node[gp node center] at (4.702,0.308) {RM$_{1B}$};
\draw[gp path] (6.392,0.616)--(6.392,0.796);
\draw[gp path] (6.392,3.246)--(6.392,3.066);
\node[gp node center] at (6.392,0.308) {RU$_{2B}$};
\draw[gp path] (1.320,3.246)--(1.320,0.616)--(8.083,0.616)--(8.083,3.246)--cycle;
\node[gp node center,rotate=-270] at (0.276,1.931) {Memory (GB)};
\gpfill{rgb color={0.000,1.000,0.000},color=.!50} (2.528,0.616)--(2.770,0.616)--(2.770,1.451)--(2.528,1.451)--cycle;
\gpfill{rgb color={0.000,1.000,0.000},color=.!50} (4.218,0.616)--(4.461,0.616)--(4.461,2.548)--(4.218,2.548)--cycle;
\def\gpfillpath{(2.769,0.616)--(3.012,0.616)--(3.012,0.805)--(2.769,0.805)--cycle}
\gpfill{color=gpbgfillcolor} \gpfillpath;
\gpfill{rgb color={0.373,0.620,0.627},gp pattern 4,pattern color=.} \gpfillpath;
\gpcolor{rgb color={0.373,0.620,0.627}}
\draw[gp path] (2.769,0.616)--(2.769,0.804)--(3.011,0.804)--(3.011,0.616)--cycle;
\def\gpfillpath{(4.460,0.616)--(4.703,0.616)--(4.703,0.842)--(4.460,0.842)--cycle}
\gpfill{color=gpbgfillcolor} \gpfillpath;
\gpfill{rgb color={0.373,0.620,0.627},gp pattern 4,pattern color=.} \gpfillpath;
\draw[gp path] (4.460,0.616)--(4.460,0.841)--(4.702,0.841)--(4.702,0.616)--cycle;
\def\gpfillpath{(6.151,0.616)--(6.393,0.616)--(6.393,1.008)--(6.151,1.008)--cycle}
\gpfill{color=gpbgfillcolor} \gpfillpath;
\gpfill{rgb color={0.373,0.620,0.627},gp pattern 4,pattern color=.} \gpfillpath;
\draw[gp path] (6.151,0.616)--(6.151,1.007)--(6.392,1.007)--(6.392,0.616)--cycle;
\def\gpfillpath{(3.011,0.616)--(3.253,0.616)--(3.253,1.436)--(3.011,1.436)--cycle}
\gpfill{color=gpbgfillcolor} \gpfillpath;
\gpfill{rgb color={0.498,0.498,0.498},gp pattern 7,pattern color=.} \gpfillpath;
\gpcolor{rgb color={0.498,0.498,0.498}}
\draw[gp path] (3.011,0.616)--(3.011,1.435)--(3.252,1.435)--(3.252,0.616)--cycle;
\def\gpfillpath{(4.702,0.616)--(4.944,0.616)--(4.944,2.879)--(4.702,2.879)--cycle}
\gpfill{color=gpbgfillcolor} \gpfillpath;
\gpfill{rgb color={0.498,0.498,0.498},gp pattern 7,pattern color=.} \gpfillpath;
\draw[gp path] (4.702,0.616)--(4.702,2.878)--(4.943,2.878)--(4.943,0.616)--cycle;
\def\gpfillpath{(3.252,0.616)--(3.495,0.616)--(3.495,2.901)--(3.252,2.901)--cycle}
\gpfill{color=gpbgfillcolor} \gpfillpath;
\gpfill{rgb color={0.294,0.000,0.510},gp pattern 6,pattern color=.} \gpfillpath;
\gpcolor{rgb color={0.294,0.000,0.510}}
\draw[gp path] (3.252,0.616)--(3.252,2.900)--(3.494,2.900)--(3.494,0.616)--cycle;
\def\gpfillpath{(4.943,0.616)--(5.186,0.616)--(5.186,3.014)--(4.943,3.014)--cycle}
\gpfill{color=gpbgfillcolor} \gpfillpath;
\gpfill{rgb color={0.294,0.000,0.510},gp pattern 6,pattern color=.} \gpfillpath;
\draw[gp path] (4.943,0.616)--(4.943,3.013)--(5.185,3.013)--(5.185,0.616)--cycle;
\def\gpfillpath{(3.494,0.616)--(3.736,0.616)--(3.736,1.601)--(3.494,1.601)--cycle}
\gpfill{color=gpbgfillcolor} \gpfillpath;
\gpfill{rgb color={1.000,0.000,0.000},gp pattern 8,pattern color=.} \gpfillpath;
\gpcolor{rgb color={1.000,0.000,0.000}}
\draw[gp path] (3.494,0.616)--(3.494,1.600)--(3.735,1.600)--(3.735,0.616)--cycle;
\gpcolor{color=gp lt color border}
\draw[gp path] (1.320,3.246)--(1.320,0.616)--(8.083,0.616)--(8.083,3.246)--cycle;
\gpdefrectangularnode{gp plot 1}{\pgfpoint{1.320cm}{0.616cm}}{\pgfpoint{8.083cm}{3.246cm}}
\end{tikzpicture}
\vspace{-10pt}
\caption{Memory consumption of each routine.}
\label{fig:mem-rmvm}
\end{subfigure}
\caption{Performance comparison on RM$_{856M}$ and RM$_{1B}$ datasets.
  Turi is unable to run on RM$_{1B}$ on our machine and only SEM routines
  are able to run on RU$_{2B}$ on our machine. Page cache size = $4$GB,
  Row cache size = $2$GB, and k = $100$.}
\label{fig:rmvm}
\end{figure}

Both in-memory and SEM modules outperform popular
frameworks on $100$GB+ datasets. Figure \ref{fig:perf-rmvm} shows that we
achieve $7$-$20$x improvement when
in-memory and $3$-$6$x improvement in SEM when compared to MLlib, H$_2$O and Turi.
As data increases in size, the performance difference between \textsf{knori} and
\textsf{knors} narrows because there is now enough data to mask I/O latency which turns
\textsf{knors} from I/O bound to computation bound. We observe
\textsf{knors} is only $3$-$4$x slower than its in-memory counterpart in such cases.

Memory capacity limits the scalability of k-means and semi-external memory
allows algorithms to scale well beyond the limits of physical memory.
The 1B point matrix (RM$_{1B}$) is the largest that fits in 1TB of memory
on our machine. Figure \ref{fig:rmvm} shows that at 2B points (RU$_{2B}$),
semi-external memory algorithms
continue to execute proportionally and all other algorithms fail.

\subsection{Distributed Execution} \label{sec:dist-eval}

We demonstrate the performance and scalability of \textsf{knord} and \textsf{knord-}.
We analyze their performance on Amazon's EC$2$ cloud.
We compare against \textit{(i)} MLlib (\textbf{MLlib-EC$2$}), \textit{(ii)} a
pure MPI implementation of our $||$Lloyd's algorithm with
MTI pruning (\textbf{MPI}), and \textit{(iii)} a pure MPI implementation of $||$Lloyd's algorithm with
pruning \textit{disabled} (\textbf{MPI-}).
H$_2$O has no distributed memory implementation and Turi
discontinued their distributed memory interface prior to our experiments.

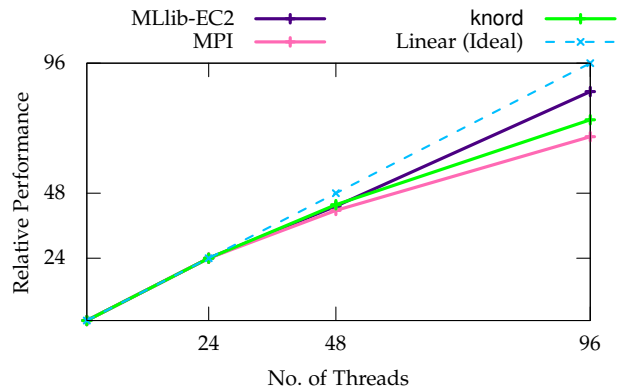
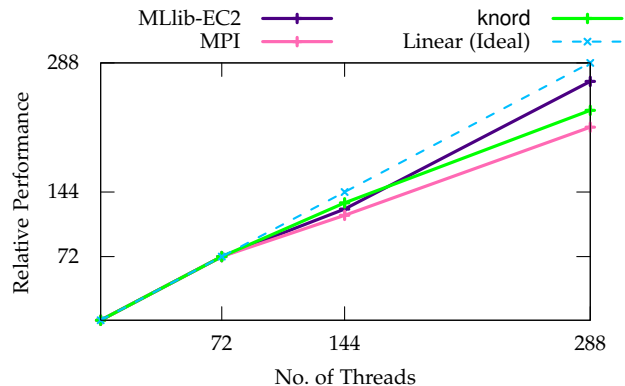
\begin{figure}[!htb]
    \centering
    \footnotesize
    \vspace{-5pt}
    \begin{subfigure}{.5\textwidth}
        \begin{tikzpicture}[gnuplot]
\path (0.000,0.000) rectangle (8.382,5.334);
\gpcolor{color=gp lt color border}
\gpsetlinetype{gp lt border}
\gpsetdashtype{gp dt solid}
\gpsetlinewidth{1.00}
\draw[gp path] (1.136,1.814)--(1.316,1.814);
\draw[gp path] (7.829,1.814)--(7.649,1.814);
\node[gp node right] at (0.952,1.814) {24};
\draw[gp path] (1.136,2.679)--(1.316,2.679);
\draw[gp path] (7.829,2.679)--(7.649,2.679);
\node[gp node right] at (0.952,2.679) {48};
\draw[gp path] (1.136,4.409)--(1.316,4.409);
\draw[gp path] (7.829,4.409)--(7.649,4.409);
\node[gp node right] at (0.952,4.409) {96};
\draw[gp path] (2.756,0.985)--(2.756,1.165);
\draw[gp path] (2.756,4.409)--(2.756,4.229);
\node[gp node center] at (2.756,0.677) {24};
\draw[gp path] (4.447,0.985)--(4.447,1.165);
\draw[gp path] (4.447,4.409)--(4.447,4.229);
\node[gp node center] at (4.447,0.677) {48};
\draw[gp path] (7.829,0.985)--(7.829,1.165);
\draw[gp path] (7.829,4.409)--(7.829,4.229);
\node[gp node center] at (7.829,0.677) {96};
\draw[gp path] (1.136,4.409)--(1.136,0.985)--(7.829,0.985)--(7.829,4.409)--cycle;
\node[gp node center,rotate=-270] at (0.292,2.697) {Relative Performance};
\node[gp node center] at (4.482,0.215) {No. of Threads};
\node[gp node right] at (3.198,5.000) {MLlib-EC2};
\gpcolor{rgb color={0.294,0.000,0.510}}
\gpsetlinewidth{3.00}
\draw[gp path] (3.382,5.000)--(4.298,5.000);
\draw[gp path] (1.136,0.985)--(2.756,1.814)--(4.447,2.506)--(7.829,4.031);
\gpsetpointsize{4.00}
\gppoint{gp mark 1}{(1.136,0.985)}
\gppoint{gp mark 1}{(2.756,1.814)}
\gppoint{gp mark 1}{(4.447,2.506)}
\gppoint{gp mark 1}{(7.829,4.031)}
\gppoint{gp mark 1}{(3.840,5.000)}
\gpcolor{color=gp lt color border}
\node[gp node right] at (3.198,4.692) {MPI};
\gpcolor{rgb color={1.000,0.412,0.706}}
\draw[gp path] (3.382,4.692)--(4.298,4.692);
\draw[gp path] (1.136,0.985)--(2.756,1.814)--(4.447,2.450)--(7.829,3.431);
\gppoint{gp mark 1}{(1.136,0.985)}
\gppoint{gp mark 1}{(2.756,1.814)}
\gppoint{gp mark 1}{(4.447,2.450)}
\gppoint{gp mark 1}{(7.829,3.431)}
\gppoint{gp mark 1}{(3.840,4.692)}
\gpcolor{color=gp lt color border}
\node[gp node right] at (7.058,5.000) {\textsf{knord}};
\gpcolor{rgb color={0.000,1.000,0.000}}
\draw[gp path] (7.242,5.000)--(8.158,5.000);
\draw[gp path] (1.136,0.985)--(2.756,1.814)--(4.447,2.526)--(7.829,3.656);
\gppoint{gp mark 1}{(1.136,0.985)}
\gppoint{gp mark 1}{(2.756,1.814)}
\gppoint{gp mark 1}{(4.447,2.526)}
\gppoint{gp mark 1}{(7.829,3.656)}
\gppoint{gp mark 1}{(7.700,5.000)}
\gpcolor{color=gp lt color border}
\node[gp node right] at (7.058,4.692) {Linear (Ideal)};
\gpcolor{rgb color={0.000,0.749,1.000}}
\gpsetdashtype{gp dt 3}
\gpsetlinewidth{2.00}
\draw[gp path] (7.242,4.692)--(8.158,4.692);
\draw[gp path] (1.136,0.985)--(2.756,1.814)--(4.447,2.679)--(7.829,4.409);
\gppoint{gp mark 2}{(1.136,0.985)}
\gppoint{gp mark 2}{(2.756,1.814)}
\gppoint{gp mark 2}{(4.447,2.679)}
\gppoint{gp mark 2}{(7.829,4.409)}
\gppoint{gp mark 2}{(7.700,4.692)}
\gpcolor{color=gp lt color border}
\gpsetdashtype{gp dt solid}
\gpsetlinewidth{1.00}
\draw[gp path] (1.136,4.409)--(1.136,0.985)--(7.829,0.985)--(7.829,4.409)--cycle;
\gpdefrectangularnode{gp plot 1}{\pgfpoint{1.136cm}{0.985cm}}{\pgfpoint{7.829cm}{4.409cm}}
\end{tikzpicture}
        \vspace{-10pt}
        \caption {Distributed speedup comparison on the Friendster-32 dataset.}
        \label{fig:dist-speedupfr32}
    \end{subfigure}

    \begin{subfigure}{.5\textwidth}
        \begin{tikzpicture}[gnuplot]
\path (0.000,0.000) rectangle (8.382,5.334);
\gpcolor{color=gp lt color border}
\gpsetlinetype{gp lt border}
\gpsetdashtype{gp dt solid}
\gpsetlinewidth{1.00}
\draw[gp path] (1.320,1.832)--(1.500,1.832);
\draw[gp path] (7.829,1.832)--(7.649,1.832);
\node[gp node right] at (1.136,1.832) {72};
\draw[gp path] (1.320,2.691)--(1.500,2.691);
\draw[gp path] (7.829,2.691)--(7.649,2.691);
\node[gp node right] at (1.136,2.691) {144};
\draw[gp path] (1.320,4.409)--(1.500,4.409);
\draw[gp path] (7.829,4.409)--(7.649,4.409);
\node[gp node right] at (1.136,4.409) {288};
\draw[gp path] (2.930,0.985)--(2.930,1.165);
\draw[gp path] (2.930,4.409)--(2.930,4.229);
\node[gp node center] at (2.930,0.677) {72};
\draw[gp path] (4.563,0.985)--(4.563,1.165);
\draw[gp path] (4.563,4.409)--(4.563,4.229);
\node[gp node center] at (4.563,0.677) {144};
\draw[gp path] (7.829,0.985)--(7.829,1.165);
\draw[gp path] (7.829,4.409)--(7.829,4.229);
\node[gp node center] at (7.829,0.677) {288};
\draw[gp path] (1.320,4.409)--(1.320,0.985)--(7.829,0.985)--(7.829,4.409)--cycle;
\node[gp node center,rotate=-270] at (0.292,2.697) {Relative Performance};
\node[gp node center] at (4.574,0.215) {No. of Threads};
\node[gp node right] at (3.290,5.000) {MLlib-EC2};
\gpcolor{rgb color={0.294,0.000,0.510}}
\gpsetlinewidth{3.00}
\draw[gp path] (3.474,5.000)--(4.390,5.000);
\draw[gp path] (1.320,0.985)--(2.930,1.832)--(4.563,2.464)--(7.829,4.163);
\gpsetpointsize{4.00}
\gppoint{gp mark 1}{(1.320,0.985)}
\gppoint{gp mark 1}{(2.930,1.832)}
\gppoint{gp mark 1}{(4.563,2.464)}
\gppoint{gp mark 1}{(7.829,4.163)}
\gppoint{gp mark 1}{(3.932,5.000)}
\gpcolor{color=gp lt color border}
\node[gp node right] at (3.290,4.692) {MPI};
\gpcolor{rgb color={1.000,0.412,0.706}}
\draw[gp path] (3.474,4.692)--(4.390,4.692);
\draw[gp path] (1.320,0.985)--(2.930,1.832)--(4.563,2.381)--(7.829,3.554);
\gppoint{gp mark 1}{(1.320,0.985)}
\gppoint{gp mark 1}{(2.930,1.832)}
\gppoint{gp mark 1}{(4.563,2.381)}
\gppoint{gp mark 1}{(7.829,3.554)}
\gppoint{gp mark 1}{(3.932,4.692)}
\gpcolor{color=gp lt color border}
\node[gp node right] at (7.150,5.000) {\textsf{knord}};
\gpcolor{rgb color={0.000,1.000,0.000}}
\draw[gp path] (7.334,5.000)--(8.250,5.000);
\draw[gp path] (1.320,0.985)--(2.930,1.832)--(4.563,2.547)--(7.829,3.778);
\gppoint{gp mark 1}{(1.320,0.985)}
\gppoint{gp mark 1}{(2.930,1.832)}
\gppoint{gp mark 1}{(4.563,2.547)}
\gppoint{gp mark 1}{(7.829,3.778)}
\gppoint{gp mark 1}{(7.792,5.000)}
\gpcolor{color=gp lt color border}
\node[gp node right] at (7.150,4.692) {Linear (Ideal)};
\gpcolor{rgb color={0.000,0.749,1.000}}
\gpsetdashtype{gp dt 3}
\gpsetlinewidth{2.00}
\draw[gp path] (7.334,4.692)--(8.250,4.692);
\draw[gp path] (1.320,0.985)--(2.930,1.832)--(4.563,2.691)--(7.829,4.409);
\gppoint{gp mark 2}{(1.320,0.985)}
\gppoint{gp mark 2}{(2.930,1.832)}
\gppoint{gp mark 2}{(4.563,2.691)}
\gppoint{gp mark 2}{(7.829,4.409)}
\gppoint{gp mark 2}{(7.792,4.692)}
\gpcolor{color=gp lt color border}
\gpsetdashtype{gp dt solid}
\gpsetlinewidth{1.00}
\draw[gp path] (1.320,4.409)--(1.320,0.985)--(7.829,0.985)--(7.829,4.409)--cycle;
\gpdefrectangularnode{gp plot 1}{\pgfpoint{1.320cm}{0.985cm}}{\pgfpoint{7.829cm}{4.409cm}}
\end{tikzpicture}
        \vspace{-10pt}
        \caption{Distributed speedup comparison on the RM$_{1B}$ dataset.}
        \label{fig:dist-speeduprm1b}
    \end{subfigure}

    \caption {Speedup experiments are normalized to each implementation's
			serial performance. MLlib-EC$2$'s serial performance is over $5$
            times slower than that of \textsf{knord}. This leads the speedup of
            MLlib-EC$2$ to improve faster than other routines with respect to
            \textit{itself}, but MLlib-EC$2$ remains an order of magnitude
            slower than \textsf{knord}. Each machine has 18 physical cores
            with 1 thread per core.}
		\label{fig:dist-speedup}
        \vspace{-5pt}
\end{figure}

\begin{figure}[!htb]
    \centering
    \begin{subfigure}{.5\textwidth}
        \vspace{-10pt}
        \begin{center}
            \begin{tikzpicture}[gnuplot]
\gpcolor{color=gp lt color border}
\gpsetlinetype{gp lt border}
\gpsetlinewidth{1.00}
\node[gp node right] at (3.198,4.492) {\textsf{knord}};
\gpfill{rgb color={0.000,1.000,0.000},color=.!50} (3.382,4.415)--(4.298,4.415)--(4.298,4.569)--(3.382,4.569)--cycle;
\node[gp node right] at (3.198,4.184) {MPI};
\gpfill{rgb color={1.000,0.412,0.706},color=.!50} (3.382,4.107)--(4.298,4.107)--(4.298,4.261)--(3.382,4.261)--cycle;
\node[gp node right] at (3.198,3.876) {\textsf{knord-}};
\def\gpfillpath{(3.382,3.799)--(4.298,3.799)--(4.298,3.953)--(3.382,3.953)--cycle}
\gpfill{color=gpbgfillcolor} \gpfillpath;
\gpfill{rgb color={0.498,1.000,0.831},gp pattern 3,pattern color=.} \gpfillpath;
\node[gp node right] at (6.138,4.492) {MPI-};
\gpfill{rgb color={1.000,0.000,1.000},color=.!50} (6.322,4.415)--(7.238,4.415)--(7.238,4.569)--(6.322,4.569)--cycle;
\node[gp node right] at (6.138,4.184) {MLlib-EC2};
\def\gpfillpath{(6.322,4.107)--(7.238,4.107)--(7.238,4.261)--(6.322,4.261)--cycle}
\gpfill{color=gpbgfillcolor} \gpfillpath;
\gpfill{rgb color={0.294,0.000,0.510},gp pattern 6,pattern color=.} \gpfillpath;
\gpcolor{rgb color={0.294,0.000,0.510}}
\gpsetlinetype{gp lt plot 4}
\draw[gp path] (6.322,4.107)--(7.238,4.107)--(7.238,4.261)--(6.322,4.261)--cycle;
\gpcolor{color=gp lt color border}
\end{tikzpicture}
        \end{center}
        \vspace{-10pt}
    \end{subfigure}

    \begin{subfigure}{.5\textwidth}
        \begin{tikzpicture}[gnuplot]
\path (0.000,0.000) rectangle (9.144,4.826);
\gpcolor{color=gp lt color border}
\gpsetlinetype{gp lt border}
\gpsetlinewidth{1.00}
\draw[gp path] (0.952,0.985)--(1.132,0.985);
\draw[gp path] (4.019,0.985)--(3.839,0.985);
\node[gp node right] at (0.768,0.985) {$0$};
\draw[gp path] (0.952,1.427)--(1.132,1.427);
\draw[gp path] (4.019,1.427)--(3.839,1.427);
\node[gp node right] at (0.768,1.427) {$1$};
\draw[gp path] (0.952,1.868)--(1.132,1.868);
\draw[gp path] (4.019,1.868)--(3.839,1.868);
\node[gp node right] at (0.768,1.868) {$2$};
\draw[gp path] (0.952,2.310)--(1.132,2.310);
\draw[gp path] (4.019,2.310)--(3.839,2.310);
\node[gp node right] at (0.768,2.310) {$3$};
\draw[gp path] (0.952,2.751)--(1.132,2.751);
\draw[gp path] (4.019,2.751)--(3.839,2.751);
\node[gp node right] at (0.768,2.751) {$4$};
\draw[gp path] (0.952,3.193)--(1.132,3.193);
\draw[gp path] (4.019,3.193)--(3.839,3.193);
\node[gp node right] at (0.768,3.193) {$5$};
\draw[gp path] (0.952,3.634)--(1.132,3.634);
\draw[gp path] (4.019,3.634)--(3.839,3.634);
\node[gp node right] at (0.768,3.634) {$6$};
\draw[gp path] (0.952,4.076)--(1.132,4.076);
\draw[gp path] (4.019,4.076)--(3.839,4.076);
\node[gp node right] at (0.768,4.076) {$7$};
\draw[gp path] (0.952,4.517)--(1.132,4.517);
\draw[gp path] (4.019,4.517)--(3.839,4.517);
\node[gp node right] at (0.768,4.517) {$8$};
\draw[gp path] (1.974,0.985)--(1.974,1.165);
\draw[gp path] (1.974,4.517)--(1.974,4.337);
\node[gp node center] at (1.974,0.677) {48};
\draw[gp path] (2.997,0.985)--(2.997,1.165);
\draw[gp path] (2.997,4.517)--(2.997,4.337);
\node[gp node center] at (2.997,0.677) {64};
\draw[gp path] (0.952,4.517)--(0.952,0.985)--(4.019,0.985)--(4.019,4.517)--cycle;
\node[gp node center,rotate=-270] at (0.276,2.751) {Time/iter (sec)};
\node[gp node center] at (2.485,0.215) {No. Cores};
\gpfill{rgb color={0.000,1.000,0.000},color=.!50} (1.682,0.985)--(1.829,0.985)--(1.829,1.191)--(1.682,1.191)--cycle;
\gpfill{rgb color={0.000,1.000,0.000},color=.!50} (2.705,0.985)--(2.852,0.985)--(2.852,1.129)--(2.705,1.129)--cycle;
\gpfill{rgb color={1.000,0.412,0.706},color=.!50} (1.828,0.985)--(1.975,0.985)--(1.975,1.339)--(1.828,1.339)--cycle;
\gpfill{rgb color={1.000,0.412,0.706},color=.!50} (2.851,0.985)--(2.998,0.985)--(2.998,1.299)--(2.851,1.299)--cycle;
\def\gpfillpath{(1.974,0.985)--(2.121,0.985)--(2.121,1.845)--(1.974,1.845)--cycle}
\gpfill{color=gpbgfillcolor} \gpfillpath;
\gpfill{rgb color={0.498,1.000,0.831},gp pattern 3,pattern color=.} \gpfillpath;
\gpcolor{rgb color={0.498,1.000,0.831}}
\draw[gp path] (1.974,0.985)--(1.974,1.844)--(2.120,1.844)--(2.120,0.985)--cycle;
\def\gpfillpath{(2.997,0.985)--(3.144,0.985)--(3.144,1.702)--(2.997,1.702)--cycle}
\gpfill{color=gpbgfillcolor} \gpfillpath;
\gpfill{rgb color={0.498,1.000,0.831},gp pattern 3,pattern color=.} \gpfillpath;
\draw[gp path] (2.997,0.985)--(2.997,1.701)--(3.143,1.701)--(3.143,0.985)--cycle;
\gpfill{rgb color={1.000,0.000,1.000},color=.!50} (2.120,0.985)--(2.267,0.985)--(2.267,2.378)--(2.120,2.378)--cycle;
\gpfill{rgb color={1.000,0.000,1.000},color=.!50} (3.143,0.985)--(3.290,0.985)--(3.290,2.100)--(3.143,2.100)--cycle;
\def\gpfillpath{(2.266,0.985)--(2.413,0.985)--(2.413,4.128)--(2.266,4.128)--cycle}
\gpfill{color=gpbgfillcolor} \gpfillpath;
\gpfill{rgb color={0.294,0.000,0.510},gp pattern 6,pattern color=.} \gpfillpath;
\gpcolor{rgb color={0.294,0.000,0.510}}
\draw[gp path] (2.266,0.985)--(2.266,4.127)--(2.412,4.127)--(2.412,0.985)--cycle;
\def\gpfillpath{(3.289,0.985)--(3.436,0.985)--(3.436,3.768)--(3.289,3.768)--cycle}
\gpfill{color=gpbgfillcolor} \gpfillpath;
\gpfill{rgb color={0.294,0.000,0.510},gp pattern 6,pattern color=.} \gpfillpath;
\draw[gp path] (3.289,0.985)--(3.289,3.767)--(3.435,3.767)--(3.435,0.985)--cycle;
\gpcolor{color=gp lt color border}
\draw[gp path] (0.952,4.517)--(0.952,0.985)--(4.019,0.985)--(4.019,4.517)--cycle;
\gpdefrectangularnode{gp plot 1}{\pgfpoint{0.952cm}{0.985cm}}{\pgfpoint{4.019cm}{4.517cm}}
\draw[gp path] (5.400,0.985)--(5.580,0.985);
\draw[gp path] (8.591,0.985)--(8.411,0.985);
\node[gp node right] at (5.216,0.985) {$0$};
\draw[gp path] (5.400,1.338)--(5.580,1.338);
\draw[gp path] (8.591,1.338)--(8.411,1.338);
\node[gp node right] at (5.216,1.338) {$1$};
\draw[gp path] (5.400,1.691)--(5.580,1.691);
\draw[gp path] (8.591,1.691)--(8.411,1.691);
\node[gp node right] at (5.216,1.691) {$2$};
\draw[gp path] (5.400,2.045)--(5.580,2.045);
\draw[gp path] (8.591,2.045)--(8.411,2.045);
\node[gp node right] at (5.216,2.045) {$3$};
\draw[gp path] (5.400,2.398)--(5.580,2.398);
\draw[gp path] (8.591,2.398)--(8.411,2.398);
\node[gp node right] at (5.216,2.398) {$4$};
\draw[gp path] (5.400,2.751)--(5.580,2.751);
\draw[gp path] (8.591,2.751)--(8.411,2.751);
\node[gp node right] at (5.216,2.751) {$5$};
\draw[gp path] (5.400,3.104)--(5.580,3.104);
\draw[gp path] (8.591,3.104)--(8.411,3.104);
\node[gp node right] at (5.216,3.104) {$6$};
\draw[gp path] (5.400,3.457)--(5.580,3.457);
\draw[gp path] (8.591,3.457)--(8.411,3.457);
\node[gp node right] at (5.216,3.457) {$7$};
\draw[gp path] (5.400,3.811)--(5.580,3.811);
\draw[gp path] (8.591,3.811)--(8.411,3.811);
\node[gp node right] at (5.216,3.811) {$8$};
\draw[gp path] (5.400,4.164)--(5.580,4.164);
\draw[gp path] (8.591,4.164)--(8.411,4.164);
\node[gp node right] at (5.216,4.164) {$9$};
\draw[gp path] (5.400,4.517)--(5.580,4.517);
\draw[gp path] (8.591,4.517)--(8.411,4.517);
\node[gp node right] at (5.216,4.517) {$10$};
\draw[gp path] (6.198,0.985)--(6.198,1.165);
\draw[gp path] (6.198,4.517)--(6.198,4.337);
\node[gp node center] at (6.198,0.677) {48};
\draw[gp path] (6.996,0.985)--(6.996,1.165);
\draw[gp path] (6.996,4.517)--(6.996,4.337);
\node[gp node center] at (6.996,0.677) {96};
\draw[gp path] (7.793,0.985)--(7.793,1.165);
\draw[gp path] (7.793,4.517)--(7.793,4.337);
\node[gp node center] at (7.793,0.677) {126};
\draw[gp path] (5.400,4.517)--(5.400,0.985)--(8.591,0.985)--(8.591,4.517)--cycle;
\node[gp node center] at (6.995,0.215) {No. Cores};
\gpfill{rgb color={0.000,1.000,0.000},color=.!50} (5.970,0.985)--(6.085,0.985)--(6.085,1.179)--(5.970,1.179)--cycle;
\gpfill{rgb color={0.000,1.000,0.000},color=.!50} (6.768,0.985)--(6.883,0.985)--(6.883,1.106)--(6.768,1.106)--cycle;
\gpfill{rgb color={0.000,1.000,0.000},color=.!50} (7.565,0.985)--(7.680,0.985)--(7.680,1.066)--(7.565,1.066)--cycle;
\gpfill{rgb color={1.000,0.412,0.706},color=.!50} (6.084,0.985)--(6.199,0.985)--(6.199,1.193)--(6.084,1.193)--cycle;
\gpfill{rgb color={1.000,0.412,0.706},color=.!50} (6.882,0.985)--(6.997,0.985)--(6.997,1.119)--(6.882,1.119)--cycle;
\gpfill{rgb color={1.000,0.412,0.706},color=.!50} (7.679,0.985)--(7.794,0.985)--(7.794,1.096)--(7.679,1.096)--cycle;
\def\gpfillpath{(6.198,0.985)--(6.313,0.985)--(6.313,2.812)--(6.198,2.812)--cycle}
\gpfill{color=gpbgfillcolor} \gpfillpath;
\gpfill{rgb color={0.498,1.000,0.831},gp pattern 3,pattern color=.} \gpfillpath;
\gpcolor{rgb color={0.498,1.000,0.831}}
\draw[gp path] (6.198,0.985)--(6.198,2.811)--(6.312,2.811)--(6.312,0.985)--cycle;
\def\gpfillpath{(6.996,0.985)--(7.110,0.985)--(7.110,2.050)--(6.996,2.050)--cycle}
\gpfill{color=gpbgfillcolor} \gpfillpath;
\gpfill{rgb color={0.498,1.000,0.831},gp pattern 3,pattern color=.} \gpfillpath;
\draw[gp path] (6.996,0.985)--(6.996,2.049)--(7.109,2.049)--(7.109,0.985)--cycle;
\def\gpfillpath{(7.793,0.985)--(7.908,0.985)--(7.908,1.838)--(7.793,1.838)--cycle}
\gpfill{color=gpbgfillcolor} \gpfillpath;
\gpfill{rgb color={0.498,1.000,0.831},gp pattern 3,pattern color=.} \gpfillpath;
\draw[gp path] (7.793,0.985)--(7.793,1.837)--(7.907,1.837)--(7.907,0.985)--cycle;
\gpfill{rgb color={1.000,0.000,1.000},color=.!50} (6.312,0.985)--(6.427,0.985)--(6.427,3.720)--(6.312,3.720)--cycle;
\gpfill{rgb color={1.000,0.000,1.000},color=.!50} (7.109,0.985)--(7.224,0.985)--(7.224,2.639)--(7.109,2.639)--cycle;
\gpfill{rgb color={1.000,0.000,1.000},color=.!50} (7.907,0.985)--(8.022,0.985)--(8.022,2.098)--(7.907,2.098)--cycle;
\def\gpfillpath{(6.426,0.985)--(6.541,0.985)--(6.541,4.168)--(6.426,4.168)--cycle}
\gpfill{color=gpbgfillcolor} \gpfillpath;
\gpfill{rgb color={0.294,0.000,0.510},gp pattern 6,pattern color=.} \gpfillpath;
\gpcolor{rgb color={0.294,0.000,0.510}}
\draw[gp path] (6.426,0.985)--(6.426,4.167)--(6.540,4.167)--(6.540,0.985)--cycle;
\def\gpfillpath{(7.223,0.985)--(7.338,0.985)--(7.338,2.594)--(7.223,2.594)--cycle}
\gpfill{color=gpbgfillcolor} \gpfillpath;
\gpfill{rgb color={0.294,0.000,0.510},gp pattern 6,pattern color=.} \gpfillpath;
\draw[gp path] (7.223,0.985)--(7.223,2.593)--(7.337,2.593)--(7.337,0.985)--cycle;
\def\gpfillpath{(8.021,0.985)--(8.136,0.985)--(8.136,2.367)--(8.021,2.367)--cycle}
\gpfill{color=gpbgfillcolor} \gpfillpath;
\gpfill{rgb color={0.294,0.000,0.510},gp pattern 6,pattern color=.} \gpfillpath;
\draw[gp path] (8.021,0.985)--(8.021,2.366)--(8.135,2.366)--(8.135,0.985)--cycle;
\gpcolor{color=gp lt color border}
\draw[gp path] (5.400,4.517)--(5.400,0.985)--(8.591,0.985)--(8.591,4.517)--cycle;
\gpdefrectangularnode{gp plot 2}{\pgfpoint{5.400cm}{0.985cm}}{\pgfpoint{8.591cm}{4.517cm}}
\end{tikzpicture}
        \vspace{-10pt}
        \caption{Friendster8 (left) and Friendster32 (right) datasets
        per iteration runtime for $k=100$.}
		\label{fig:dist-perf-frX}
	\end{subfigure}

    \begin{subfigure}{.5\textwidth}
        \begin{tikzpicture}[gnuplot]
\path (0.000,0.000) rectangle (9.652,4.826);
\gpcolor{color=gp lt color border}
\gpsetlinetype{gp lt border}
\gpsetlinewidth{1.00}
\draw[gp path] (1.136,0.985)--(1.316,0.985);
\draw[gp path] (4.273,0.985)--(4.093,0.985);
\node[gp node right] at (0.952,0.985) {0};
\draw[gp path] (1.136,1.490)--(1.316,1.490);
\draw[gp path] (4.273,1.490)--(4.093,1.490);
\node[gp node right] at (0.952,1.490) {10};
\draw[gp path] (1.136,1.994)--(1.316,1.994);
\draw[gp path] (4.273,1.994)--(4.093,1.994);
\node[gp node right] at (0.952,1.994) {20};
\draw[gp path] (1.136,2.499)--(1.316,2.499);
\draw[gp path] (4.273,2.499)--(4.093,2.499);
\node[gp node right] at (0.952,2.499) {30};
\draw[gp path] (1.136,3.003)--(1.316,3.003);
\draw[gp path] (4.273,3.003)--(4.093,3.003);
\node[gp node right] at (0.952,3.003) {40};
\draw[gp path] (1.136,3.508)--(1.316,3.508);
\draw[gp path] (4.273,3.508)--(4.093,3.508);
\node[gp node right] at (0.952,3.508) {50};
\draw[gp path] (1.136,4.012)--(1.316,4.012);
\draw[gp path] (4.273,4.012)--(4.093,4.012);
\node[gp node right] at (0.952,4.012) {60};
\draw[gp path] (1.136,4.517)--(1.316,4.517);
\draw[gp path] (4.273,4.517)--(4.093,4.517);
\node[gp node right] at (0.952,4.517) {70};
\draw[gp path] (1.920,0.985)--(1.920,1.165);
\draw[gp path] (1.920,4.517)--(1.920,4.337);
\node[gp node center] at (1.920,0.677) {72};
\draw[gp path] (2.705,0.985)--(2.705,1.165);
\draw[gp path] (2.705,4.517)--(2.705,4.337);
\node[gp node center] at (2.705,0.677) {144};
\draw[gp path] (3.489,0.985)--(3.489,1.165);
\draw[gp path] (3.489,4.517)--(3.489,4.337);
\node[gp node center] at (3.489,0.677) {288};
\draw[gp path] (1.136,4.517)--(1.136,0.985)--(4.273,0.985)--(4.273,4.517)--cycle;
\node[gp node center,rotate=-270] at (0.276,2.751) {Time/iter (sec)};
\node[gp node center] at (2.704,0.215) {No. Cores};
\gpfill{rgb color={0.000,1.000,0.000},color=.!50} (1.696,0.985)--(1.809,0.985)--(1.809,1.416)--(1.696,1.416)--cycle;
\gpfill{rgb color={0.000,1.000,0.000},color=.!50} (2.480,0.985)--(2.593,0.985)--(2.593,1.207)--(2.480,1.207)--cycle;
\gpfill{rgb color={0.000,1.000,0.000},color=.!50} (3.265,0.985)--(3.378,0.985)--(3.378,1.123)--(3.265,1.123)--cycle;
\gpfill{rgb color={1.000,0.412,0.706},color=.!50} (1.808,0.985)--(1.921,0.985)--(1.921,1.665)--(1.808,1.665)--cycle;
\gpfill{rgb color={1.000,0.412,0.706},color=.!50} (2.592,0.985)--(2.706,0.985)--(2.706,1.585)--(2.592,1.585)--cycle;
\gpfill{rgb color={1.000,0.412,0.706},color=.!50} (3.377,0.985)--(3.490,0.985)--(3.490,1.444)--(3.377,1.444)--cycle;
\def\gpfillpath{(1.920,0.985)--(2.033,0.985)--(2.033,1.326)--(1.920,1.326)--cycle}
\gpfill{color=gpbgfillcolor} \gpfillpath;
\gpfill{rgb color={0.498,1.000,0.831},gp pattern 3,pattern color=.} \gpfillpath;
\gpcolor{rgb color={0.498,1.000,0.831}}
\draw[gp path] (1.920,0.985)--(1.920,1.325)--(2.032,1.325)--(2.032,0.985)--cycle;
\def\gpfillpath{(2.705,0.985)--(2.818,0.985)--(2.818,1.174)--(2.705,1.174)--cycle}
\gpfill{color=gpbgfillcolor} \gpfillpath;
\gpfill{rgb color={0.498,1.000,0.831},gp pattern 3,pattern color=.} \gpfillpath;
\draw[gp path] (2.705,0.985)--(2.705,1.173)--(2.817,1.173)--(2.817,0.985)--cycle;
\def\gpfillpath{(3.489,0.985)--(3.602,0.985)--(3.602,1.083)--(3.489,1.083)--cycle}
\gpfill{color=gpbgfillcolor} \gpfillpath;
\gpfill{rgb color={0.498,1.000,0.831},gp pattern 3,pattern color=.} \gpfillpath;
\draw[gp path] (3.489,0.985)--(3.489,1.082)--(3.601,1.082)--(3.601,0.985)--cycle;
\gpfill{rgb color={1.000,0.000,1.000},color=.!50} (2.032,0.985)--(2.145,0.985)--(2.145,1.706)--(2.032,1.706)--cycle;
\gpfill{rgb color={1.000,0.000,1.000},color=.!50} (2.817,0.985)--(2.930,0.985)--(2.930,1.540)--(2.817,1.540)--cycle;
\gpfill{rgb color={1.000,0.000,1.000},color=.!50} (3.601,0.985)--(3.714,0.985)--(3.714,1.453)--(3.601,1.453)--cycle;
\def\gpfillpath{(2.144,0.985)--(2.257,0.985)--(2.257,4.328)--(2.144,4.328)--cycle}
\gpfill{color=gpbgfillcolor} \gpfillpath;
\gpfill{rgb color={0.294,0.000,0.510},gp pattern 6,pattern color=.} \gpfillpath;
\gpcolor{rgb color={0.294,0.000,0.510}}
\draw[gp path] (2.144,0.985)--(2.144,4.327)--(2.256,4.327)--(2.256,0.985)--cycle;
\def\gpfillpath{(2.929,0.985)--(3.042,0.985)--(3.042,2.449)--(2.929,2.449)--cycle}
\gpfill{color=gpbgfillcolor} \gpfillpath;
\gpfill{rgb color={0.294,0.000,0.510},gp pattern 6,pattern color=.} \gpfillpath;
\draw[gp path] (2.929,0.985)--(2.929,2.448)--(3.041,2.448)--(3.041,0.985)--cycle;
\def\gpfillpath{(3.713,0.985)--(3.826,0.985)--(3.826,1.670)--(3.713,1.670)--cycle}
\gpfill{color=gpbgfillcolor} \gpfillpath;
\gpfill{rgb color={0.294,0.000,0.510},gp pattern 6,pattern color=.} \gpfillpath;
\draw[gp path] (3.713,0.985)--(3.713,1.669)--(3.825,1.669)--(3.825,0.985)--cycle;
\gpcolor{color=gp lt color border}
\draw[gp path] (1.136,4.517)--(1.136,0.985)--(4.273,0.985)--(4.273,4.517)--cycle;
\gpdefrectangularnode{gp plot 1}{\pgfpoint{1.136cm}{0.985cm}}{\pgfpoint{4.273cm}{4.517cm}}
\draw[gp path] (5.654,0.985)--(5.834,0.985);
\draw[gp path] (9.099,0.985)--(8.919,0.985);
\node[gp node right] at (5.470,0.985) {0};
\draw[gp path] (5.654,1.574)--(5.834,1.574);
\draw[gp path] (9.099,1.574)--(8.919,1.574);
\node[gp node right] at (5.470,1.574) {5};
\draw[gp path] (5.654,2.162)--(5.834,2.162);
\draw[gp path] (9.099,2.162)--(8.919,2.162);
\node[gp node right] at (5.470,2.162) {10};
\draw[gp path] (5.654,2.751)--(5.834,2.751);
\draw[gp path] (9.099,2.751)--(8.919,2.751);
\node[gp node right] at (5.470,2.751) {15};
\draw[gp path] (5.654,3.340)--(5.834,3.340);
\draw[gp path] (9.099,3.340)--(8.919,3.340);
\node[gp node right] at (5.470,3.340) {20};
\draw[gp path] (5.654,3.928)--(5.834,3.928);
\draw[gp path] (9.099,3.928)--(8.919,3.928);
\node[gp node right] at (5.470,3.928) {25};
\draw[gp path] (5.654,4.517)--(5.834,4.517);
\draw[gp path] (9.099,4.517)--(8.919,4.517);
\node[gp node right] at (5.470,4.517) {30};
\draw[gp path] (6.802,0.985)--(6.802,1.165);
\draw[gp path] (6.802,4.517)--(6.802,4.337);
\node[gp node center] at (6.802,0.677) {144};
\draw[gp path] (7.951,0.985)--(7.951,1.165);
\draw[gp path] (7.951,4.517)--(7.951,4.337);
\node[gp node center] at (7.951,0.677) {288};
\draw[gp path] (5.654,4.517)--(5.654,0.985)--(9.099,0.985)--(9.099,4.517)--cycle;
\node[gp node center] at (7.376,0.215) {No. Cores};
\gpfill{rgb color={0.000,1.000,0.000},color=.!50} (6.474,0.985)--(6.639,0.985)--(6.639,1.132)--(6.474,1.132)--cycle;
\gpfill{rgb color={0.000,1.000,0.000},color=.!50} (7.623,0.985)--(7.788,0.985)--(7.788,1.089)--(7.623,1.089)--cycle;
\gpfill{rgb color={1.000,0.412,0.706},color=.!50} (6.638,0.985)--(6.803,0.985)--(6.803,1.179)--(6.638,1.179)--cycle;
\gpfill{rgb color={1.000,0.412,0.706},color=.!50} (7.787,0.985)--(7.952,0.985)--(7.952,1.136)--(7.787,1.136)--cycle;
\def\gpfillpath{(6.802,0.985)--(6.967,0.985)--(6.967,1.483)--(6.802,1.483)--cycle}
\gpfill{color=gpbgfillcolor} \gpfillpath;
\gpfill{rgb color={0.498,1.000,0.831},gp pattern 3,pattern color=.} \gpfillpath;
\gpcolor{rgb color={0.498,1.000,0.831}}
\draw[gp path] (6.802,0.985)--(6.802,1.482)--(6.966,1.482)--(6.966,0.985)--cycle;
\def\gpfillpath{(7.951,0.985)--(8.116,0.985)--(8.116,1.312)--(7.951,1.312)--cycle}
\gpfill{color=gpbgfillcolor} \gpfillpath;
\gpfill{rgb color={0.498,1.000,0.831},gp pattern 3,pattern color=.} \gpfillpath;
\draw[gp path] (7.951,0.985)--(7.951,1.311)--(8.115,1.311)--(8.115,0.985)--cycle;
\gpfill{rgb color={1.000,0.000,1.000},color=.!50} (6.966,0.985)--(7.131,0.985)--(7.131,2.436)--(6.966,2.436)--cycle;
\gpfill{rgb color={1.000,0.000,1.000},color=.!50} (8.115,0.985)--(8.280,0.985)--(8.280,1.777)--(8.115,1.777)--cycle;
\def\gpfillpath{(7.130,0.985)--(7.295,0.985)--(7.295,4.077)--(7.130,4.077)--cycle}
\gpfill{color=gpbgfillcolor} \gpfillpath;
\gpfill{rgb color={0.294,0.000,0.510},gp pattern 6,pattern color=.} \gpfillpath;
\gpcolor{rgb color={0.294,0.000,0.510}}
\draw[gp path] (7.130,0.985)--(7.130,4.076)--(7.294,4.076)--(7.294,0.985)--cycle;
\def\gpfillpath{(8.279,0.985)--(8.444,0.985)--(8.444,2.719)--(8.279,2.719)--cycle}
\gpfill{color=gpbgfillcolor} \gpfillpath;
\gpfill{rgb color={0.294,0.000,0.510},gp pattern 6,pattern color=.} \gpfillpath;
\draw[gp path] (8.279,0.985)--(8.279,2.718)--(8.443,2.718)--(8.443,0.985)--cycle;
\gpcolor{color=gp lt color border}
\draw[gp path] (5.654,4.517)--(5.654,0.985)--(9.099,0.985)--(9.099,4.517)--cycle;
\gpdefrectangularnode{gp plot 2}{\pgfpoint{5.654cm}{0.985cm}}{\pgfpoint{9.099cm}{4.517cm}}
\end{tikzpicture}
        \vspace{-10pt}
        \caption{RM$_{856M}$ (left) and RM$_{1B}$ (right) datasets
        per iteration runtime for $k=10$.}
		\label{fig:dist-perf-rmX}
	\end{subfigure}

	\caption {Distributed performance comparison of \textsf{knord}, MPI and MLlib
		on Amazon's EC$2$ cloud. Each machine has 18 physical cores with 1 thread per core.}
   \label{fig:dist-perf}
\end{figure}
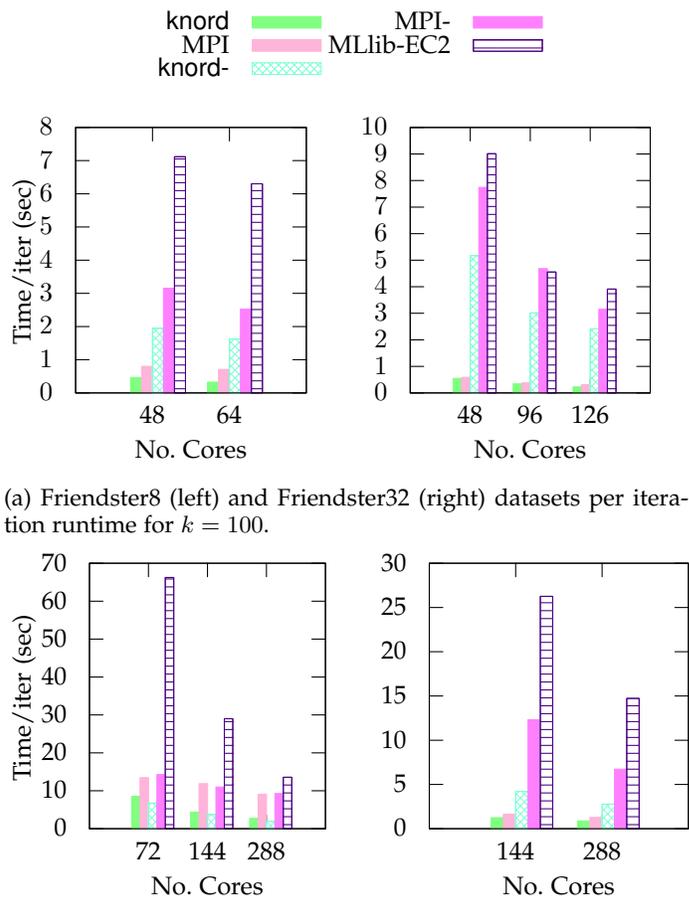

Figures \ref{fig:dist-speedup} and \ref{fig:dist-perf} reveal several fundamental
and important results. Figure \ref{fig:dist-speedup} shows that \textsf{knord} scales
well to very large numbers of machines, performing within a constant factor of
linear performance. This is a necessity today as many organizations push big-data
computation to the cloud. Figure \ref{fig:dist-perf} shows that in a cluster,
\textsf{knord}, even with TI \textit{disabled}, outperforms MLlib by a factor of
$5$ or more. This means we can often use fractions of the hardware required
by MLlib to perform equivalent tasks. Figure \ref{fig:dist-perf} demonstrates
that \textsf{knord} also benefits from our in-memory NUMA optimizations because
we outperform a NUMA-oblivious MPI routine by $20$-$50$\%.
Finally, Figure \ref{fig:dist-perf} shows that
MTI remains a low-overhead, effective method to reduce computation
even in the distributed setting.

\subsection{Semi-External Memory in the Cloud}

We measure the performance of \textsf{knors} on a
single 32 core i3.16xlarge machine with 8 SSDs on Amazon EC2 compared
to \textsf{knord}, MLlib and an optimized MPI routine running in a cluster.
We run \textsf{knors} with 48 threads, with extra parallelism coming
from symmetric multiprocessing.
We run all other implementations with the same number of processes/threads as physical cores.

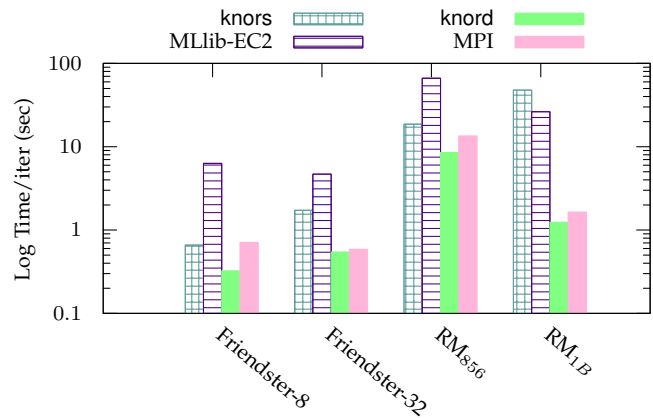
\begin{figure}[!htb]
\centering
\footnotesize
\begin{tikzpicture}[gnuplot]
\path (0.000,0.000) rectangle (9.144,6.096);
\gpcolor{color=gp lt color border}
\gpsetlinetype{gp lt border}
\gpsetlinewidth{1.00}
\draw[gp path] (1.320,1.845)--(1.500,1.845);
\draw[gp path] (8.591,1.845)--(8.411,1.845);
\node[gp node right] at (1.136,1.845) {$0.1$};
\draw[gp path] (1.320,2.179)--(1.410,2.179);
\draw[gp path] (8.591,2.179)--(8.501,2.179);
\draw[gp path] (1.320,2.374)--(1.410,2.374);
\draw[gp path] (8.591,2.374)--(8.501,2.374);
\draw[gp path] (1.320,2.512)--(1.410,2.512);
\draw[gp path] (8.591,2.512)--(8.501,2.512);
\draw[gp path] (1.320,2.620)--(1.410,2.620);
\draw[gp path] (8.591,2.620)--(8.501,2.620);
\draw[gp path] (1.320,2.708)--(1.410,2.708);
\draw[gp path] (8.591,2.708)--(8.501,2.708);
\draw[gp path] (1.320,2.782)--(1.410,2.782);
\draw[gp path] (8.591,2.782)--(8.501,2.782);
\draw[gp path] (1.320,2.846)--(1.410,2.846);
\draw[gp path] (8.591,2.846)--(8.501,2.846);
\draw[gp path] (1.320,2.903)--(1.410,2.903);
\draw[gp path] (8.591,2.903)--(8.501,2.903);
\draw[gp path] (1.320,2.954)--(1.500,2.954);
\draw[gp path] (8.591,2.954)--(8.411,2.954);
\node[gp node right] at (1.136,2.954) {$1$};
\draw[gp path] (1.320,3.287)--(1.410,3.287);
\draw[gp path] (8.591,3.287)--(8.501,3.287);
\draw[gp path] (1.320,3.483)--(1.410,3.483);
\draw[gp path] (8.591,3.483)--(8.501,3.483);
\draw[gp path] (1.320,3.621)--(1.410,3.621);
\draw[gp path] (8.591,3.621)--(8.501,3.621);
\draw[gp path] (1.320,3.729)--(1.410,3.729);
\draw[gp path] (8.591,3.729)--(8.501,3.729);
\draw[gp path] (1.320,3.816)--(1.410,3.816);
\draw[gp path] (8.591,3.816)--(8.501,3.816);
\draw[gp path] (1.320,3.891)--(1.410,3.891);
\draw[gp path] (8.591,3.891)--(8.501,3.891);
\draw[gp path] (1.320,3.955)--(1.410,3.955);
\draw[gp path] (8.591,3.955)--(8.501,3.955);
\draw[gp path] (1.320,4.012)--(1.410,4.012);
\draw[gp path] (8.591,4.012)--(8.501,4.012);
\draw[gp path] (1.320,4.062)--(1.500,4.062);
\draw[gp path] (8.591,4.062)--(8.411,4.062);
\node[gp node right] at (1.136,4.062) {$10$};
\draw[gp path] (1.320,4.396)--(1.410,4.396);
\draw[gp path] (8.591,4.396)--(8.501,4.396);
\draw[gp path] (1.320,4.591)--(1.410,4.591);
\draw[gp path] (8.591,4.591)--(8.501,4.591);
\draw[gp path] (1.320,4.730)--(1.410,4.730);
\draw[gp path] (8.591,4.730)--(8.501,4.730);
\draw[gp path] (1.320,4.837)--(1.410,4.837);
\draw[gp path] (8.591,4.837)--(8.501,4.837);
\draw[gp path] (1.320,4.925)--(1.410,4.925);
\draw[gp path] (8.591,4.925)--(8.501,4.925);
\draw[gp path] (1.320,4.999)--(1.410,4.999);
\draw[gp path] (8.591,4.999)--(8.501,4.999);
\draw[gp path] (1.320,5.064)--(1.410,5.064);
\draw[gp path] (8.591,5.064)--(8.501,5.064);
\draw[gp path] (1.320,5.120)--(1.410,5.120);
\draw[gp path] (8.591,5.120)--(8.501,5.120);
\draw[gp path] (1.320,5.171)--(1.500,5.171);
\draw[gp path] (8.591,5.171)--(8.411,5.171);
\node[gp node right] at (1.136,5.171) {$100$};
\draw[gp path] (2.774,1.845)--(2.774,2.025);
\draw[gp path] (2.774,5.171)--(2.774,4.991);
\node[gp node left,rotate=-40] at (2.774,1.661) {Friendster-8};
\draw[gp path] (4.228,1.845)--(4.228,2.025);
\draw[gp path] (4.228,5.171)--(4.228,4.991);
\node[gp node left,rotate=-40] at (4.228,1.661) {Friendster-32};
\draw[gp path] (5.683,1.845)--(5.683,2.025);
\draw[gp path] (5.683,5.171)--(5.683,4.991);
\node[gp node left,rotate=-40] at (5.683,1.661) {RM$_{856}$};
\draw[gp path] (7.137,1.845)--(7.137,2.025);
\draw[gp path] (7.137,5.171)--(7.137,4.991);
\node[gp node left,rotate=-40] at (7.137,1.661) {RM$_{1B}$};
\draw[gp path] (1.320,5.171)--(1.320,1.845)--(8.591,1.845)--(8.591,5.171)--cycle;
\node[gp node center,rotate=-270] at (0.276,3.508) {Log Time/iter (sec)};
\node[gp node right] at (3.671,5.762) {\textsf{knors}};
\def\gpfillpath{(3.855,5.685)--(4.771,5.685)--(4.771,5.839)--(3.855,5.839)--cycle}
\gpfill{color=gpbgfillcolor} \gpfillpath;
\gpfill{rgb color={0.373,0.620,0.627},gp pattern 4,pattern color=.} \gpfillpath;
\gpcolor{rgb color={0.373,0.620,0.627}}
\draw[gp path] (3.855,5.685)--(4.771,5.685)--(4.771,5.839)--(3.855,5.839)--cycle;
\def\gpfillpath{(2.411,1.845)--(2.654,1.845)--(2.654,2.754)--(2.411,2.754)--cycle}
\gpfill{color=gpbgfillcolor} \gpfillpath;
\gpfill{rgb color={0.373,0.620,0.627},gp pattern 4,pattern color=.} \gpfillpath;
\draw[gp path] (2.411,1.845)--(2.411,2.753)--(2.653,2.753)--(2.653,1.845)--cycle;
\def\gpfillpath{(3.865,1.845)--(4.108,1.845)--(4.108,3.217)--(3.865,3.217)--cycle}
\gpfill{color=gpbgfillcolor} \gpfillpath;
\gpfill{rgb color={0.373,0.620,0.627},gp pattern 4,pattern color=.} \gpfillpath;
\draw[gp path] (3.865,1.845)--(3.865,3.216)--(4.107,3.216)--(4.107,1.845)--cycle;
\def\gpfillpath{(5.319,1.845)--(5.562,1.845)--(5.562,4.363)--(5.319,4.363)--cycle}
\gpfill{color=gpbgfillcolor} \gpfillpath;
\gpfill{rgb color={0.373,0.620,0.627},gp pattern 4,pattern color=.} \gpfillpath;
\draw[gp path] (5.319,1.845)--(5.319,4.362)--(5.561,4.362)--(5.561,1.845)--cycle;
\def\gpfillpath{(6.773,1.845)--(7.017,1.845)--(7.017,4.816)--(6.773,4.816)--cycle}
\gpfill{color=gpbgfillcolor} \gpfillpath;
\gpfill{rgb color={0.373,0.620,0.627},gp pattern 4,pattern color=.} \gpfillpath;
\draw[gp path] (6.773,1.845)--(6.773,4.815)--(7.016,4.815)--(7.016,1.845)--cycle;
\gpcolor{color=gp lt color border}
\node[gp node right] at (3.671,5.454) {MLlib-EC2};
\def\gpfillpath{(3.855,5.377)--(4.771,5.377)--(4.771,5.531)--(3.855,5.531)--cycle}
\gpfill{color=gpbgfillcolor} \gpfillpath;
\gpfill{rgb color={0.294,0.000,0.510},gp pattern 6,pattern color=.} \gpfillpath;
\gpcolor{rgb color={0.294,0.000,0.510}}
\draw[gp path] (3.855,5.377)--(4.771,5.377)--(4.771,5.531)--(3.855,5.531)--cycle;
\def\gpfillpath{(2.653,1.845)--(2.896,1.845)--(2.896,3.841)--(2.653,3.841)--cycle}
\gpfill{color=gpbgfillcolor} \gpfillpath;
\gpfill{rgb color={0.294,0.000,0.510},gp pattern 6,pattern color=.} \gpfillpath;
\draw[gp path] (2.653,1.845)--(2.653,3.840)--(2.895,3.840)--(2.895,1.845)--cycle;
\def\gpfillpath{(4.107,1.845)--(4.351,1.845)--(4.351,3.698)--(4.107,3.698)--cycle}
\gpfill{color=gpbgfillcolor} \gpfillpath;
\gpfill{rgb color={0.294,0.000,0.510},gp pattern 6,pattern color=.} \gpfillpath;
\draw[gp path] (4.107,1.845)--(4.107,3.697)--(4.350,3.697)--(4.350,1.845)--cycle;
\def\gpfillpath{(5.561,1.845)--(5.805,1.845)--(5.805,4.974)--(5.561,4.974)--cycle}
\gpfill{color=gpbgfillcolor} \gpfillpath;
\gpfill{rgb color={0.294,0.000,0.510},gp pattern 6,pattern color=.} \gpfillpath;
\draw[gp path] (5.561,1.845)--(5.561,4.973)--(5.804,4.973)--(5.804,1.845)--cycle;
\def\gpfillpath{(7.016,1.845)--(7.259,1.845)--(7.259,4.528)--(7.016,4.528)--cycle}
\gpfill{color=gpbgfillcolor} \gpfillpath;
\gpfill{rgb color={0.294,0.000,0.510},gp pattern 6,pattern color=.} \gpfillpath;
\draw[gp path] (7.016,1.845)--(7.016,4.527)--(7.258,4.527)--(7.258,1.845)--cycle;
\gpcolor{color=gp lt color border}
\node[gp node right] at (6.611,5.762) {\textsf{knord}};
\gpfill{rgb color={0.000,1.000,0.000},color=.!50} (6.795,5.685)--(7.711,5.685)--(7.711,5.839)--(6.795,5.839)--cycle;
\gpfill{rgb color={0.000,1.000,0.000},color=.!50} (2.895,1.845)--(3.139,1.845)--(3.139,2.411)--(2.895,2.411)--cycle;
\gpfill{rgb color={0.000,1.000,0.000},color=.!50} (4.350,1.845)--(4.593,1.845)--(4.593,2.662)--(4.350,2.662)--cycle;
\gpfill{rgb color={0.000,1.000,0.000},color=.!50} (5.804,1.845)--(6.047,1.845)--(6.047,3.986)--(5.804,3.986)--cycle;
\gpfill{rgb color={0.000,1.000,0.000},color=.!50} (7.258,1.845)--(7.501,1.845)--(7.501,3.058)--(7.258,3.058)--cycle;
\node[gp node right] at (6.611,5.454) {MPI};
\gpfill{rgb color={1.000,0.412,0.706},color=.!50} (6.795,5.377)--(7.711,5.377)--(7.711,5.531)--(6.795,5.531)--cycle;
\gpfill{rgb color={1.000,0.412,0.706},color=.!50} (3.138,1.845)--(3.381,1.845)--(3.381,2.789)--(3.138,2.789)--cycle;
\gpfill{rgb color={1.000,0.412,0.706},color=.!50} (4.592,1.845)--(4.835,1.845)--(4.835,2.698)--(4.592,2.698)--cycle;
\gpfill{rgb color={1.000,0.412,0.706},color=.!50} (6.046,1.845)--(6.290,1.845)--(6.290,4.206)--(6.046,4.206)--cycle;
\gpfill{rgb color={1.000,0.412,0.706},color=.!50} (7.500,1.845)--(7.744,1.845)--(7.744,3.193)--(7.500,3.193)--cycle;
\draw[gp path] (1.320,5.171)--(1.320,1.845)--(8.591,1.845)--(8.591,5.171)--cycle;
\gpdefrectangularnode{gp plot 1}{\pgfpoint{1.320cm}{1.845cm}}{\pgfpoint{8.591cm}{5.171cm}}
\end{tikzpicture}
\vspace{-15pt}
\caption{Performance comparison of \textsf{knors} to distributed packages.
\textsf{knors} uses one i3.16xlarge machine with 32 physical cores.
\textsf{knord}, MLlib-EC2 and MPI use 3 c4.8xlarge with a total of 48 physical
cores for all datasets other than RU$_{1B}$ where they use 8 c4.8xlarge with a
total of 128 physical cores.}
\label{fig:sem-ec2}
\vspace{-5pt}
\end{figure}

\textsf{knors} has comparable performance to both MPI and \textsf{knord},
leading to our assertion that the SEM scale-up model should be considered
prior to moving to the distributed setting.
Figure \ref{fig:sem-ec2} highlights that \textsf{knors} often outperforms MLlib
even when MLLib runs in a cluster that contains more physical CPU cores.

\subsection{Application Evaluation} \label{subsec:apps-eval}

\begin{figure}[!tb]
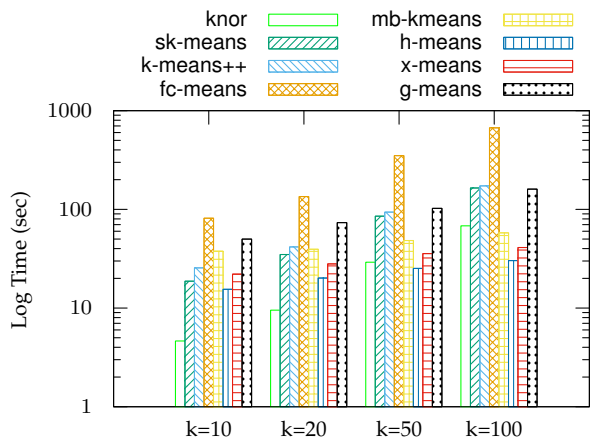

    \centering
    \footnotesize
    \include{./charts/perf.all.friendster-32}
    \vspace{-10pt}
    \caption{In-memory performance of \textsf{clusterNOR}
        benchmark applications on the Friendster-32 dataset. We fix the
        number of iterations to $20$ for all applications and use a mini-batch
        size of $20\%$ of the data size for \textsf{mb-kmeans}.}
    \label{fig:app-im-eval}
    \vspace{-5pt}
\end{figure}

We benchmark the performance of the nine applications developed using
\textsf{clusterNOR} (Section \ref{sec:apps}). We present only results for in-memory execution due to
space limitations. The relative performance in other settings, SEM and distributed memory,
track in-memory results closely.
Figure \ref{fig:app-im-eval} demonstrates
that for applications with similar computational complexity as
k-means, \textsf{clusterNOR} achieves comparable performance to
\textsf{knor}. More complex algorithms exhibit overhead in proportion to their
data access patterns and increased computation.

We are unable to provide meaningful comparisons to other implementations, because
there are no efficient parallel implementations for most MM algorithms.
These algorithms have not seen the effort and attention to optimize performance for
big data sets as has k-means.  To our knowledge, there exist no other open-source
large-scale parallel clustering libraries; the \textsf{clusterNOR} benchmark
applications enable scientific experimentation at a scale previously unavailable.
For overall performance, we use the k-means results of Figure \ref{fig:sem-ec2}
as a proxy that indicates that these implementations achieve scalable, good performance.

Figure \ref{fig:app-im-eval} demonstrates that applications with similar
algorithmic complexity to k-means perform comparably to \textsf{knor}.
This is a strong demonstration that \textsf{clusterNOR} optimizations
apply to a wide range of MM algorithms.
For mini-batch k-means (\textsf{mbk-means}), we set the batch size, $B$, to $20\%$ of the dataset size.
This is roughly twice the value used in experiments by Sculley \cite{mbkmeans}
 in his seminal work describing the algorithm. We
highlight that even though \textsf{mbk-means} performs several factors fewer
distance computations compared to batched k-means (e.g., \textsf{knor}), its computation time
can be greater due to the algorithmically serial gradient step (Equation \ref{eqn:mbk}). Furthermore,
we note that the computation time of fuzzy c-means can be up to an order of magnitude slower than that of k-means. This
is due to \textsf{fc-means} performing a series of linear algebraic operations, some of which must be
performed outside the confines of the parallel constructs provided by the framework. As such,
the application's performance is bound by the computation of updates to the cluster \textit{contribution matrix},
an $\mathcal{O}(kn)$ data structure containing the probability of a data point being in a cluster.

\begin{figure}[!tb]
    \centering
    \footnotesize
    \begin{tikzpicture}[gnuplot]
\path (0.000,0.000) rectangle (8.382,4.826);
\gpcolor{color=gp lt color border}
\gpsetlinetype{gp lt border}
\gpsetlinewidth{1.00}
\draw[gp path] (1.320,1.135)--(1.500,1.135);
\draw[gp path] (7.829,1.135)--(7.649,1.135);
\node[gp node right] at (1.136,1.135) {$0.1$};
\draw[gp path] (1.320,1.530)--(1.500,1.530);
\draw[gp path] (7.829,1.530)--(7.649,1.530);
\node[gp node right] at (1.136,1.530) {$0.2$};
\draw[gp path] (1.320,1.925)--(1.500,1.925);
\draw[gp path] (7.829,1.925)--(7.649,1.925);
\node[gp node right] at (1.136,1.925) {$0.3$};
\draw[gp path] (1.320,2.320)--(1.500,2.320);
\draw[gp path] (7.829,2.320)--(7.649,2.320);
\node[gp node right] at (1.136,2.320) {$0.4$};
\draw[gp path] (1.320,2.716)--(1.500,2.716);
\draw[gp path] (7.829,2.716)--(7.649,2.716);
\node[gp node right] at (1.136,2.716) {$0.5$};
\draw[gp path] (1.320,3.111)--(1.500,3.111);
\draw[gp path] (7.829,3.111)--(7.649,3.111);
\node[gp node right] at (1.136,3.111) {$0.6$};
\draw[gp path] (1.320,3.506)--(1.500,3.506);
\draw[gp path] (7.829,3.506)--(7.649,3.506);
\node[gp node right] at (1.136,3.506) {$0.7$};
\draw[gp path] (1.320,3.901)--(1.500,3.901);
\draw[gp path] (7.829,3.901)--(7.649,3.901);
\node[gp node right] at (1.136,3.901) {$0.8$};
\draw[gp path] (3.490,1.135)--(3.490,1.315);
\draw[gp path] (3.490,3.901)--(3.490,3.721);
\node[gp node left,rotate=-40] at (3.490,0.951) {X-means};
\draw[gp path] (5.659,1.135)--(5.659,1.315);
\draw[gp path] (5.659,3.901)--(5.659,3.721);
\node[gp node left,rotate=-40] at (5.659,0.951) {G-means};
\draw[gp path] (1.320,3.901)--(1.320,1.135)--(7.829,1.135)--(7.829,3.901)--cycle;
\node[gp node center,rotate=-270] at (0.276,2.518) {Relative Performance};
\node[gp node right] at (3.290,4.492) {k=10};
\def\gpfillpath{(3.474,4.415)--(4.390,4.415)--(4.390,4.569)--(3.474,4.569)--cycle}
\gpfill{color=gpbgfillcolor} \gpfillpath;
\gpfill{rgb color={0.580,0.000,0.827},gp pattern 0,pattern color=.} \gpfillpath;
\gpcolor{rgb color={0.580,0.000,0.827}}
\draw[gp path] (3.474,4.415)--(4.390,4.415)--(4.390,4.569)--(3.474,4.569)--cycle;
\def\gpfillpath{(2.947,1.135)--(3.310,1.135)--(3.310,3.515)--(2.947,3.515)--cycle}
\gpfill{color=gpbgfillcolor} \gpfillpath;
\gpfill{rgb color={0.580,0.000,0.827},gp pattern 0,pattern color=.} \gpfillpath;
\draw[gp path] (2.947,1.135)--(2.947,3.514)--(3.309,3.514)--(3.309,1.135)--cycle;
\def\gpfillpath{(5.117,1.135)--(5.480,1.135)--(5.480,1.964)--(5.117,1.964)--cycle}
\gpfill{color=gpbgfillcolor} \gpfillpath;
\gpfill{rgb color={0.580,0.000,0.827},gp pattern 0,pattern color=.} \gpfillpath;
\draw[gp path] (5.117,1.135)--(5.117,1.963)--(5.479,1.963)--(5.479,1.135)--cycle;
\gpcolor{color=gp lt color border}
\node[gp node right] at (3.290,4.184) {k=20};
\def\gpfillpath{(3.474,4.107)--(4.390,4.107)--(4.390,4.261)--(3.474,4.261)--cycle}
\gpfill{color=gpbgfillcolor} \gpfillpath;
\gpfill{rgb color={0.000,0.620,0.451},gp pattern 1,pattern color=.} \gpfillpath;
\gpcolor{rgb color={0.000,0.620,0.451}}
\draw[gp path] (3.474,4.107)--(4.390,4.107)--(4.390,4.261)--(3.474,4.261)--cycle;
\def\gpfillpath{(3.309,1.135)--(3.671,1.135)--(3.671,3.583)--(3.309,3.583)--cycle}
\gpfill{color=gpbgfillcolor} \gpfillpath;
\gpfill{rgb color={0.000,0.620,0.451},gp pattern 1,pattern color=.} \gpfillpath;
\draw[gp path] (3.309,1.135)--(3.309,3.582)--(3.670,3.582)--(3.670,1.135)--cycle;
\def\gpfillpath{(5.479,1.135)--(5.841,1.135)--(5.841,1.825)--(5.479,1.825)--cycle}
\gpfill{color=gpbgfillcolor} \gpfillpath;
\gpfill{rgb color={0.000,0.620,0.451},gp pattern 1,pattern color=.} \gpfillpath;
\draw[gp path] (5.479,1.135)--(5.479,1.824)--(5.840,1.824)--(5.840,1.135)--cycle;
\gpcolor{color=gp lt color border}
\node[gp node right] at (5.494,4.492) {k=50};
\def\gpfillpath{(5.678,4.415)--(6.594,4.415)--(6.594,4.569)--(5.678,4.569)--cycle}
\gpfill{color=gpbgfillcolor} \gpfillpath;
\gpfill{rgb color={0.337,0.706,0.914},gp pattern 2,pattern color=.} \gpfillpath;
\gpcolor{rgb color={0.337,0.706,0.914}}
\draw[gp path] (5.678,4.415)--(6.594,4.415)--(6.594,4.569)--(5.678,4.569)--cycle;
\def\gpfillpath{(3.670,1.135)--(4.033,1.135)--(4.033,3.538)--(3.670,3.538)--cycle}
\gpfill{color=gpbgfillcolor} \gpfillpath;
\gpfill{rgb color={0.337,0.706,0.914},gp pattern 2,pattern color=.} \gpfillpath;
\draw[gp path] (3.670,1.135)--(3.670,3.537)--(4.032,3.537)--(4.032,1.135)--cycle;
\def\gpfillpath{(5.840,1.135)--(6.203,1.135)--(6.203,1.715)--(5.840,1.715)--cycle}
\gpfill{color=gpbgfillcolor} \gpfillpath;
\gpfill{rgb color={0.337,0.706,0.914},gp pattern 2,pattern color=.} \gpfillpath;
\draw[gp path] (5.840,1.135)--(5.840,1.714)--(6.202,1.714)--(6.202,1.135)--cycle;
\gpcolor{color=gp lt color border}
\node[gp node right] at (5.494,4.184) {k=100};
\def\gpfillpath{(5.678,4.107)--(6.594,4.107)--(6.594,4.261)--(5.678,4.261)--cycle}
\gpfill{color=gpbgfillcolor} \gpfillpath;
\gpfill{rgb color={0.902,0.624,0.000},gp pattern 3,pattern color=.} \gpfillpath;
\gpcolor{rgb color={0.902,0.624,0.000}}
\draw[gp path] (5.678,4.107)--(6.594,4.107)--(6.594,4.261)--(5.678,4.261)--cycle;
\def\gpfillpath{(4.032,1.135)--(4.395,1.135)--(4.395,3.653)--(4.032,3.653)--cycle}
\gpfill{color=gpbgfillcolor} \gpfillpath;
\gpfill{rgb color={0.902,0.624,0.000},gp pattern 3,pattern color=.} \gpfillpath;
\draw[gp path] (4.032,1.135)--(4.032,3.652)--(4.394,3.652)--(4.394,1.135)--cycle;
\def\gpfillpath{(6.202,1.135)--(6.564,1.135)--(6.564,1.486)--(6.202,1.486)--cycle}
\gpfill{color=gpbgfillcolor} \gpfillpath;
\gpfill{rgb color={0.902,0.624,0.000},gp pattern 3,pattern color=.} \gpfillpath;
\draw[gp path] (6.202,1.135)--(6.202,1.485)--(6.563,1.485)--(6.563,1.135)--cycle;
\gpcolor{color=gp lt color border}
\draw[gp path] (1.320,3.901)--(1.320,1.135)--(7.829,1.135)--(7.829,3.901)--cycle;
\gpdefrectangularnode{gp plot 1}{\pgfpoint{1.320cm}{1.135cm}}{\pgfpoint{7.829cm}{3.901cm}}
\end{tikzpicture}
    \vspace{-15pt}
    \caption{The relative performance of hierarchical algorithms in comparison
    to H-means, the baseline hierarchical cluster application on the
    Friendster-32 dataset}
    \label{fig:hclust}
\end{figure}
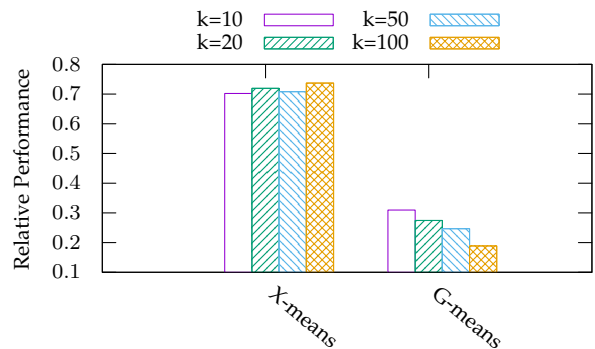

Hierarchical clustering algorithms also perform well in comparison to \textsf{knor},
despite requiring heavier computation between iterations. To benchmark H-means, X-means
and G-means we perform $20$ iterations of k-means between each divisive cluster-splitting
step i.e., the \texttt{SplitStep} (Appendix \ref{subsec:hclust}).
We recognize that the computation cost of the
hierarchical algorithms for one iteration is lower than that of k-means, but argue
that performing the same number of iterations at each level of the hierarchy provides
a comparable measure of computation. Furthermore, X-means requires the computation of BIC
and G-means requires the computation of the Anderson-Darling statistic between \texttt{SplitStep}s. This increases the cost of hierarchical clustering over
H-means (Figure \ref{fig:hclust}), in which X-means and G-means perform at about
$70\%$ and $30\%$ of the performance of H-means.

We present the result of the k-medoids experiment (Table \ref{tbl:medoids})
on a $250$ thousand subsampling of the Friendster-32 dataset.
We subsample because the complexity of k-medoids is significantly higher than that of
all other applications making it infeasible for even our smallest dataset.
Nevertheless, k-medoids demonstrates the programming flexibility of our framework.
We observe that as the number of clusters, $k$, increases the computational
overhead reduces. This is due to the size of each cluster generally decreasing as
data points are spread across more clusters. \textsf{clusterNOR} ensures that the
degree of parallelism achieved is independent of the number of clusters.
The most intensive medoid \textit{swap} procedure now requires less
inter-cluster computation leading to reduced computation time. We
vary the degree to which we subsample within the \textit{swap} procedure from $20\%$
up to $100\%$ to highlight the observed phenomenon.

\begin{table}[!htb]
\caption{The performance of k-medoids on a 250 thousand random sampling of the
    Friendster-32 dataset run for $20$ iterations.}
\vspace{-10pt}
\begin{center}
\footnotesize
\begin{tabular}{|c|c|c|c|c|}
\hline
\textbf{Sample \%} & $k=10$ & $k=20$ & $k=50$ & $k=100$\\
\hline
    $20$ & $455.95s$ & $679.52s$ & $262.42s$ & $134.46s$\\
\hline
    $50$ & $2003.74s$ & $1652.90s$ & $717.19s$ & $342.34s$\\
\hline
    $100$ & $2154.81s$ & $2616.57s$ & $1801.56s$ & $761.98s$\\
\hline
\end{tabular}
\normalsize
\end{center}
\label{tbl:medoids}
\end{table}


\section{Discussion} \label{sec:discuss}

\textsf{clusterNOR} demonstrates large performance benefits
associated with NUMA optimizations for clustering.
Data locality optimizations, such as NUMA-node thread binding,
NUMA-aware task scheduling, and NUMA-aware memory allocation schemes,
provide several times speedup for MM algorithms.
Many of the optimizations within \textsf{clusterNOR} are applicable to
data processing frameworks built for non-specialized commodity hardware.

For technical accomplishments, we accelerate k-means and its derived
algorithms by over an order of
magnitude by rethinking Lloyd's algorithm for modern multiprocessor
NUMA architectures through the minimization of critical regions.
Additionally, we formulate
a minimal triangle inequality (MTI) pruning algorithm
that further boosts the performance of k-means on real-world billion point datasets
by over $100$x when compared to some popular frameworks. MTI does so without
significantly increasing memory consumption.

Finally, \textsf{clusterNOR} provides an extensible unified framework for in-memory,
semi-external memory and distributed MM algorithm development.
The \textsf{clusterNOR} benchmark applications provide a scalable,
state-of-the-art clustering library. Bindings to the open source library
are accessible within `CRAN', the R Programming Language \cite{R}
package manager, under the name \textit{clusternor}.
We are an open source project available at \\
\href{https://github.com/flashxio/knor}{https://github.com/flashxio/knor}.
Our flagship \textsf{knor} application, on which this work is based,
receives hundreds of downloads monthly on both CRAN and \texttt{pip},
the Python package manager.

\ifCLASSOPTIONcompsoc
\section*{Acknowledgments}
\else
\section*{Acknowledgment}
\fi
This work is partially supported by DARPA GRAPHS N66001-14-1-4028
and DARPA SIMPLEX program through SPAWAR contract N66001-15-C-4041.
We thank Nikita Ivkin for discussions that assisted immensely in
realizing this work.

\bibliographystyle{abbrv}
\bibliography{clusterNOR}

\appendices
\section{Application Programming \\Interface (API)} \label{sec:api}

\textsf{clusterNOR} provides a C++ API on which users may
define their own algorithms. There are two core components:

\begin{itemize}
\item the base iterative interface, \texttt{base}.
\item the hierarchical iterative interface, \texttt{hclust}.
\end{itemize}

\noindent, in addition to two API extensions:

\begin{itemize}
\item the semi-external memory interface, \texttt{sem}.
\item the distributed interface, \texttt{dist}.
\end{itemize}

\subsection{\texttt{\large base}}

The \texttt{base} interface provides developers with abstract
methods that can be overridden to implement a variety of algorithms, such
as k-means, mini-batch k-means, fuzzy C-means,
and k-mediods (Sections \ref{apps:kmeans}, \ref{apps:mbkmeans}, \ref{apps:fcm},
and \ref{apps:kmedoids}).

\begin{itemize}
    \item \texttt{run()}: Defines algorithm specific steps for a particular
        application. This generally follows the serial algorithm.
    \item \texttt{MMStep()}: Used when both MM steps
        can be performed simultaneously
        and reduces the effect of the barrier between the two steps.
    \item \texttt{M1Step()}: Used when the Majorize or Minorize step must be
        performed independently from the Minimization or Maximization step.
    \item \texttt{M2Step()}: Used in conjunction with \texttt{M1Step} as the
         Minimization or Maximization step of the algorithm.
\end{itemize}

\subsection{\texttt{\large hclust}} \label{subsec:hclust}

The \texttt{hclust} interface extends \texttt{base} and is used to develop
algorithms in which clustering is performed in a hierarchical fashion, such as
H-means, X-means, and G-means (Sections \ref{apps:hmeans}, \ref{apps:xmeans},
and \ref{apps:gmeans}). For performance reasons, this interface is iterative rather
than recursive. We discuss this design decision and its merits in Section
\ref{sec:hclust-design}. \texttt{hclust} provides the following additional abstract
methods for user definition:

\begin{itemize}
    \item \texttt{SplitStep()}: Used to determine when a cluster should split.
    \item \texttt{HclustUpdate()}: Used to update the hierarchical
        global state from one iteration to the next.
\end{itemize}

\subsection{\texttt{\large sem}}

The SEM interface builds upon \texttt{base} and \texttt{hclust}
and incorporates a modified FlashGraph \cite{flashgraph}
API that we extend to support matrices and iterative clustering algorithms. The
interface provides an abstraction over an asynchronous I/O model in which data are
requested from disk and computation is overlapped with I/O transparently to users:
\begin{itemize}
    \item \texttt{request(ids[])}: Issues I/O requests to the
        underlying storage media for the feature-vectors associated with the
        entries in \texttt{ids[]}.
\end{itemize}

\subsection{\texttt{\large dist}}

The distributed interface builds upon \texttt{base} and \texttt{hclust} creating
infrastructure to support distributed processing. As is common with distributed
memory, there also exist \textit{optional} primitives for data
synchronization, scattering and gathering, if necessary. Mandatory methods
pertain to organizing state before and after computation and are
abstractions above MPI calls:

\begin{itemize}
    \item \texttt{OnComputeStart()}: Pass state or configuration details
    to processes when an algorithm begins.
    \item \texttt{OnComputeEnd()}: Extract state or organize algorithmic
    metadata upon completion of an algorithm.
\end{itemize}

\subsection{Code Example} \label{subsec:code}

We provide a high-level implementation of the G-means algorithm
written within \textsf{clusterNOR} to run in parallel on a standalone server.
The simple C++ interface provides an abstraction that encapsulates parallelism,
NUMA-awareness and cache friendliness. This code can be
extended to SEM and distributed memory by simply inheriting from
and implementing the required methods from \texttt{sem} and \texttt{dist}.
The example illustrates how an application that extends the \texttt{hclust}
interface also inherits the properties of \texttt{base}. Critically, users must
explicitly define the \texttt{MMstep} and \texttt{SplitStep} methods that
contain algorithm specific computation instructions.

\begin{lstlisting}[
    basicstyle=\footnotesize\ttfamily,
    language=c++,
    tabsize=2,
    ]
using namespace clusterNOR;

class gmeans : public hclust {
  void MMstep() {
    for (auto& sample : samples()) { // Data iterator
      auto best = min(Euclidean(sample, clusters()));
      JoinCluster(sample, best);
    }
  }

  void SplitStep() override {
    for (auto& sample : samples())
      if (ClusterIsActive(sample))
        AndersonDarlingStatistic(sample);
  }

  void run() override {
    while (nclust() < kmax()) {
      initialize(); // Starting conditions
      MMstep();
      SplitStep();
      Sync(); // Split clusters
      if (SteadyState())
        break; // Splits impossible
    }
  }
 \end{lstlisting}

\end{document}